\documentclass[11pt]{article}
\usepackage[english]{babel}
\usepackage[utf8]{inputenc}
\usepackage{johd}
\usepackage{xcolor}

\usepackage{amsthm,amssymb,amsmath,amsfonts,mathtools}

\usepackage{caption}
\usepackage{subcaption}
\usepackage{setspace}

\newcommand{\Prob}{\mathbb{P}}

\title{Evaluating Meta-Regression Techniques: A Simulation Study on Heterogeneity in Location and Time}

\author{Jonathan Gendron\thanks{Department of Economics, Virginia Tech, USA. ORCID ID: https://orcid.org/0009-0004-7302-4385.\newline \indent Corresponding Author: jegendron@vt.edu\newline} \and Ali Habibnia \thanks{Department of Economics, Virginia Tech, USA. ORCID ID: https://orcid.org/0000-0002-3012-435X  
\smallskip
\newline\newline We would like to extend our sincere thanks and appreciation to Brian D'Orazio, Brianna Felegi, Aris Spanos, Shamar Stewart, and Le Wang for their feedback during the different phases of the project. We would also like to share our appreciation and gratitude for their feedback from the attendees of \href{https://campecon2025.wordpress.com/}{Camp Econometrics XIX} and \href{https://appliedeconometrics.org/}{International Association for Applied Econometrics 2025}.}}

\date{}

\begin{document}
\onehalfspacing
\maketitle
\begin{abstract}
\noindent In this paper, we conduct a simulation study with subject level data to evaluate conventional meta-regression approaches (study-level random, fixed, and mixed effects) against seven methodology specifications new to meta-regressions that control joint heterogeneity in location and time (including a new one that we introduce). We systematically vary heterogeneity levels to assess statistical power, estimator bias and model robustness for each methodology specification. This assessment focuses on three aspects: performance under joint heterogeneity in location and time, the effectiveness of our proposed settings incorporating location fixed effects and study-level fixed effects with a time trend, as well as guidelines for model selection. The results show that jointly modeling heterogeneity when heterogeneity is in both dimensions improves performance compared to modeling only one type of heterogeneity.\newline



\end{abstract}

\\


\textbf{Code availability:} Github \href{https://github.com/jegendron/Evaluating-Meta-Regression-Techniques-A-Simulation-Study-on-Heterogeneity-in-Location-and-Time}{link}\newline\newline


\textbf{Competing Disclosure:} To the best of our knowledge, no competing interests exist \newline

\textbf{Funding Statement:} This research was not funded \newline

\newpage

    %

    %

\section{Introduction}\label{section1}


\noindent A meta-analysis is a useful tool that combines results from independent studies with similar design to examine whether there is consistency in the results across studies or if those results conflict \citep{mcgrath19,kulin18,rice_re-evaluation_2018,kelley_statistical_2012}. Meta-regressions are commonly referred to as meta-analyses (and vice versa), since the only difference introduced by a meta-regression is that it transcends the univariate model of a meta-analysis by including identifiable study and/or subject characteristics as covariates. {Meta-regression data looks very similar to regression data with covariates, a variable of interest, and group-level controls. The only difference is that the estimators are estimating the common effect sizes after controlling for the study-level heterogeneity, instead of a effect size for a specific data set. Data from the different studies under consideration can be at the subject level or aggregate level using the summary statistics of each study; this manuscript will focus on the subject level due to the many issues inherent in aggregate level data \citep{yamaguchi14,cooper_relative_2009,lambert02,clarke95}. Consequently, thousands of meta-analyses have been performed, primarily in the medical, pharmaceutical, ecology, and social sciences literature \citep{borenstein}. The methodology used for meta-analyses is designed to accurately estimate the effect sizes and mitigate any heterogeneity that may be present between each of the studies in the meta-analysis. The standard methodology uses study-level effects: most commonly random effects, fixed effects, and on rare occasions mixed effects \citep{altoe24,stanley23,noma21,richardson21,pateras18,stanley17}. But what about heterogeneity in location or time? It is common for meta-analyses to include decades worth of studies from many locations around the world, which introduces the likely variation in both the time and location in which each study was conducted \citep{thoni_converging_2021,van_den_akker_sex_2020,engel2011,oosterbeek}. In the experimental economic and meta-analysis simulation literature, almost all meta-analyses are conducted at the study level (dominantly using random effects), with only a small percentage controlling for location variation and none controlling for time variation.\footnote{Table \ref{table0} in Appendix A shows a representative sample of meta-regressions in experimental economics and how they control for heterogeneity: only about 30\% control for location heterogeneity, only about 15\% control for precise country-level heterogeneity, and only about 5\% control for time heterogeneity (implicitly).} This manuscript tests three hypotheses when the data follows the standard assumptions:
\begin{enumerate}
    \item When facing joint location and time heterogeneity, specifications that control both location and time will perform better than those that only control for one dimension (i.e., location or time).
    \begin{itemize}
        \item We posit this because ignoring cross-sectional variation will likely induce the omitted variable bias.
    \end{itemize}
    \item Study-level specifications will perform worse than specifications that explicitly control location and time when heterogeneity is at the location and time level.
    \begin{itemize}
        \item We propose this because if studies share a location or year, study-level controls may fail to capture that shared variation.
    \end{itemize}
    \item When location or time heterogeneity is present, even methodology specifications that control that heterogeneity in location or time will perform worse.
    \begin{itemize}
        \item We posit this because even when study-level heterogeneity exists, study-level specifications tend to under perform - consistent with meta-analysis simulation findings (see Section \ref{section2}).\newline
    \end{itemize}
\end{enumerate}



\noindent This manuscript provides three key contributions that are novel for both the meta-regression simulation literature and practitioners in any discipline that use meta-regressions. First, the performance of seven methodology specifications new to meta-regressions is tested in a simulation environment with various degrees of location and time-level heterogeneity. This is very important because meta-regressions are inherently at high risk of joint heterogeneity in location and time, and location heterogeneity as well as joint heterogeneity have not been tested in a simulation environment\footnote{The only exception are spatiotemporal heterogeneity meta-analyses, but these measure this heterogeneity very differently. Please see the literature review in Section \ref{section2} for more details}. Second, performance of two particular methodologies is tested in our simulation, location-level fixed effects from the experimental economic literature, and study-level fixed effects with a trend from the meta-analysis simulation literature. This testing is also important because it will answer the appropriateness of each methodology specification in the face of joint location and time heterogeneity. Third, since thousands of meta-regressions are regularly conducted in a large variety of literature, we have outlined steps for optimal model selection that any practitioner can follow. As the manuscript will show, model selection should be carefully considered depending on the characteristics of the data, especially whether there is joint heterogeneity in location and time or not.\newline
\par

\noindent This research aims to test which meta-regression methodology performs optimally when joint heterogeneity in the location and time between studies in a meta-regression is present in varying degrees. To achieve our research objective, we conduct a simulation study that tests the performance of ten meta-regression methodology specifications that include the standard methodology and the specifications that control for heterogeneity in location and/or time. Note that most of the methodology specifications are fixed-effect specifications since the assumption of random effects is a phenomenon that rarely holds in practice (which is required for random or mixed effect models, please see Section \ref{section5.1} for further details) \citep{clark15,wooldridge_econometric_2010,baltagi2005,hausman81,mundlak78}. In summary, our simulation found that the standard methodology performed the best and the specifications that control for joint heterogeneity performed a close second best. This excludes study-level fixed effects with a trend proposed in the meta-analysis simulation literature because this provided inaccurate trend estimates, but includes a new specification inspired by the literature and introduced by this manuscript: location-level fixed effects with a trend. Accordingly, specifications that did not jointly control for location and time heterogeneity performed the worst (including location-level fixed effects proposed in the experimental economics literature). Note that these results are for joint heterogeneity, in future research we plan to test when only location-level heterogeneity is present. The organization of this paper is as follows: Section \ref{section2} reviews the relevant literature, Section \ref{section3} shares the methodologies of a standard meta-analysis, and Section \ref{section4} shows the additional methodology specifications that control for location and/or time heterogeneity. The simulation in Section \ref{section5} tests the performance of each of the methods in the following way: simulation design, the performance criteria of each specification, and which specification performs optimally. An essential step-by-step practitioners guide is outlined in Section \ref{section6.05}. Section \ref{sectionSim2} shows the utility of example applications as well as how simulations can be extended beyond purely synthetic data by extracting parameters from real data. Lastly, Section \ref{section6} provides concluding remarks and notes on the related areas we plan to study in the future.
\par

\section{Literature Review}\label{section2}


There is dense literature that argues whether meta-analyses and meta-regressions should utilize a study-level fixed effects model or a study-level random effects model \citep{altoe24,noma21,viech10,gordon01,hunter_fixed_2000}. Random effects are the most dominant because their design handles data randomly drawn from a larger population (such as experimental subject data), can capture the sampling error of the participants, and can also capture the study-level variance of the participant sampling for each experiment \citep{kelley_statistical_2012, higgins_re-evaluation_2009, cooper_relative_2009}. However, a random effects model requires that variation in draws of the random effect are uncorrelated with variation in all other explanatory variables, which is rare in practice \citep{clark15,wooldridge_econometric_2010,baltagi2005,hausman81,mundlak78}. {The random effects model should not be used when this assumption is violated, the consequences of this model misspecification can be most easily illustrated in the inflated estimator bias \citep{aert19,spanos2019}. There is also a less common mixed effects model (also known as the generalized linear mixed model). This model combines fixed and random effects to capture both the variation within each group (with fixed effects) as well as the variation across groups (with random effects); mixed effects are commonly used for hierarchical data where data can be categorized into groups across different levels \citep{stijnen19,kulin18,stijnen10,Rviecht10,overton98}. There are important trade-offs to consider for all of these methodology specifications. For instance, one study found an overestimation of standard errors with random effects and an underestimation of standard errors with fixed effects \citep{overton98}. The tradeoffs are compounded for mixed effects since it includes the benefits of both fixed and random effects, but also the limitations of both fixed and random effects. This is related to our research because this is the standard methodology that is tested and compared to our methodology specifications that control location and/or time. To recap, do not indiscriminately use random effects; it has assumptions and limitations of their own, just like fixed effects and mixed effects.\newline
\par


\noindent This literature also covers methodology performance when varying degrees of study-level heterogeneity are present \citep{mcgrath19,kulin18,sanchez08,hedges98,tweedie97}. The common finding is that when higher heterogeneity between studies is present (compared to lower heterogeneity), neither study-level random or fixed effects perform as well. Our research extends this to varying degrees of heterogeneity at the location and time level. This is important since the literature overlooks the possibility of variation in location or time in a meta-analysis (excluding only a select few studies). The only exception are spatiotemporal meta-analyses, but the focus of this field of literature is very different since these use spatial economic models to model the heterogeneity between locations by using the distance between them along with other related covariates \citep{johnson_st_17}. In contrast, when this manuscript describes location heterogeneity, its focus is closer to two way fixed effects in panel data models: distance apart is not a factor, we solely control whether locations are different or not. Note that many of the sources that highlight spatiotemporal heterogeneity is present, don't account for it explicitly in their modeling \citep{jiang24,liu22,ren22}. Recently, some of those studies have explored controlling for location-level or time-level effects more directly than implicitly through study-level effects. In the experimental economic literature, location-level effects were controlled by country, geographic region, or continent \citep{aimone_macro-level_2023,johnson_trust_2011,oosterbeek}. Our research analyzes the performance of each specification and how well their specification captures varying degrees of location-level heterogeneity, neither of these have been done since the aforementioned meta-analyses are only empiric. \newline
\par

\noindent The meta-analysis simulation literature explores time-level effects, but each study controls or measures time differently, so there is no set consensus. Accordingly, our research aims to set a consensus by testing the relevant specifications that control for time heterogeneity. Two studies measure fixed effects with a trend for experiments that take place at different time periods, and another study introduces a meta-analysis methodology called Continuous Time Meta-Analysis (CoTiMA) that includes time lags for both within each study and between the studies inside of a meta-analysis \citep{banks23,rudolph20,dormann20}. Closest to our research is a study that introduces a meta-analysis methodology to model variation over time between the static studies inside the meta-analysis \citep{banks23}. This methodology models variation in time by testing if there are any trends in each static study across time by category. So, only the date in the meta-analysis is dynamic; there are no time lags within each study. But, either time-level fixed effects or fixed effects with a trend capture the variation of time between studies (please see Section \ref{section4} for further details). Both studies used study-level fixed effects with a trend, so we are going to test this specification in the face of joint location and time heterogeneity and compare its performance to other specifications. These other specifications include the standard methodology, location-level fixed effects with a trend, and others that control for location and/or time. Our manuscript is the first to test the following meta-regression specifications together in a simulation: location-level fixed/random effects, time-level fixed effects, and location-level fixed effects with trend. The goal of testing these methods are different from spatiotemporal heterogeneity meta-analyses because they control for the heterogeneity instead of spatially modeling it. The next section reviews the standard methodology in more detail and compares it to the additional methodology specifications in the following section. 
\par

\section{The Standard Meta-Analysis Methodology} \label{section3}
As discussed in the literature review section, a standard meta-analysis uses a study-level model with most commonly random effects, fixed effects, or less commonly mixed effects. This section reviews this standard meta-analysis methodology, while the following section compares this with the additional methodology specifications mentioned in the literature review: location-level fixed/random/mixed effects, time-level fixed effects, and fixed effects with a trend term. Each of these methodology specifications are used in the simulation to see which performs optimally when the meta-regression data being used has joint location and time heterogeneity between the studies inside of it. Before reviewing study-level fixed and random effects models, Equation \ref{EQ0} below depicts the most basic form of a meta-analysis. 

\begin{equation}
    \hat{\theta}_s=\theta_s+\varepsilon_s, \quad y_{s}=\theta+\varepsilon_{s}, \quad \varepsilon_{s} \sim N(0,\sigma_s^2)
    \label{EQ0}
\end{equation}

\noindent Individual studies are indexed by $s=(1,...,S)$, $\hat{\theta_s}$ represents the estimated effect size per study, $\theta_s$ represents the true effect size, and $\varepsilon_s$ represents a random error term per study. Note that the second equality is applied under the fixed effects assumption. Equation \ref{EQ0.1} below shows how a meta-analysis can easily be extended to a meta-regression, and this basic functional form for meta-analyses is shared for both study-level fixed and random effects models.

\begin{equation}
    \hat{\beta}_s=\beta_s+\varepsilon_s, \quad y_{s}=\mathbf{x}_s^\top\boldsymbol{\beta}+\varepsilon_{s}, \quad \varepsilon_{s} \sim N(0,\sigma_s^2)
    \label{EQ0.1}
\end{equation}

\noindent For the above equation $y_s$ represents a ($s \times 1$) vector for the dependent variable, $\mathbf{x}_s^\top=(1,x_{s1},x_{s2},...,,x_{sp})$ represents a ($s \times p$) vector of covariates per study $s$ (which can include control variables but not fixed effect controls), $\boldsymbol{\beta}=(\beta_0,\beta_1,...,\beta_p)$ represents a ($p \times 1$) vector of effect sizes (i.e. regression coefficients), and $\varepsilon_s$ represents a ($s \times 1$) vector for the error term. Note that data used in meta-analyses or meta-regressions can be either at the subject level or aggregate level \citep{altoe24,poom22,yamaguchi14,aert10,viech10,cooper_relative_2009}. This format will be shared for all the following methodology specifications in this section and the following section. Another way to show study-level fixed or random effects models is through their shared parent model called General Linear Mixed-effects Model (GLMM), this model is shown below in Equation \ref{EQ0.2}.

\begin{equation}
y_{s}=\mathbf{x}_s^\top\boldsymbol{\beta}+\boldsymbol{\alpha}_{s,FE}+\boldsymbol{\alpha}_{s,RE}+\varepsilon_{s}, \quad \varepsilon_{s} \sim N(0,\sigma_s^2)    
    \label{EQ0.2}
\end{equation}

$$\mathbf{y}=\mathbf{X}\mathbf{\beta}+\mathbf{u}+\mathbf{v}+\mathbf{\varepsilon}$$

\noindent For GLMM, $y_s$ denotes a ($s \times 1$) vector for the dependent variable, $\mathbf{x}_s^\top=(1,x_{s1},x_{s2},...,,x_{sp})$ representing a ($s \times p$) vector of covariates per study $s$, $\boldsymbol{\beta}=(\beta_0,\beta_1,...,\beta_p)$ representing a ($p \times 1$) vector of regression coefficients (i.e. effect sizes), $\boldsymbol{\alpha}_{s,FE}$ denotes the fixed effects vector, $\boldsymbol{\alpha}_{s,RE}$ denotes the random effects vector, and $\varepsilon_s$ denotes the ($s \times 1$) vector for the error term \citep{viech10}. The fixed effect can be shown as $\boldsymbol{\alpha}_{s,FE}=\gamma_s D_s$ such that $D_s$ is a vector of dummy variables per study and $\gamma_s$ is a vector of fixed effects. The random effect is drawn from $N(0,\sigma^2_{\alpha})$ and the error term $\varepsilon_s$ is $N(0,\sigma^2_{\varepsilon})$ distributed. If $\boldsymbol{\alpha}_{s,FE}$ is removed, the model will be equivalent to a random-effects model, while if $\boldsymbol{\alpha}_{s,RE}$ is removed, the model will be equivalent to a fixed-effects model. The mixed effects model is also discussed in this section when both of the fixed and random effects are present. Note that caution should be used if using the mixed effects model, there is simulation literature that found it performed poorly in certain conditions (like location heterogeneity) compared to fixed effects and/or random effects \citep{burkner23,kulin21,lebeau18,kulin18}. Note that we show both the typical syntax of the meta-regression simulation literature as well as the vector notation commonly used in the econometrics literature. The remaining equations will only show the syntax of the meta-regression simulation literature.\newline
\par

\noindent Appendix F shows the following details for each methodology specification: the meta-regression framework from statistics/econometrics, parameter estimation, and standard error estimation. Note that all methodology specifications are very similar grouped at the study-level, but their differences lie between whether they use fixed, random and mixed effects. For all the methods in Section \ref{section3} \& \ref{section4}, the parameter is estimated with the weighted least square (WLS) estimation strategy and the standard error is estimated with a weighted inverse-variance. WLS is used instead of ordinary least squares (OLS) to provide weights for each group (in this case each study), or the omitted variable bias would likely occur (due to the omitted study-level heterogeneity). This omitted variable bias would then harm performance since it makes the estimator biased \citep{wooldridge_econometric_2010,cameron2005,baltagi2005}. Note that there are some studies in the literature that suggest using WLS as both the estimation strategy and model specification, but the focus of this manuscript is a WLS estimation strategy with various random or fixed effects model specifications \citep{stanley23,aert19,stanley17}. Both approaches are designed to address heterogeneity of meta-analysis data sets, the typical form of the WLS estimator is shown below in Equation \ref{EQ1.5}.

\begin{equation}
    \hat{\boldsymbol{\beta}} = (\mathbf{X}^\top \mathbf{W} \mathbf{X})^{-1} \mathbf{X}^\top \mathbf{W} \mathbf{y}
    \label{EQ1.5}
\end{equation}

\noindent For WLS estimators, $\mathbf{X}$ is the design matrix of covariates ($s \times p$), $\mathbf{W}$ is the diagonal of a $(s \times s)$ matrix $\mathbf{W} = \text{diag}(w_1, w_2, \dots, w_s)$, and $\mathbf{y} = (y_1, y_2, \dots, y_s)^\top$ is a $(s \times 1)$ vector. The weights in the $\mathbf{W}$ matrix were calculated with the inverse-variance matrix as $w_s=\frac{1}{\sigma^2_s}$. The weighted inverse-variance matrix is commonly used in meta-analyses to provide weights for each group, giving higher weights to groups that have higher precision to minimize variance of estimates \citep{lee_comparison_2016}.\newline
\par





\noindent Overall, this section has presented the standard methodology used for conducting a meta-analysis. Equation \ref{EQ0} shows the general form of a meta-analysis while Equation \ref{EQ0.1} shows how to extend a meta-analysis into a meta-regression. WLS and a weighted inverse-variance matrix are used to account for the weight of each grouping for all methodology specifications. Most commonly random effects are used in the literature, but fixed or mixed effects can also be used. This decision is not trivial, fixed and random effects are designed differently, so they have different assumptions, which means that the inferences each model can make are different. Note that if the random effects assumption does not hold, then random effects and mixed effects should not be used. The following section reviews the additional specifications for handling variation in location and time. In the simulation, these additional specifications will be compared with the standard methodology that was reviewed in this section. 
\par



\section{Additional Meta-Analysis Methodologies: Location \& Time}\label{section4}

In the following subsections, we briefly discuss methodology specifications that control for location and/or time variation: location-level random/fixed/mixed effects are presented in Section \ref{section4.1}, time-level fixed effects are presented in Section \ref{section4.2}, combined location and time-level fixed effects are presented in Section \ref{section4.3}, and specifications that model time with a trend combined with study-level and location-level fixed effects are presented in Section \ref{section4.4}. Location and study-level fixed effects with a trend term have been introduced in the literature, and the other specifications have been added to ensure that all relevant variations of the methodology specification are tested. These effects help to account for differences across groups (i.e. geographic locations) or across time periods that may otherwise bias the estimates of the treatment effect. By including these effects, we can obtain more accurate and unbiased estimates of the relationships between covariates and effect sizes by accounting for heterogeneity with precision beyond study-level. Note that there are only a few random and mixed effect specifications since our simulation is designed so the random effects assumption does not hold (since it is rare in practice). All of these methodologies will be compared to the standard meta-regression methodology, and then all of these methods will be tested in the simulation in the next section with data that have joint location and time heterogeneity between the studies in a meta-regression. The performance criteria in Section \ref{section5.2} will dictate which methodology specification performs best in each situation. Hypothesis 1 will be tested by comparing two joint heterogeneity methodology specifications with two non-joint heterogeneity methodology specifications. Hypothesis 2 will be tested by comparing the standard method to the joint heterogeneity methodology specifications. Hypothesis 3 will be tested with all of the non-standard specifications. Appendix G shows the following details for each methodology specification in this section: the meta-regression framework from statistics/econometrics, parameter estimation, and standard error estimation.\newline
\par 



\subsection{Location-level Effects: Fixed, Random, and Mixed Effects}\label{section4.1}

When key factors of a group of studies vary within each location, a location-level effects model can be used to control the variation between locations. This operates nearly in the same way as the study-level effect model does with the same estimation strategy and underlying assumptions/constraints (please see Section \ref{section3} for further details). The only difference is the subscripts that are now at the location-level, $l$, instead of at the study-level, $s$. An important note is that a location-level model should be used instead of a study-level model when there is specific heterogeneity in locations. The above description applies to fixed effects, random effects, or mixed effects. Note that random effects and mixed effects specifications are only appropriate when the random effects assumption holds and subjects are randomly drawn or represent the full population. In contrast, only fixed effects would be appropriate if the random effects assumption does not hold. Note that the variation between locations is omitted, but within-location variation is retained. Also, note that homogeneity in time is still assumed if these location-level fixed effects are not combined with time controls (either a trend or time-level fixed effect).\newline
\par

\subsection{Time-level Fixed Effects}\label{section4.2}

When key factors in a group of studies change over time, a fixed effect model at the time level can be used to control the variation between different time periods. This could be due to factors such as technological advancements, economic conditions, or changes in standards of care over time. This operates almost in the same way as fixed effects at the study level using the same estimation strategy and underlying assumptions/constraints (see Appendix G for further details), but Equation \ref{EQ11} below shows a few slight differences. 

\begin{equation}
y_{t}=\mathbf{x}_t^\top\boldsymbol{\beta}+\boldsymbol{\alpha}_{t,FE}+\varepsilon_{t}, \quad \varepsilon_{t} \sim N(0,\sigma_t^2)
    \label{EQ11}
\end{equation}

\noindent The only difference is the subscripts that are now at the time-level, $t$, instead of at the study-level, $s$. Each time period is indexed by $t=1,...,T$, $y_t$ is a $(t \times 1)$ vector for the variable of interest, $\mathbf{x}_t^\top=(1,x_{t1},x_{t2},...,,x_{tp})$ represents a $(t \times p)$ vector of covariates per time period $t$, $\boldsymbol{\beta}=(\beta_0,\beta_1,...,\beta_p)$ represents a $(p \times 1)$ vector of effect sizes (i.e. regression coefficients), $\varepsilon_t$ is a $(t \times 1)$ vector for the error term. Note that $\boldsymbol{\alpha}_{t,FE}$ denotes the fixed effect where $\boldsymbol{\alpha}_{t,FE}=\gamma_t D_t$ such that $D_t$ is a vector of dummy variables per time period and $\gamma_t$ is a vector of fixed effects for time period $t$ (which represents the difference in effect size between each time period $t$).\newline
\par 

\noindent This specification would be most appropriate when heterogeneity is present at only the time level. A limitation to note is that variation across time cannot be accounted for because comparisons are made within each year but not between. Also, if a time-level fixed effect is not combined with a location-level fixed effect, this model then assumes homogeneity in locations, and that homogeneity would need to be tested. Similar to Section \ref{section4.1}, the only difference between the standard methodology used to conduct a meta-analysis and the time-level methodology is the respective use of study-level grouping versus time-level grouping. This holds for both fixed or random effects, and the only difference in the math notation is the $s$ subscripts are changed with $t$ subscripts. An important note is that a time-level model should be used instead of a study-level model when there is specific heterogeneity in time periods.\newline
\par

\subsection{Combined Location and Time-level Fixed Effects}\label{section4.3}


Location and time-level fixed effects, if used separately, do not account for the heterogeneity in the other dimension. So, the simulation of this paper will also test this new specification of combined location and time-level fixed effects. A typical meta-analysis pools as many relevant studies available, so it is important to address heterogeneity jointly in the location and time of each study. This is especially important if the meta-analysis includes decades of data from many countries like those in the experimental economics literature \citep{thoni_converging_2021,van_den_akker_sex_2020,engel2011,oosterbeek}. Equation \ref{EQ14} below shows how the new specification combines the benefits of location and time level fixed effects to account for both forms of heterogeneity by combining the fixed effects from Section \ref{section4.1} \& \ref{section4.2}. 

\begin{equation}
y_i = \mathbf{x}_i^\top\boldsymbol{\beta} + \sum_{l=1}^{L-1} \delta_l d_{il} + \sum_{t=1}^{T-1} \theta_t d_{it} + \varepsilon_i
\label{EQ14}
\end{equation}

\noindent Each location is indexed by $l=1,2,...,L,$ each time period is indexed by $t=1,2,...,T$, and each observation is indexed by $i=1,2,...,(L\cdot T)$. For example, if there are five locations and each location has one study per year for five years, then the $i$ index will be $i=1,2,...,25$. This $i$ index is important to note since each observation includes both location and time but should not be indexed as panel data since each observation is only observed in one particular time period. $y_{i}$ is a $(i \times 1)$ vector for the variable of interest, $\mathbf{x}_i^\top=(1,x_{1},x_{2},...,,x_{p})$ representing a $(i \times p)$ vector of covariates per observation, $\boldsymbol{\beta}=(\beta_0,\beta_1,...,\beta_p)$ representing a $(p \times 1)$ vector of effect sizes (i.e. regression coefficients), and $\varepsilon_i$ is a $(i \times 1)$ vector for the error term. Note that fixed effects are shown differently to match the $i$ indexing but are equivalent to the previous $\mathbf{\alpha}_{l,FE}$ and $\mathbf{\alpha}_{t,FE}$ syntax. The location-level fixed effect is shown as $\sum_{l=1}^{L-1} \delta_l d_{il}$ where $d_{il}$ is the dummy variable for location $l$ and $\delta_l$ is the fixed effect for location $l$. Lastly, the time-level fixed effect is shown as $\sum_{t=1}^{T-1} \theta_t d_{it}$ where $d_{it}$ is the dummy variable for time period $t$ and $\theta_t$ is the fixed effect for time period $t$.\newline
\par 

\noindent Note that when both location and time-level fixed effects are combined, multicollinearity needs to be tested since any location may have a direct link to a specific time. For instance, what if there is only one study in the meta-analysis that was conducted in Antarctica? What if this same study was the only one conducted in 2024? In this hypothetical example, the dummy variable for 2024 is perfectly collinear with Antarctica. Also note that this methodology specification has the same underlying assumptions/constraints of study-level fixed effects in Appendix G since it also uses fixed effects.\newline
\par



\subsection{Modeling Trend with Fixed Effects: Study or Location Level}\label{section4.4}

In addition to controlling for time heterogeneity with time-level fixed effects, a trend term could instead be used. Not only to control time heterogeneity but also to model it. The most appropriate circumstance to use this specification would be when heterogeneity is present jointly at the location and time level, and you want to estimate the overall time effect. For example, if there is a world-wide trend over time as well as additional differences between countries, a location-level fixed effects model with a trend methodology specification would be the most appropriate. The study-level with a trend specification has been introduced in the meta-analysis simulation literature, this inspired the location-level with a trend specification newly introduced by this manuscript. This was introduced since we suspect that the study-level with a trend methodology specification controls for time heterogeneity twice which could likely lead to inaccuracies in said specification. \newline
\par

\noindent Either methodology specification (study or location level), is achieved by adding a rescaled trend polynomial term $t_0=\frac{2t-n-1}{n-1}$ that rescales to $[-1,1]$. The number of time periods is represented by $n$ and each step of a time period is represented by $t=\{1,2,...,\}$. This monotonic transformation is used to combat the near-multicollinearity likely present in trend polynomials, especially when high degrees are included \citep{spanos2019b,seber03,spanos2002}. Trend polynomials have also been found to be unstable due to an ill-conditioned matrix in estimation unless they are rescaled \citep{spanos2020,seber03}. This trend, $t_0$, is simply one of the covariates, so it is nested inside the same $x_i$ for study-level fixed effects in Appendix G and Section \ref{section4.4} for location-level fixed effects. To illustrate this trend polynomial transformation suppose your yearly data spans 2020-2024. The values for $t_0$ will accordingly be $\{-1,-0.5,0,0.5,1\}$ by using the following values of $t=\{1,2,3,4,5\}$. Now the magnitude is removed so the only factor considered are the time intervals themselves.
\par



\section{Simulation}\label{section5}

To reveal which meta-regression methodology performs optimally when there is joint location and time level heterogeneity between studies in a meta-regression, a standard Monte Carlo simulation is performed. This simulation generates subject-level meta-regression data and then models it with the standard methodology (study-level random, fixed and mixed effects) as well as seven other specification variations designed to control location or time heterogeneity (including the proposed location-level fixed effects and study-level fixed effects with a trend methodology specifications). The remaining specifications were designed to ensure all relevant methodology specification variations are tested. For reference, the following abbreviations were used: $RE_s$ (study-level random effects), $FE_s$ (study-level fixed effects), $ME_s$ (study-level mixed effects), $FE_{lt}$ (two-way fixed effects), $FE_t$ (time-level fixed effects), $FE_l$ (location-level fixed effects), $RE_l$ (location-level random effects), $ME_l$ (location-level mixed effects), $FE_{s,Trend}$ (study-level fixed effects with a trend), \& $FE_{l,Trend}$ (location-level fixed effects with a trend). In the simulation, a variety of parameters are tested, including the sample size in each study, the number of covariates in each study, the size of location and time effects, the number of locations, and the number of studies. These parameters signal whether particular cases or variations affect the performance of each meta-regression methodology being compared. Each of the subsections within review the mechanics of this simulation including further details on the simulation design, what the simulation performance criteria are, and then show the results. The performance criteria specify which methodology specification performed best. Figure \ref{flowchart} below illustrates the mechanics of this simulation.\newline
\par

\subsection{Design}\label{section5.1}

\noindent The standard Monte Carlo simulation is conducted with 10,000 iterations. For each Monte Carlo iteration, the meta-regression data has the following parameters as specified in Table \ref{table1}: the sample size in each study (50, 100 \& 150), the number of covariates (i.e. independent variables) in each study (1, 3 \& 5), the number of locations (5, 9, 15), and the number of studies (25, 45, 75) in the meta-regression. Each location in the meta-regression include 1 study per year for 5 years, which results in either 25, 45, or 75 studies due to the number of locations (5, 9, or 15). Excluding the number of locations, these are common parameters tested in the literature with values varying depending on the research question or the measure that is being tested \citep{richardson21,noma21,dormann20,kulin21,kulin18,field01}. Note that the meta-regression data does not have any overlap across studies, which means that no data from the same study is double counted in the meta-regression. This is the most common case when conducting a meta-regression \citep{bom20,johnson20,lin09}. Practitioners need to check if there is overlap or not though because if there is, this could introduce confounding factors in the model because of a duplicate study in the meta-regression. Also note that no data points are missing in the meta-regression data. This is unlikely in practice and will be explored in future research. {To better visualize this type of dataset, suppose there are five time periods for five locations, say 2020-2024. This would mean that there are $25$ studies total in the meta-regression since $L=5$ and $T=5$. Since each study includes subject-level data, the compiled meta-regression data set has sample size $N=25n$ where n is the number of subjects in the study. For instance, if $n=100$, $N=2,500$.\newline
\par 

\begin{figure}[htp!]
    \centering
    \includegraphics[width=16cm]{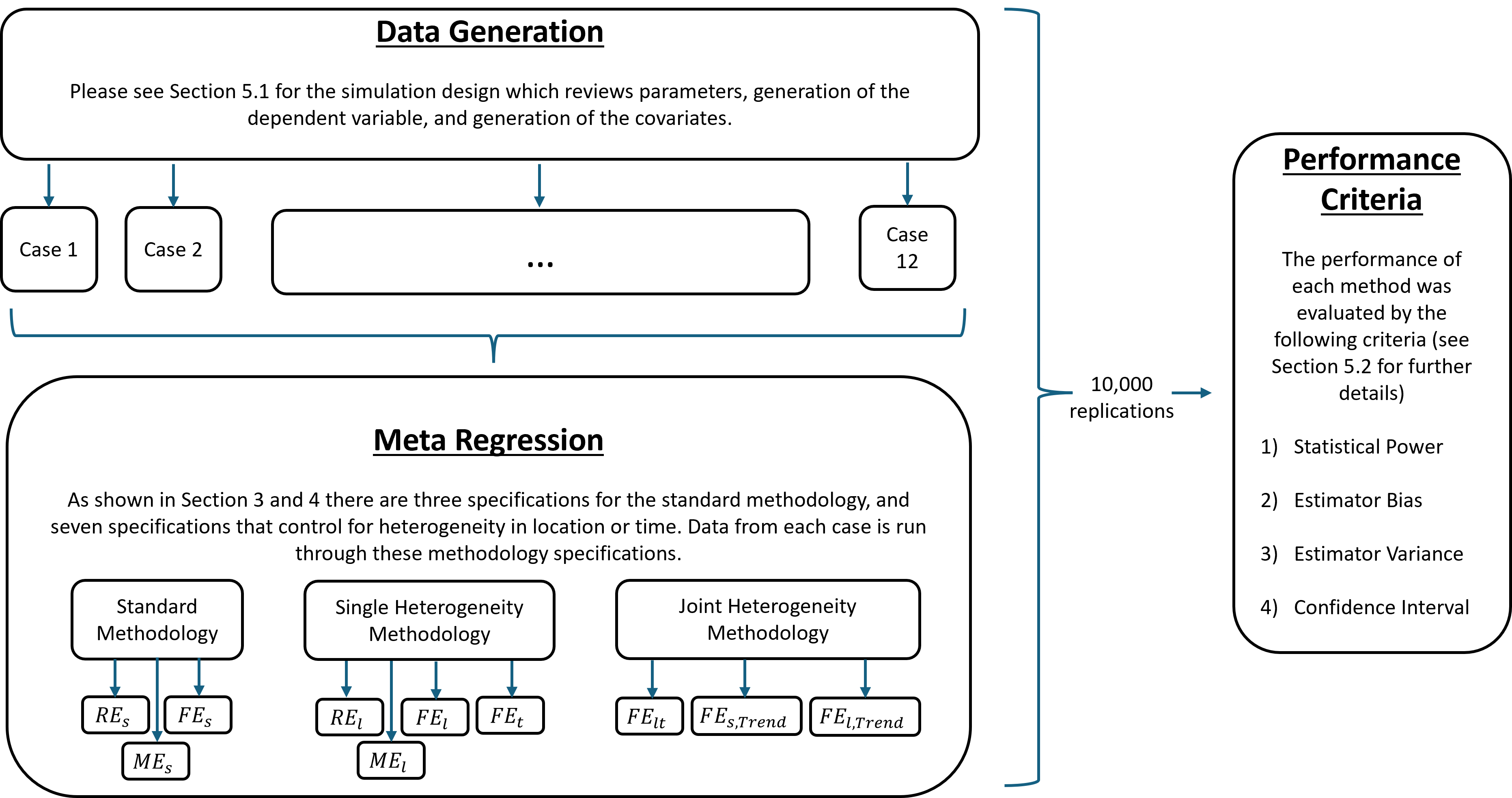}
    \caption{Flowchart of the Simulation: This flowchart shows detailed steps of how the simulation was conducted (from data generation to the performance criteria)}
    \label{flowchart}
\end{figure}

\noindent Each parameter always has a range of values to test a variety of situations, and the ranges of values we selected are very commonly used in the literature. The purpose of these parameters is to capture if variation in each of these parameters impacts the performance of each method. To simulate location and time level heterogeneity, our simulation includes twelve cases that embed varying degrees of location and time level heterogeneity into the mean $\mu_Y$ when generating the simulated data for the dependent variable $Y$. Table \ref{table2} shows the differences between all cases, which are as follows. Cases 1-6 include a small time effect by adding 0.1 to $\mu_Y$ each year, while Cases 7-12 include a large time effect by adding 0.5 to $\mu_Y$ each year. This means that 0.1 (or 0.5) is added once to $\mu_Y$ in the second year, but added a total of four times in the fifth year, resulting in 0.4 (or 2.0) added to $\mu_Y$ over the five years. Cases 1-2 \& 7-8 include 5 locations, Cases 3-4 \& 9-10 include 9 locations, while the remaining cases include 15 locations. The odd numbered cases include a small location effect by assigning each location with a different $\mu_Y$; 5 locations are assigned one element each from $\mu_Y=\{-2,-1,0,1,2\}$, 9 locations are assigned one element each from $\mu_Y=\{-2,-1.5,-1,-0.5,0,0.5,1,1.5,2\}$, and 15 locations are assigned one element from $\mu_Y=\{0,\pm(0.15, 0.33, 0.66, 1, 1.33, 1.66, 2)\}$. In contrast, the even-numbered cases include a large location effect by assigning each location with a different $\mu_Y$; 5 locations are assigned one element from $\mu_Y=\{-10,-5,0,5,10\}$, 9 locations are assigned one element from $\mu_Y=\{-10,-7.5,-5,-2.5,0,2.5,5,7.5,10\}$, and 15 locations are assigned one element from $\mu_Y=\{0,\pm(1, 2.5, 4, 5.5, 7, 8.5, 10)\}$. Note that the $\mu_Y$ values are not based on real data, they are solely illustrative of small versus large differences in the means of the dependent variable in each location. The large versus small differences in the mean of each location simulate large versus small effects of location-level heterogeneity. Similarly, the smaller versus larger trend added to the mean simulates weaker versus more powerful effects of time level heterogeneity. Note that both trends are designed to be positive, constant over time, and at the same rate for all locations. These differences in generating the dependent variable test how well each of the methodology specifications performs in the face of varying degrees of both location and time level heterogeneity. Section \ref{section5.2} defines specific model performance criteria. \newline
\par

            


\noindent When generating the covariates for each of the studies in the meta-regression, we generate them as Normal, Independent, and Identically-Distributed with a pre-specified variance-covariance matrix. By generating data with a specified variance-covariance matrix, the true parameter estimates can be calculated (which is needed to test the accuracy of each model's predictions), and our covariates will be statistically significant ensuring that our estimators have enough power \citep{noma21,dormann20,spanos2001,field01}. Note that the Normal distribution was chosen since most software packages analyze random effects and fixed effects models by assuming the error term or random effect are normally distributed to be able to hypothesis test and generate confidence intervals via the central limit theorem \citep{pateras18,aert10,wooldridge_econometric_2010,sanchez08,cameron2005}. The literature has demonstrated that estimators that assume a normal distribution (i.e. majority of estimators) perform very poorly when the data are non-normally distributed, even when the sample size is large \citep{botella23,mcgrath23,poudyal_model_2022,rubio18,spanos2010,bohnet2007}. The mean of each covariate is assigned a value close to the range of $[0,1]$ following the suggested design from the literature \citep{botella23,richardson21,mcgrath19,pateras18,sanchez08}. Variances of 1 are also suggested by the literature for simplicity. These parameters can be linked to real meta-regression data sets if they are substituted with extracted parameters from said data. Most studies extract parameters from one particular meta-regression to ensure the parameters are consistent as much as possible \citep{schwarzer24,dormann20,aert10}. To analyze beyond the sample, one study extracted parameters from 12,894 meta-analyses in the Cochrane Database of Systematic Reviews \citep{langan18}. Figure \ref{figure0} in Appendix B presents the summary statistics for all variables (five covariates) as box and whisker plots. This is done to visualize the data generation process. Since $\mu_Y$ varies per study, each study in the meta-regression has a different true intercept coefficient, $\beta_0$, calculated by Equation \ref{EQ17} below.

\begin{equation}
    \beta_0 = \mu_Y-\left(\begin{array}{ccc} \beta_1 & \cdots & \beta_n \end{array}\right)\left(\begin{array}{c} \mu_{X_1} \\ \vdots \\ \mu_{X_n} \end{array}\right)
    \label{EQ17}\newline
\end{equation}

\noindent In the above calculation, $\mu_{X_i}$ are the means of each covariate and $\beta_i$ are the true coefficients when $i = (1,2,...,n)$. On the other hand, the true covariate coefficients and variance do not vary per study in the meta-regression since they are calculated from the set variance-covariance matrix as seen in Equation \ref{EQ18} below. 

\begin{equation}
    \left(\begin{array}{c} \beta_1 \\ \vdots \\ \beta_n \end{array}\right) =  \left(\begin{array}{ccc} \sigma_{22} & \cdots & \sigma_{n2} \\ \vdots & \ddots & \vdots \\ \sigma_{2n} & ... & \sigma_{nn} \end{array}\right)^{-1}\left(\begin{array}{c} \sigma_{12} \\ \vdots \\ \sigma_{1n} \end{array}\right)
    \label{EQ18}
\end{equation}

$$\sigma^2=\sigma_{11}-\left(\begin{array}{ccc} \sigma_{12} & \cdots & \sigma_{1n} \end{array}\right)\left(\begin{array}{ccc} \sigma_{22} & \cdots & \sigma_{n2} \\ \vdots & \ddots & \vdots \\ \sigma_{2n} & ... & \sigma_{nn} \end{array}\right)^{-1}\left(\begin{array}{c} \sigma_{12} \\ \vdots \\ \sigma_{1n} \end{array}\right)$$
\par

\noindent Variance is represented by $\sigma_{ii}$ when $i=(1,2,...,n)$, while covariance is represented by $\sigma_{ij}$ when $i=(1,2,...,n)$ $\&$ $j=(1,2,...,n)$ $\forall i\neq j$. Note that each element of the covariance matrix is within [0.2,0.5], so the random effects assumption of 0 correlation is not designed to hold.\newline
\par

\noindent To recap the design of this standard Monte Carlo simulation, Figure \ref{flowchart} shows detailed steps of how the simulation was conducted starting from data generation and ending in how the results are evaluated with the performance criteria. Table \ref{table1} shows the parameters used when generating the meta-regression data, and Table \ref{table2} shows how small or large heterogeneity in location or time are generated through the mean of the variable of interest ($\mu_Y$) when generating the data. when generating the meta-regression data we generate our variables using a Normal, Independent and Identically-Distributed joint distribution with a pre-specified variance-covariance. This allows us to calculate the true parameter value which is needed to test the accuracy of each model's predictions, and since our covariates will be statistically significant our estimators will have appropriate power. The mean and variance values are respectively within $[0.1]$ and 1 as suggested by the literature, but note that these mean and variance parameters can be linked to real meta-regression data sets if they are substituted with extracted parameters from the real data sets \citep{botella23,richardson21,mcgrath19,pateras18,sanchez08}. The following subsections show in detail how the true value of the coefficients are calculated for when there are one, three, or five covariates.\newline
\par

\subsubsection{Bivariate Data Generation Process - 1 Covariate}\label{section5.1.1}

One of the simulation parameters is the number of covariates, which changes the variance-covariance matrix and changes the true parameter values. Equation \ref{EQ19} below shows both the joint distribution specified for the bivariate case and then how the true parameter values (coefficient $\beta$ and variance $\sigma^2$) were calculated from the said joint distribution. $NIID$ is an abbreviation for a Normal, Independent, and Identically Distributed distribution, which specifies the mean of each variable at the left, and the variance-covariance matrix at the right. The diagonal elements of the variance-covariance matrix are the variances of each variable, while the remaining elements are the covariances between variables. Since this is the data-generating process for each study in the meta-regression, these true parameters do not change at the meta-regression level. In the multivariate instances below, all of the details mentioned above will be the same: details of the data-generating process, details of the joint distribution, and how to read the elements of the joint distribution. \newline
\par

\begin{equation}
    \left(\begin{array}{c} Y \\ X_1\end{array}\right) \sim NIID \left(\left(\begin{array}{c} \mu_Y \\ 1 \end{array}\right),\left(\begin{array}{cc} 1 & 0.5 \\ 0.5 & 1 \end{array}\right)\right)
    \label{EQ19}
\end{equation}

$$\beta_1=0.5=(1)^{-1}(0.5), \quad \quad \sigma^2=0.75=1-(0.5)(1)^{-1}(0.5)$$

\subsubsection{Multivariate Data Generation Process - 3 or 5 Covariates}\label{section5.1.2}

For the multivariate covariate cases (three or five), the variance-covariance matrix changes compared to Section \ref{section5.1.1}, so the true parameter values also change. Equation \ref{EQ20} shows the updated joint distribution for three covariates, Equation \ref{EQ21} shows the updated joint distribution for five covariates, and both equations show how this changes the values of the true parameters. Both equations can be found in Appendix C. The details of the data-generating process, details of the joint distribution, and the reading of the elements of the joint distribution below are detailed in Section \ref{section5.1.1} above.\newline
\par

\subsection{Performance Criteria}\label{section5.2}

In the literature on meta-analysis simulations, there is variation on which criteria are considered to be evidence of optimal model performance, typically due to the particular measure or method that is the focus of the research. This study includes the following as performance criteria that dictate which of the tested methods performed optimally in the following hierarchy: an optimal balance of power (i.e., neither under-powered nor over-powered), estimator bias, the precision of variance, and the precision of confidence intervals. Typically, confidence intervals are only included when new confidence interval estimators are being developed, but we include this since it can capture precision of the estimate \citep{pateras18,langan18,aert19,noma2019,richardson21}. {Although precision of variance is not typically included in the performance criteria of {meta-analysis simulation literature, we include it in our performance criteria since mean square error (MSE) is regularly used to either measure bias, or complement another bias measure \citep{richardson21,mcgrath19,langan18,stanley17}. Variance and bias are both elements of MSE via $MSE()=Var()+Bias()^2$, but we disentangle them from MSE to assess both the estimator bias as well as the variance in the estimates \citep{botella23}. Note that if the estimator is unbiased, then $MSE\approx Var()$ since $Bias\approx0$. Goodness of fit measures are rarely used in the meta-analysis simulation literature, the only exception found in our literature review was one seminal study that used Akaike Information Criteria (AIC) \citep{stanley23}. We accordingly carefully considered whether it might be appropriate for our simulation or not, but do not include any goodness of fit measures for the following reasons. First, $R^2$ is designed differently for random and fixed effects, and the way $R^2$ is measured is different as well. When testing our base cases, we observed stark differences in $R^2$, even when all of our other performance criteria were identical between random and fixed effects). To add further complications, $\bar{R}^2$ would need to be used instead since the number of covariates is one of our parameters that vary. Second, the simulation literature does not typically consider information criteria since it does not explicitly capture model fit, it only captures whether more variables should be added to the model or not by adding a penalty to each variable added \citep{noma21,pateras18,akaike}. Its consistency has also been disproven in simulations \citep{spanosAIC2010}. \newline
\par

\noindent Our performance criteria is made up of four measures assessed in a particular order. Power is assessed first because if an estimator has insufficient power, it is very likely estimator bias will increase and predictive accuracy will decrease \citep{altoe24,schwarzer24,spanos2019b,pateras18,wooldridge_econometric_2010}. When reviewing our results we saw exactly this; when power was poor, all of the other performance criteria uniformly performed substantially worse. This is because if an estimator has insufficient power to detect discrepancies from the null hypothesis, then it is more likely to be biased. This bias harms the accuracy of coefficient estimates, which accordingly harms the confidence intervals, model robustness, and predictive accuracy. Suppose, however, that power is strong for all methodology specifications, then estimator bias would be assessed next to determine which specification provides the most accurate effect estimation. {This is commonly used in the meta-regression simulation literature, but there are variations on how this is measured, so we define the measure we use in Section \ref{section5.2.2}. If both power and biases of the estimator are equal, then estimator variance would be assessed next. Lastly, if all of the above performance criterion are equal, then confidence intervals are compared to assess the precision of the estimates \citep{richardson21,aert19,noma2019,pateras18,langan18}. \newline
\par

\subsubsection{Statistical Power}\label{section5.2.1}
\noindent The most dominant performance criterion for this simulation is an optimal balance of statistical power (i.e. neither underpowered nor overpowered). Power to detect discrepancies from the null hypothesis (also known as 1-Type II error) is a common performance criterion in the meta-analysis simulation literature, it is typically measured as the percent of iterations with statistically significant results in the simulation literature \citep{altoe24,schwarzer24,pateras18}. To measure power more comprehensively, this study calculated power by finding the area under the non-central t distribution to the left of the Type I error \citep{cameron2005}. In addition, the figures in this manuscript show the power to detect a range of discrepancies from the null hypothesis via the x-axis \citep{lehmann05}. This goes further than typical power measures since it ensures the effects of small and large discrepancies from the null hypothesis on the estimator's power can be precisely captured (i.e. small and large effect sizes respectively) \citep{dorazio,stock20,spanos2019,greene18}. To calculate the power to measure varying discrepancies from the null hypothesis, a series of alternate hypotheses $\beta_{H_A}^{(i)}=\beta+0.1\cdot i$ for $i=(1,2,...,100)$ are used. Then, the non centrality parameter $\delta^{(i)}$ is calculated for each alternate hypothesis, which is then used to calculate the power of the estimator for each alternate hypothesis as follows in the below equation:

\begin{equation}
    Power^{(i)} = 1 - \Prob(t_{df,\delta^{(i)}}<c) \quad s.t. \quad \delta^{(i)}=\frac{|\beta-\beta_{H_A}^{(i)}|}{\hat{\sigma}}    
\end{equation}

\noindent The test statistic is represented as $t$, the critical value is denoted as $c$, and $\hat{\sigma}$ represents the estimators standard error. The estimators power increases when the discrepancy from the null hypothesis increases and the power decreases when the discrepancy decreases. The Type I error was adjusted when the sample size of the simulated meta-regression was increased by either a larger sample size or more countries. This was done to combat the Large N problem which has been shown to overpower models when the sample size is too large, which can lead to erroneous inferences due to erroneous results that appear statistically significant \citep{spanos2024,myors23}. Testing the power of each methodology specification with a variety of parameters shows the capacity of each methodology tested in the simulation to detect discrepancies from the null, and how each parameter can affect this capacity. For instance, if one of the ten methodology specifications has much lower power than the others when there is a large time effect ceterus paribus, then it can be inferred that one methodology does not have as much power when large time-level heterogeneity is present.\newline
\par

\subsubsection{Estimator Bias}\label{section5.2.2}

\noindent If power is equal between methodologies, the next performance criteria to consider is the estimator bias of the covariates. Estimator bias is captured by which methodology specification provides the smallest discrepancy from the true values. This measure is the most dominant performance criteria in the simulation literature because it measures how close the estimates are to actual values, but is measured in various ways in the literature. We accordingly use the following measure $Bias=\frac{1}{N}\sum_{i=1}^N\hat{\beta}_i-\beta$ from the literature where $N$ is the total number of iterations in the Monte Carlo simulation, but we additionally use $Bias_i=\hat{\beta_i}-\beta$ to capture variation in each iteration within the simulation \citep{botella23,viech14}. The combinations of parameters and the varying levels of location and time heterogeneity specified in Section \ref{section5.1} allow us to see how accurate the estimates of the coefficients are for each methodology tested in the simulation. For instance, if one of the ten specifications has a much higher bias than the others when there is a large location effect ceteris paribus, then it can be inferred that one methodology does not handle large magnitudes of location-level heterogeneity well when estimating the covariate. Note that this is not the same thing as mean squared error (MSE) not only because the penalty is squared, but also because it captures both bias and variance in one combined measure (see Appendix H for further details on MSE and similar measures). \newline
\par

\subsubsection{Estimator Variance}\label{section5.2.3}

When both power and estimator bias are equal between methodologies, the next performance criteria to consider is the variance of the estimator. This is calculated as $Var=\sigma^2=(\sqrt{n}\cdot \hat{\sigma})^2$ from the equation $\hat{\sigma}=\frac{\sigma}{\sqrt{n}}$ \citep{aert23,wooldridge23,water17,viech14,higgins_re-evaluation_2009}. An ideal estimator will have minimal variance, which means that there will be minimal variation in the values of the estimate. This is important to capture since an estimator with minimal variance is indicative of stability and consistency of estimated values \citep{wooldridge_econometric_2010,cameron2005,baltagi2005}. In fact, estimates can be highly distorted with large variance even if their estimator is unbiased. This is especially important for data sets with smaller sample sizes. Note that variance is independent of accuracy of the estimate, it is possible for an estimate to not be accurate (i.e. biased) but have a small variance or for an estimate to be accurate (i.e. unbiased) but have a large variance. Also note that this is not necessarily identical to MSE because MSE captures both bias and variance in one combined measure.\newline
\par

\noindent To illustrate how this performance criteria will be used in assessing our simulation results, suppose one of the ten specifications has a much larger variance than the others when there is a large time effect ceteris paribus. It can then be inferred that one methodology does not handle large magnitudes of time-level heterogeneity well since the variation in the estimate is higher. This inference is valid even if the estimators bias is not different, which highlights why measuring bias and variance separately provides more information than MSE. In the meta-analysis simulation literature, there are measures for between-study variance, but since most of the methodology specifications use fixed effects rather than random effects, we do not use them. Instead, the standard variance measure shown above is employed since group level heterogeneity is controlled for in each methodology specification. \newline
\par



\subsubsection{Confidence Interval}\label{section5.2.4}
If power, estimator bias and variance are equal between methodology specifications, the next performance criteria to consider are confidence intervals. This has been extensively covered in the meta-analysis simulation literature with the focus of capturing estimate precision \citep{richardson21,aert19,noma2019,pateras18,langan18}. The primary focus is to ensure the coefficient estimate is within the estimated confidence interval, and then the focus is on the confidence interval having the smallest spread possible (i.e. highest precision). For instance, even if all methodology specifications include the true value in their coefficient estimate, a method that has the smallest range in its confidence interval would be considered best (since it is the most precise estimate).\newline
\par

\subsubsection{Effects of Parameters on Performance Criteria}\label{section5.2.5}
\par

\noindent This subsection will review the common effects that variations in the simulation parameters have on each of the performance criteria according to the meta-analysis simulation literature. The magnitude of power increases when sample size is larger with all other parameters equal \citep{botella23,richardson21,kulin21,noma2019,spanos2019b,wooldridge_econometric_2010}. This increase in power due to sample size increasing also decreases estimator bias and variance \citep{altoe24,schwarzer24,spanos2019b,pateras18,wooldridge_econometric_2010}. In comparison, the magnitude of power decreases when sample size is small with all other parameters equal which increase estimator bias and variance. In the simulation design, the number of studies and locations are linked to the sample size since the number of studies is driven by the number of locations, and when there are more locations the sample size of the meta-regression is larger. The magnitude of power increases with the number of covariates being higher with all other parameters equal \citep{stock20,greene18,lehmann05}. This increase in power from more covariates decreases the estimator bias and variance \citep{altoe24,schwarzer24,spanos2019b,pateras18,wooldridge_econometric_2010}. In contrast, when the number of covariates is small, the magnitude of power decreases and estimator bias increases and variance . To recap, the effects of parameters on our performance criteria have been well documented in the meta-analysis simulation literature, these can be seen in Table \ref{table3} below.\newline
\par

\subsection{Results}\label{section5.3}

\noindent Figures \ref{resultFigure1} - \ref{resultFigure2.5} in Appendix D show performance criteria from a sample result to illustrate what happens when more covariates are added by comparing one versus five covariates (ceteris paribus). It can be seen for all cases and methodology specifications that each specification performs slightly better when it has more covariates. Figure \ref{resultFigure1} reveals that power does slightly increase, and Figures \ref{resultFigure2} \& \ref{resultFigure2.5} reveal that although the estimator bias does not change, the estimator variance does uniformly lower for all specifications with more covariates. Note that the two specifications that model time heterogeneity with a trend term add value by quantifying the overall trend over time (rather than being simply controlled). Figure \ref{resultFigure3} in Appendix D shows no difference between the one and five covariate case. Note that the estimates for the covariate coefficients are accurate for all methodology specifications for each case, highlighting why solely coefficient accuracy is an insufficient performance criterion.\newline
\par

\noindent To summarize our overall simulation results beyond the number of covariates, the standard methodology performed the best and the specifications that control joint heterogeneity performed the second best. This includes the $FE_{l,Trend}$ specification we introduced in this manuscript, but excludes $FE_{s,Trend}$ proposed in the meta-analysis simulation literature since it consistently provided biased trend estimates. The worst-performing specifications were those that did not jointly control for location and time heterogeneity (including location-level fixed effects proposed in the experimental economics literature). Note that for all cases the mean estimator was found to be unbiased, but at the Monte Carlo iteration level there were small levels of positive of negative bias. Accordingly, this means that on average $MSE\approx Var$, and increases/decreases in estimator bias discussed in this section are relative. Further details are broken down into the following subsections: Section \ref{section5.3.1} covers the methodology specifications that are best controlled for time heterogeneity, as well as the impact of a small time effect (compared to a large time effect). Section \ref{section5.3.2} covers the impact of a small location effect (compared to a large location effect), while Section \ref{section5.3.3} covers the impact when the number of locations differ. Lastly, Section \ref{section5.3.4} briefly summarizes the results and highlights the key takeaways.\newline
\par




\subsubsection{Impact of Smaller versus Larger Time Effect}\label{section5.3.1}

\noindent Between all methodology specifications, the standard methodology performed best when faced with a large time effect, but when faced with a small time effect $FE_{l,Trend}$ and $FE_{lt}$ performed just as well as the standard methodology. Figures \ref{resultFigure4} - \ref{resultFigure5.5} in Appendix D show performance criteria from a sample result that illustrates this point. Figure \ref{resultFigure4} in Appendix D shows that when time heterogeneity is larger, power is substantially worse for the location-level methodology specifications $ME_l$, $RE_l$ \& $FE_l$, power is equally poor for $FE_t$ and power remains high and unchanged for the remaining specifications. The estimator bias, variance and MSE are accordingly worse for the location-level specifications as seen in Figures \ref{resultFigure5} \& \ref{resultFigure5.5} in Appendix D. Figure \ref{resultFigure6} also in Appendix D shows that in the face of higher time heterogeneity, the bias in the biased trend estimate for $FE_{s,Trend}$ remains unchanged while the bias in the unbiased trend estimate for $FE_{l,Trend}$ slightly increases. Precision is unchanged for both specifications. These results support the first hypothesis since $ME_l$, $RE_l$, $FE_l$ \& $FE_t$ did not perform as well as the other specifications that jointly control for time and location variation. This also means that the results do not support the second hypothesis. Note that when time effects are large, the $FE_{l,Trend}$ \& $FE_{lt}$ specifications performed second best. Time heterogeneity was controlled for with $FE_t$ \& $FE_{lt}$, and measured with $FE_{s,Trend}$ \& $FE_{l,Trend}$. The first two control for time with fixed effects, while the last two both control and capture the effect of any time trends shared by all locations in the meta-regression. Immediately, the results of $FE_t$ show that time-level fixed effects should not be used for data similar to the simulated data since this methodology has substantially less power compared to all the other methodologies tested for the twelve cases. Consequently, the estimator bias, variance, and MSE perform substantially worse. These criteria perform worse when large time effects were present, the performance especially suffered for $FE_t$ \& $FE_{lt}$. Note that the estimates for the covariate coefficients are accurate for all methodology specifications for each case, highlighting why solely coefficient accuracy is an insufficient performance criterion.\newline
\par


\noindent The trend for the two methodology specifications with a trend ($FE_{s,Trend}$ \& $FE_{l,Trend}$) have a rescaled trend term of $[-1,-0.5,0,0.5,1]$ when $t=5$ \& $n=5$ (where $t$ is the number of time periods and $n$ is the number of locations). When $n=9$ \& $t=5$, the rescaled trend term is $[-1,-0.75,-0.5,-0.25,0]$, while $n=15$ \& $t=5$ the rescaled trend term is $[-1,-0.86,-0.71,-0.57,-0.43]$. The trends' sample size is so small because this simulation includes 5, 9, or 15 locations, and each location has 5 studies over 5 years. The two specifications with this trend ($FE_{s,Trend}$ \& $FE_{l,Trend}$) both perform very well in terms of power. However, the key difference is that $FE_{s,Trend}$ provides very inaccurate estimates for the trend coefficient, while $FE_{l,Trend}$ has almost perfect accuracy for the trend coefficient. Figure \ref{resultFigure6} in Appendix D shows a sample result that illustrates exactly this point. Only when there is a large time effect is the power slightly worse for $FE_{l,Trend}$ (compared to a small time effect). This causes the estimate of the trend coefficient to be slightly worse (but still far better than $FE_{s,Trend}$). Note that all specifications perform slightly worse when the time effect is large, supporting the third hypothesis. When large time effects are present, the performance also suffers in the same way for $RE_l$ \& $FE_l$. Note that the estimator bias is very small for all cases and methodology specifications, but note that the estimator for $FE_t$ has a much larger variance.\newline
\par



\subsubsection{Impact of Smaller versus Larger Location Effect}\label{section5.3.2}

When the location effect is large and the number of locations is held constant, all methodology specifications performed just as well (excluding $FE_t$ which faced worse power, and $FE_{s,Trend}$ where the trend estimate worsened). Note that these specifications performed poorly in Sections \ref{section5.3} and \ref{section5.3.1} though, so these differences are trivial. Figures \ref{resultFigure10}, \ref{resultFigure11} \& \ref{resultFigure11.5} in Appendix D illustrate this result since power, estimator bias, and estimator variance are unchanged. Note that Figure \ref{resultFigure12} in the same Appendix D shows that in the face of higher location heterogeneity, the already biased trend estimate for $FE_{s,Trend}$ explodes while the unbiased trend estimate for $FE_{l,Trend}$ remains unbiased and its precision increases. If there are no confounding factors, this means that no matter how large the differences are between each location in the studies within meta-regression data, both joint and non-joint location-level heterogeneity methodology specifications do well in controlling location-level heterogeneity (excluding the specifications $FE_t$ \& $FE_{s,Trend}$). Note that the third hypothesis contradicts with this result based on the common finding in the meta-analysis simulation literature. Further analysis is needed to ensure there are no confounding factors and that the scale of location heterogeneity is not too small.\newline
\par



\subsubsection{Impact of Number of Locations}\label{section5.3.3}

Unexpectedly, in the face of more locations non-joint methodology specifications performed just as well on average as joint methodology specifications. The only specifications that had issues in performance were $FE_t$ due to power and $FE_{s,Trend}$ due to bias in the trend estimator. Power is shown in Figure \ref{resultFigure7} in Appendix D, when there are more locations power for all methodology specifications raises a bit higher. Figure \ref{resultFigure9} in the same in Appendix D shows that when there are more locations, the already biased trend estimate for $FE_{s,Trend}$ remains unchanged in both bias and precision while bias and precision slightly increase for the unbiased trend estimate for $FE_{l,Trend}$. Sample size is likely the key to performance improvements since increasing from 5 to 9 countries almost doubles the sample size, and increasing from 5 to 15 countries triples the sample size (since we design each location to have an identical sample size). Note that the estimates for the covariate coefficients are accurate for all methodology specifications for each case, highlighting why solely coefficient accuracy is an insufficient performance criterion. The only specification that faced accuracy problems was the estimator for $FE_t$, which although it had very low bias it also had a much larger variance. If there are no confounding factors, this means that the more diverse the studies are within a meta-analysis, the stronger the estimators will be (as long as the joint heterogeneity is appropriately accounted for in the methodology specification).\newline
\par










\subsubsection{Summary of Results}\label{section5.3.4}


\noindent Since there are many parameters and cases to capture all the nuances of meta-regression data sets, we look at each parameter and case ceteris paribus. First, Section \ref{section5.3} shows that all methodology specifications perform better when more covariates are present (only excluding $RE_l$). This is a common result in the meta-analysis simulation literature, and was found by the uniform increase in power of each covariate. Second, Section \ref{section5.3.1} shows that in the face of a large time effect the standard methodology performed best, second best were the $FE_{lt}$ and the newly posed $FE_{l,Trend}$ joint methodologies, and lastly the worst performing were the remaining non-joint methodologies. This supports the first hypothesis. 
Note that the accuracy of trend estimates lower but precision increases. 
Third, Section \ref{section5.3.2} shows that in the face of a large location effect, joint and non-joint specifications perform just as well (excluding $FE_t$ and $FE_{s,Trend}$. This does not support the first or third hypotheses, unlike all other scopes reported in the subsections of Section \ref{section5.3} (i.e. number of covariates, time effects, and number of locations). Lastly, Section \ref{section5.3.3} shows two differences when the number of locations change. One difference is that all methodology specifications have more power, and the variance of the estimator is smaller. This means that any model will perform better when more locations are present, likely because the sample size of the respective meta-regression will be larger. 
Another difference is that for $FE_{l,Trend}$, the variance of the trend estimate improves, but the precision worsens. The precision only worsens marginally in this simulation, but caution should be used regardless when there are additional locations in a meta-regression. 
These two differences support the third hypothesis. Tables \ref{table4} \& \ref{table5} in Appendix E summarize how increases or decreases in each type of heterogeneity can affect performance criteria as increases, decreases or little to no change. Note that these tables solely indicate how well each methodology specification handles larger heterogeneity compared to smaller heterogeneity, these should not be used as indicators of methodology performance. Overall, in all of these scopes we can see the standard methodology performed best, and any of the joint methodology specifications performed a very close second best (including $FE_{lt}$ and the newly introduced $FE_{l,Trend}$, but excluding $FE_{s,Trend}$ due to poor trend estimates). This also supports the first hypothesis, but does not support the second hypothesis. Note that the estimates for the covariate coefficients are accurate for all methodology specifications for each case, highlighting why solely coefficient accuracy is an insufficient performance criterion.
\par




\section{Practitioners Guide for Model Selection}\label{section6.05}

Before selecting a model, there are three essential things a practitioner needs to do. First, review the meta-regression data to ensure it is ready to model. For instance, there should be no data overlap across studies. If there is, disentangle it, but if there is no possible way to disentangle it, then consider the approach in the literature \citep{bom20}. In addition, if there are any extreme values from rare events, these need to be handled with either controlling for these outliers or other processes described in the literature \citep{gunhan20,thom20}. Also, if there is any missing data, then that needs to be handled in whichever way is most appropriate for your data. The most common ways to handle missing data are to either drop any rows with missing values, exclude variables that have too many missing values, or impute missing values using a variety of methods \citep{lee23,rubin19,buuren18,enders10,graham09}. Note that each approach to handling missing data has unique tradeoffs \footnote{Dropping any rows with missing values ensures potential explanatory power is not lost in the model, but there is also potentially large losses in data. On the other hand, excluding variables with missing values assures data loss will not be as large, but potential explanatory power will be lost in the model. As for imputing missing values this ensures data and explanatory power will not be lost, but there is the risk that the imputed data is not accurate.}. Second, test for heterogeneity in location as well as time, even if one does not suspect the presence of this in the data. Before conducting statistical tests, you can first perform a quick visual test on your data as follows. Location heterogeneity can be checked by calculating the summary statistics of the data by each location to assess if there are differences (or not). Time heterogeneity can be checked by comparing graphs over time per location to determine if any noticeable trends exist. To formally support this, there are statistical tests that can also be conducted to detect heterogeneity including a $\chi^2$ test, Breusch–Pagan test, or $I^2$ test specific to meta-analyses \citep{wooldridge_econometric_2010,aert10,borenstein,cameron2005,baltagi2005}. These tests need be used with care though due to their key limitations such as the vague interpretation of $I^2$ values, and the Breusch-Pagan test only captures linear heterogeneity for normally distributed errors. Also note that this manuscript and its recommendations only cover joint location and time heterogeneity. So, if your data does not match this, you will likely need to use a different model to better match the heterogeneity present in your data. Third, confirm your data is $NIID$. The meta-analysis simulation literature has shown that when data is not $NIID$ there are issues with the accuracy of estimation, which invalidate any inferences that are made \citep{mcgrath23,botella23,mcgrath19,aert10,spanos2001}. Once both are set, then we would recommend either $RE_s$, $FE_s$, or the newly posed $FE_{l,Trend}$. The data will dictate which of these three methodology specifications are most appropriate. For instance, does the random effects assumption hold? If not, that immediately excludes $RE_s$ \& $ME_s$. Also, do you want to estimate the overall time effect instead of just controlling for it? If so, then only the newly introduced $FE_{l,Trend}$ would be appropriate. The next subsection discusses the detailed plans for future research.
\par

\begin{figure}[htp!]
    \centering
    \includegraphics[width=10cm]{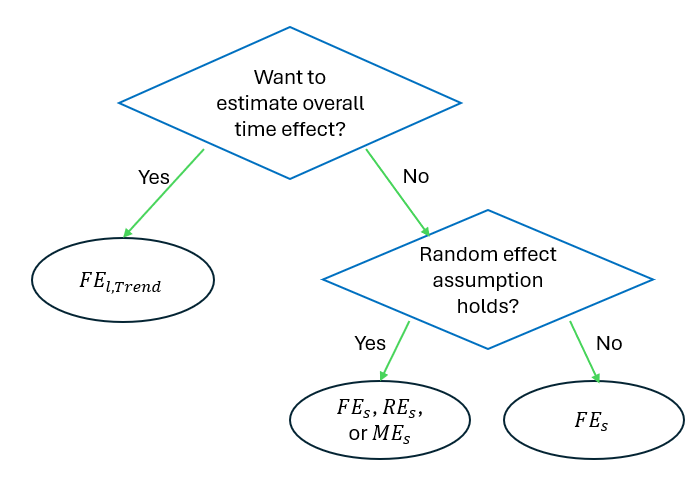}
    \caption{Basic Model Selection Decision Framework: This decision tree shows the basic algorithm on how practitioners can easily decide on which model would be best to use in the face of joint location and time heterogeneity.}
    \label{MSDF}
\end{figure}

\section{Applications \& Parameter Extraction for Simulations}\label{sectionSim2}
As mentioned in Section \ref{section5.1}, these parameters can be linked to real meta-regression data sets if they are substituted with extracted parameters from said data. Most studies in the meta-analysis simulation literature extract parameters from one particular meta-regression to ensure the consistency of parameters \citep{schwarzer24,dormann20,aert10}. Before discussing parameter extraction, this section will review where these meta-regressions can be applied. Meta-regressions are most commonly used to analyze the consistency of medical laboratory study results, but they are also regularly used in many other fields of study as well. Section \ref{section2} showed that meta-regressions are used in experimental economics data sets to evaluate the consistency of behavioral laboratory study results. Note that data in this field typically observe variations in both the number of locations and the number of static time periods per study. Meta-regressions have also been used to analyze the consistency of various types of financial data including market data, banking data, and corporate data \citep{guo23,ta20,goldberg18}. Note that this body literature uses data that observe variations in both the number of locations and the number of time periods. If there are studies that have multiple time periods within, these need to be handled with a different methodology specification that additionally accounts for time series within each study. The most notable meta-analysis in the finance literature is a collaboration between the central banks of 17 countries, where each country has time-series data \citep{goldberg18}. Subfields of economics that measure willingness to pay and contingent valuation (such as natural resource and environmental economics) also use meta-regressions to assess the consistency of these measures across similar studies \citep{afesorgbor22,vesco20,wehkamp18,nelson09}. \newline 
\par




\noindent There are steps any practitioner can follow to extract parameters from their data, in order to conduct simulations to assess which methodology specification would perform best for their data. With the extracted means, variances, and covariances, we can then generate synthetic data following the same simulation design described in Section \ref{section5.1.2}. Note that this assumes that the data is normally distributed and independent, if either are not true estimators will perform very poorly as discussed in Section \ref{section5.1} \citep{botella23,mcgrath23,poudyal_model_2022,rubio18,spanos2010,bohnet2007}. The data can then be modeled using a Monte Carlo simulation using the same performance criteria defined in Section \ref{section5.2}, but now we also have the option to compare the simulation results with the findings of the meta-analysis. This provides value in two ways: first, this can also be used to assess the replicability of results using linked simulation data. Second, the performance of the initial methodology specification can be compared with others to dictate which specification is optimal for the data in question (as illustrated in the previous sections simulation). Note that before doing either of these, a baseline needs to be established by ensuring consistency between model results from the real data and the linked simulation data. This extension can be applied with ease to simulation conducted in Section \ref{section5}, and both of the location and time effects could also easily be calibrated.
\par

\begin{enumerate}
    \item Ensure that the way each study records and/or measures a variable are the same between all of the studies in the meta-regression.
    \begin{itemize}
        \item In the sample data for instance, the dependent variable is recorded the same way in each study, but some of the studies measure it in a slightly different scale. 
        \begin{itemize}
            \item In this example, the raw values can be adjusted to a percentage.
        \end{itemize}
    \end{itemize}
    \item Make sure that the meta-regression data is ready to model.
    \begin{itemize}
        \item There should be no data overlap across studies.
            \begin{itemize}
                \item If so, disentangle it, and less preferably if this cannot be disentangled, then consider the approach in the literature \citep{bom20}.
            \end{itemize}
        \item Any missing data needs to be handled in whichever way is most appropriate \citep{lee23,rubin19,buuren18,enders10,graham09}.
        \item Any extreme values from rare events need to be handled in whichever way is most appropriate \citep{gunhan20,thom20}.
    \end{itemize}
    \item Extract the mean and variance of each variable.
    \item Estimate the covariances between all variables, including the dependent variable $y$
    \begin{itemize}
        \item This step is optional but highly encouraged to generate synthetic data as close to the real data as possible.
        \item $Cov(x,y)=\frac{\sum((x_i-\bar{x})(\sum((y_i-\bar{y}))}{n-1}$\newline
    \end{itemize}
\end{enumerate}


\noindent To recap, this section provides a step by step example of how to extract parameters from a real meta-regression dataset and showcases a variety of fields of literature that conduct meta-regressions. First, check that each variable is measured the same in each study to ensure additional confounding factors are not introduced into the data. Second, on a similar note, the data needs to be ready to model, which means that there should be no duplicates and that missing data needs to be handled appropriately. Third, now that the data ready for modeling, extract the mean and variance of each variable. Lastly, an optional but highly encouraged step is to estimate the covariances between all variables. This way, the synthetic data is as close to the real data as possible. With these steps done, synthetic data can then be generated by following the same simulation design shown in Section \ref{section5.1.2}. There are two benefits to generating synthetic data based on the parameters of a real dataset: replicability of results can be tested and which methodology specification is optimal for the data in question can be determined by the performance of each specification.
\par

\section{Conclusion}\label{section6}

The purpose of this simulation study is to determine which meta-regression methodology performs optimally when joint location and time heterogeneity are present between studies within a meta-regression. This provides three contributions that are essential for both the meta-regression simulation literature, and practitioners in any discipline that use meta-regressions. First, the standard methodology only controls for heterogeneity at the study level, so we test the performance of seven methodology specifications new to meta-regressions in the face of varying degrees of location and time level heterogeneity. Second, the efficacy of the newly proposed location-level fixed effects in the experimental economics literature and study-level fixed effects with a trend specifications in the meta-regression simulation literature are tested. Third, steps have been outlined for any practitioner to select the optimal model for their data. This is essential to ensure that the selected model matches the type of heterogeneity in the data. Our simulation embeds both location and time heterogeneity in the dependent variable and tests various parameter combinations (sample size, number of covariates, number of locations and magnitude of location and time effects). We evaluate ten methodology specifications, including the standard study-level random, fixed, and mixed effects, using performance criteria such as statistical power, estimator bias, estimator variance and confidence interval precision.\newline
\par

\noindent Results show that the standard methodology performed best, closely followed by methodologies which jointly control the location and time heterogeneity. Specifications that controlled for only one type of heterogeneity performed worse, including the location-level fixed effects proposed in experimental economics. Study-level fixed effects with a trend produced inaccurate trend estimates, while time-level fixed effects suffered from poor power. Across all methodology specifications, performance improved with more covariates and more locations, likely due to increased sample size. Importantly, when time effects were small, most specifications performed equally well, but when time or location effects were large, performance declined (especially for location-level fixed and random effects). These findings emphasize the importance of selecting models depending on the characteristics of the data, instead of selecting study-level random effects by default.\newline
\par 

\noindent There are two key areas of future study currently being developed to further this research. First, we plan to test the performance of specifications that control only location when only location heterogeneity is present. In the face of joint heterogeneity, the present manuscript found concerns with the location fixed effects methodology specification posed in the experimental economics literature, so this will test if this methodology is appropriate when there is only heterogeneity in location. Second, we plan to test the ten meta-regression methods we tested with machine learning approaches to both perform a robustness check and indicate potential improvements for each method. In the meta-analysis literature, machine learning has only been used to optimize the collection of studies used in a meta-analysis, thus expanding the use to model selection for analysis will be a novel contribution.\newline
\par

\newpage
\clearpage 

\bibliographystyle{johd}
\bibliography{bib}
\newpage

\section*{Appendix -Tables}

\begin{table}[hbt!] 
    \centering
    \caption{Simulation Parameters: The following parameters are part of a typical meta-analysis simulation}
    \begin{tabular}{|c|c|c|}
        \hline
        Parameter & Symbol & Values\\
        \hline
        Sample size (per study in meta-regression) & $n$ & 50, 100, 150\\
        Number of covariates & $k$ & 1, 3, 5\\
        Number of locations (in meta-regression) & $n_l$ & 5, 9, 15\\
        Number of studies (in meta-regression) & $n_s$ & 25, 45, 75\\
        \hline
    \end{tabular}
    \label{table1}
\end{table}

\begin{table}[hbt!] 
    \centering
    \caption{Simulation Parameters for Location \& Time Heterogeneity: The following parameters correspond to small versus large time effects (Case 1 to 6 versus 7 to 12) and location effects (odd versus even case numbers)}
    \begin{tabular}{|c|c|c|l|}
        \hline
            Case & Locations & Time Effect & Location Effect \\
            \hline
            1 & 5 & $\mu_Y=\mu_Y+0.1\cdot t$& $\mu_Y = \{-2,-1,0,1,2\}$\\
            2 & 5 & $\mu_Y=\mu_Y+0.1\cdot t$& $\mu_Y = \{-10,-5,0,5,10\}$\\
            \hline
            3 & 9 & $\mu_Y=\mu_Y+0.1\cdot t$& $\mu_Y = \{-2,-1.5,-1,0.5,0,0.5,1,1.5,2\}$\\
            4 & 9 & $\mu_Y=\mu_Y+0.1\cdot t$& $\mu_Y = \{-10,-7.5,-5,-2.5,0,2.5,5,7.5,10\}$\\
            \hline
            5 & 15 & $\mu_Y=\mu_Y+0.1\cdot t$& $\mu_Y = \{
            0,\pm(0.15, 0.33, 0.66, 1, 1.33, 1.66, 2)\}$\\
            6 & 15 & $\mu_Y=\mu_Y+0.1\cdot t$& $\mu_Y = \{0,\pm(1, 2.5, 4, 5.5, 7, 8.5, 10)\}$\\
            \hline
            
            7 & 5 & $\mu_Y=\mu_Y+0.5\cdot t$& $\mu_Y = \{-2,-1,0,1,2\}$\\
            8 & 5 & $\mu_Y=\mu_Y+0.5\cdot t$& $\mu_Y = \{-10,-5,0,5,10\}$\\
            \hline
            9 & 9 & $\mu_Y=\mu_Y+0.5\cdot t$& $\mu_Y = \{-2,-1.5,-1,0.5,0,0.5,1,1.5,2\}$\\
            10 & 9 & $\mu_Y=\mu_Y+0.5\cdot t$& $\mu_Y = \{-10,-7.5,-5,-2.5,0,2.5,5,7.5,10\}$\\
            \hline
            11 & 15 & $\mu_Y=\mu_Y+0.5\cdot t$& $\mu_Y = \{0,\pm(0.15, 0.33, 0.66, 1, 1.33, 1.66, 2)\}$\\
            12 & 15 & $\mu_Y=\mu_Y+0.5\cdot t$& $\mu_Y = \{0,\pm(1, 2.5, 4, 5.5, 7, 8.5, 10)\}$\\
            \hline
    \end{tabular}
    \label{table2}
\end{table}

\begin{table}[hbt!] 
    \centering
    \caption{Effect of Parameters on Our Performance Criteria: This table summarizes how increases or decreases in the sample size and/or the number of covariates change our performance criteria in the meta-analysis simulation literature.}
    \begin{tabular}{|r|c|c|c|}
        \hline
        & Power & Estimator Bias & Estimator Variance\\
        \hline
        $\uparrow n$ & $\uparrow$ & $\downarrow$ & $\downarrow$\\
        \hline
        $\downarrow n$ & $\downarrow$ & $\uparrow$ & $\uparrow$\\
        \hline
        $\uparrow$ Covariates &  $\uparrow$ & $\downarrow$ & $\downarrow$\\
        \hline
        $\downarrow$ Covariates & $\downarrow$ & $\uparrow$ & $\uparrow$\\
        \hline
    \end{tabular}
    \label{table3}
\end{table}

\newpage

\section*{Appendix A - Survey of Meta-Analyses}\label{appA}

\begin{table}[hbt!] 
    \centering
    \caption{Survey of Meta-Analyses in Experimental Economics: The following shows if meta-analyses in experimental economics control for heterogeneity in time and location (or not).} 
    \begin{tabular}{|c|c|c|c|c|l|}
        \hline
        & Range & Time & Number of & Location & \\ 
        Model & of Years & Controls? & Countries & Controls? & Paper\\ \hline
        $FE_l$ & 21 & Y (implicitly) & 50 & Y & \citep{aimone_macro-level_2023}\\ \hline
        $FE_l$ & 16 & N & 35 & Y & \citep{johnson_trust_2011}\\ \hline
        $FE_l$ & 21 & N & 25 & Y & \citep{oosterbeek}\\ \hline
        $RE_s$ & 23 & N & 31 & Y (indirectly) & \citep{lane}\\ \hline
        $RE_s$ & 6 & N & 12 & Y (continent) & \citep{owens}\\ \hline
        $RE_s$ & 11 & N & 46 & N (unused) & \citep{abeler2019}\\ \hline
        $RE_s$ & 17 & N & 32 & N & \citep{engel2011}\\ \hline
        $RE_s$ & 52 & N & 19 & N & \citep{fiala}\\ \hline
        $RE_s$ & 46 & N & 6 & N & \citep{mengel}\\ \hline
        $RE_s$ & 21 & N & 24 & N & \citep{thoni_converging_2021}\\ \hline
        $RE_s$ & 16 & N & 24 & N & \citep{van_den_akker_sex_2020}\\ \hline
        $RE_s$ & 20 & N & 17 & N & \citep{bilen_are_2021}\\ \hline
        
        $RE_s$ & 19 & N & 7 & N & (Acuff et al., 2020) \\ \hline
        $RE_s$ & 14 & N & 12 & N & \citep{alm}\\ \hline
        $RE_s$ & 16 & N & 10 & N & \citep{tracy}\\ \hline

        $OLS$ & 22 & N & 12 & N & \citep{irsova}\\ \hline
        $WLS$ & 18 & N & 6 & N & \citep{zelmer}\\ \hline   
        \textbf{Avg} & 21.12 & n/a & 21.65 & n/a & n/a \\ \hline
    \end{tabular}
    \label{table0}
\end{table}

\newpage

\section*{Appendix B - Summary Statistics}\label{appB}

\begin{figure}[htp!]
    \centering
    \includegraphics[width=5cm]{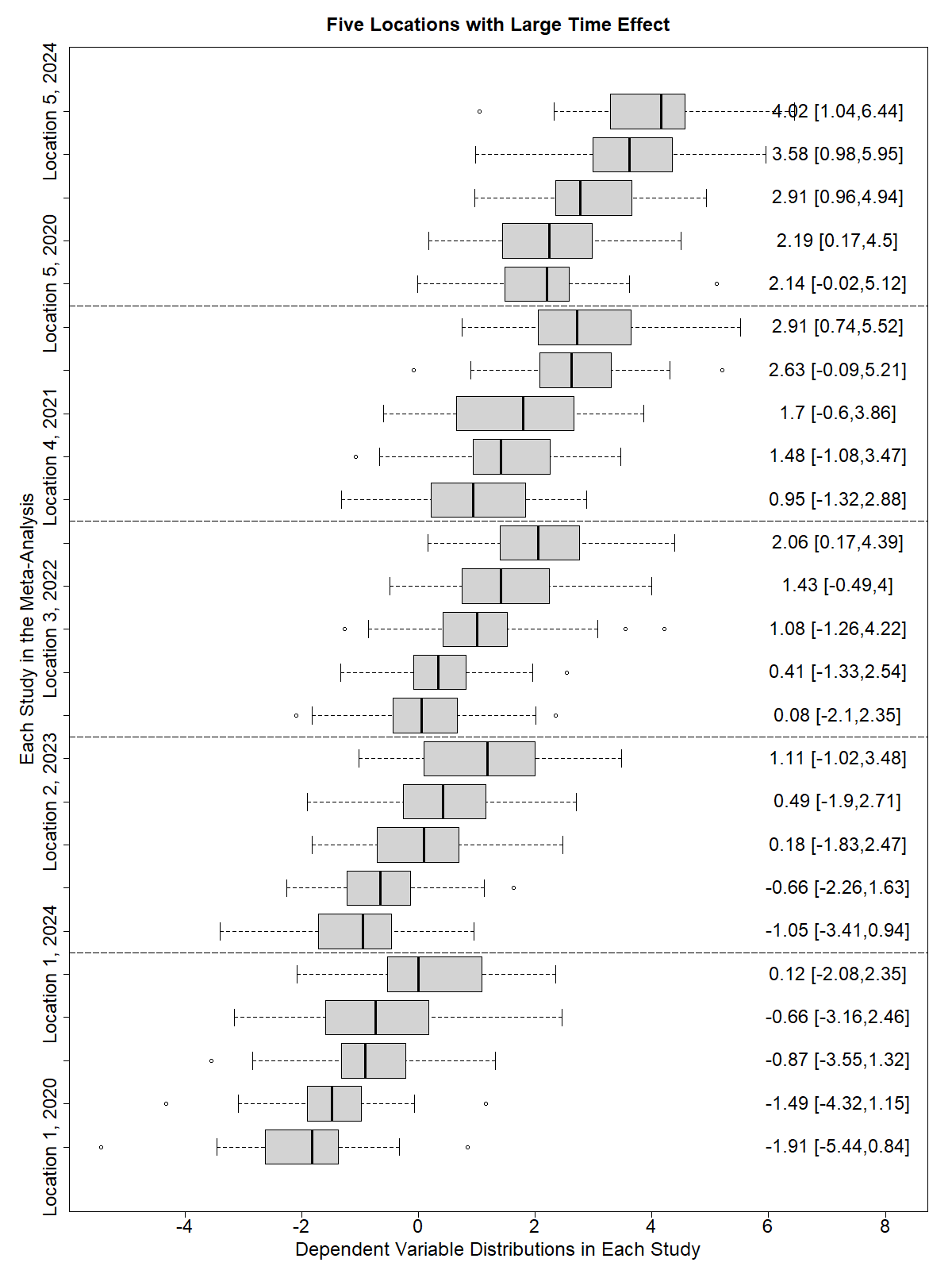}
    \includegraphics[width=5cm]{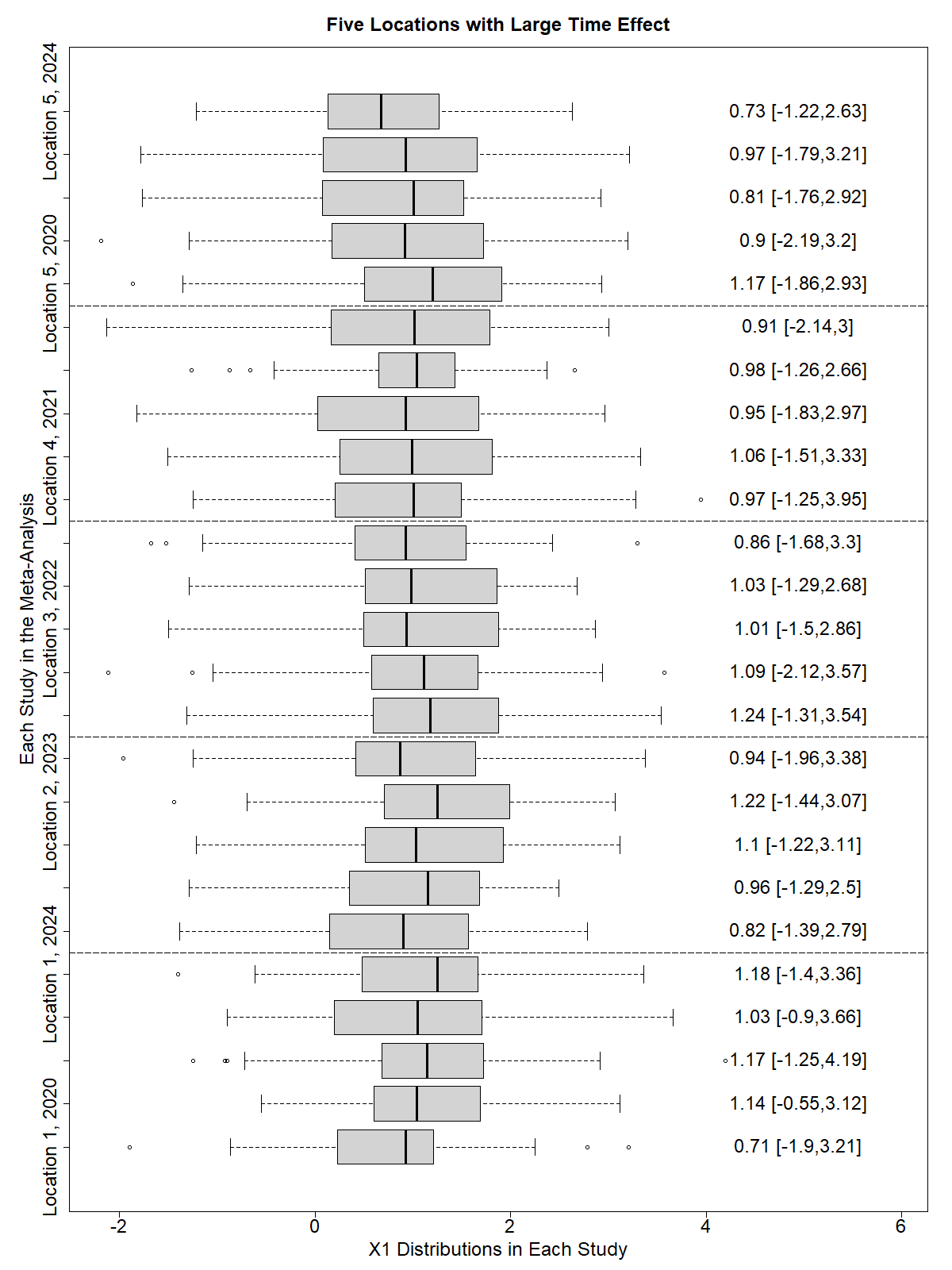}
    \includegraphics[width=5cm]{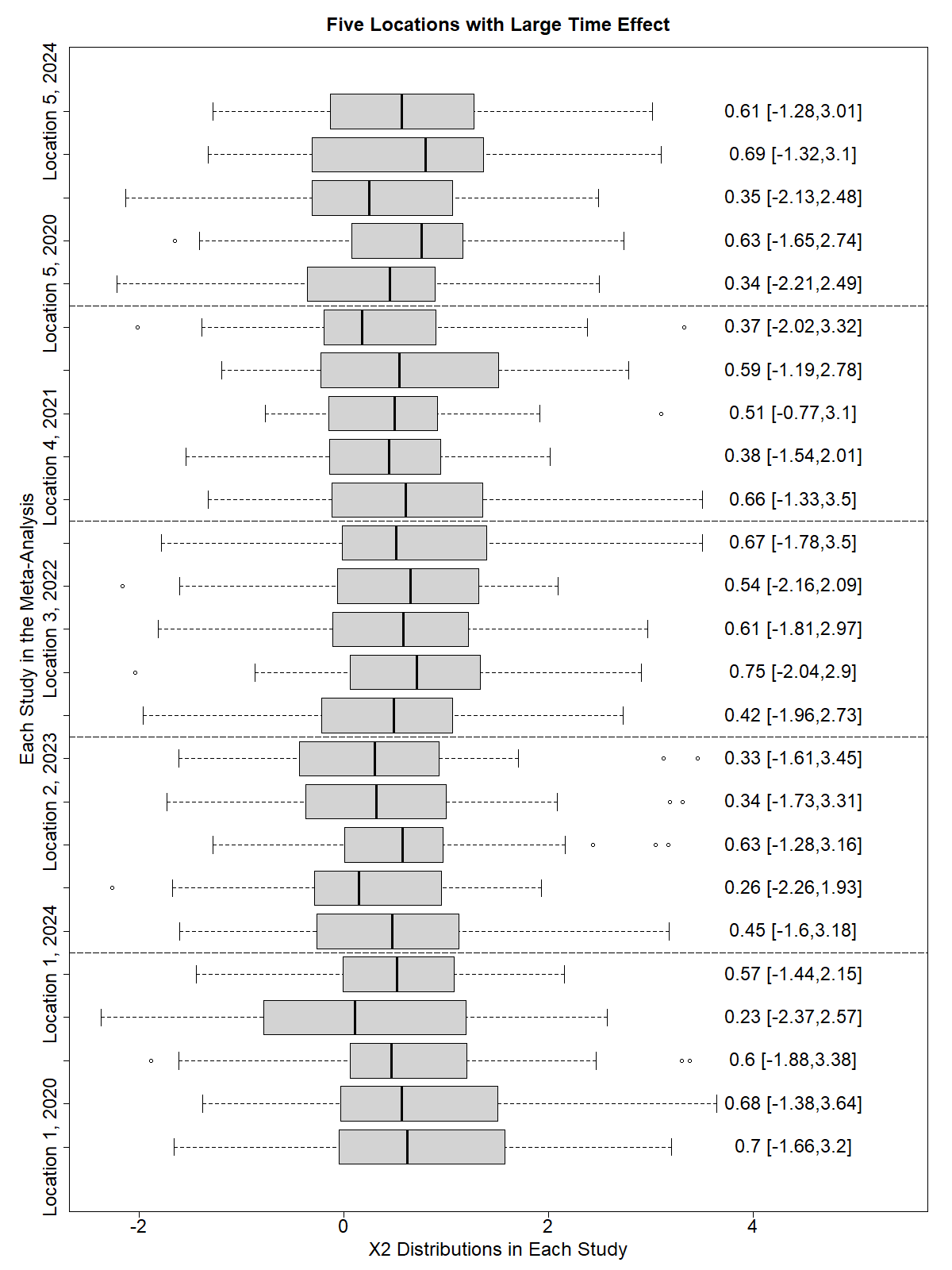}
    \caption{Study-level Variable Distributions: Each plot illustrates the distribution of each variable for each study within a meta-regression. In the meta-regression, there are five locations, a large time effect, and a small location effect.}
    \label{figure0}
\end{figure}

\begin{figure}[htp!]
    \centering
    \includegraphics[width=5cm]{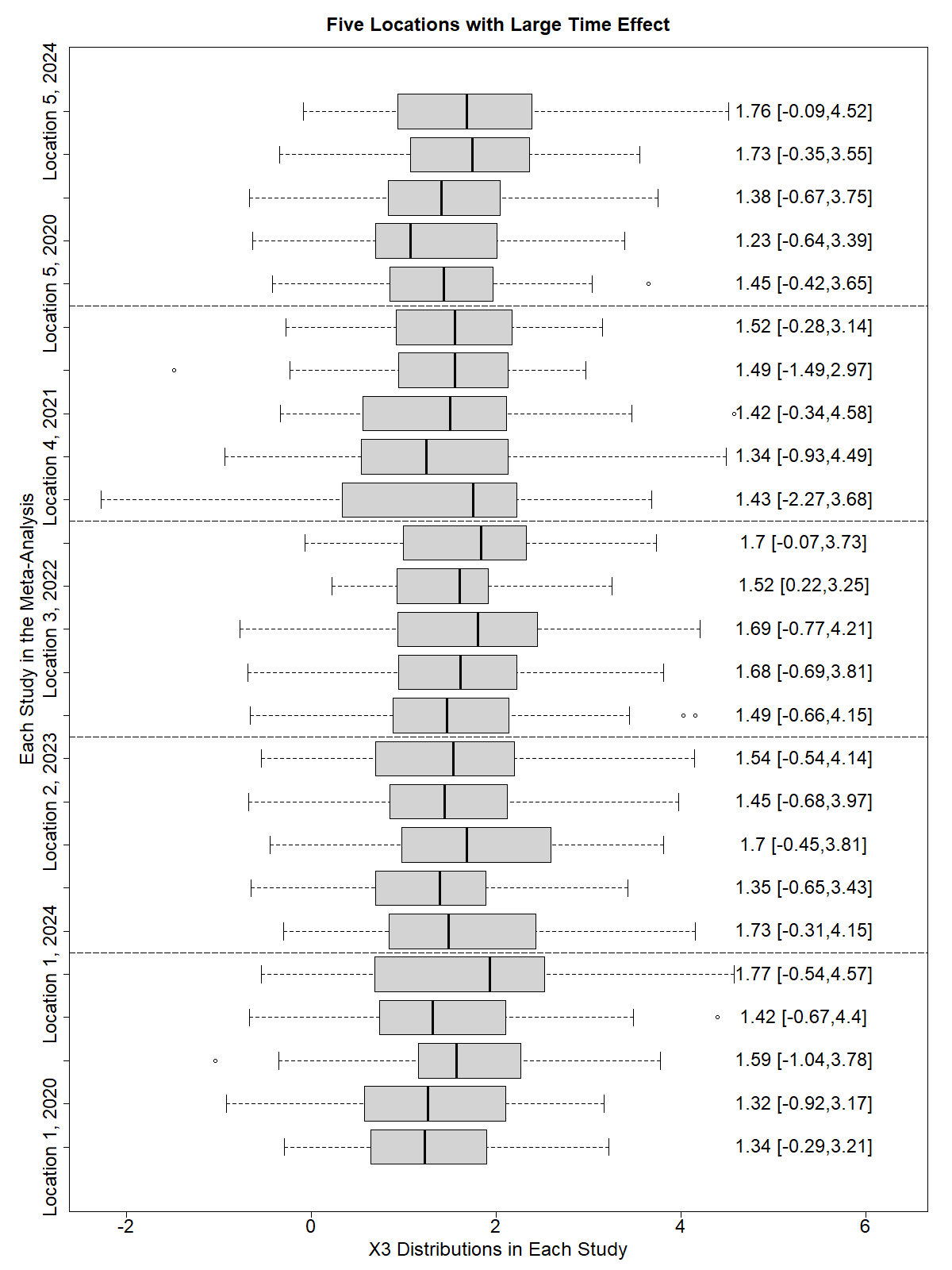}
    \includegraphics[width=5cm]{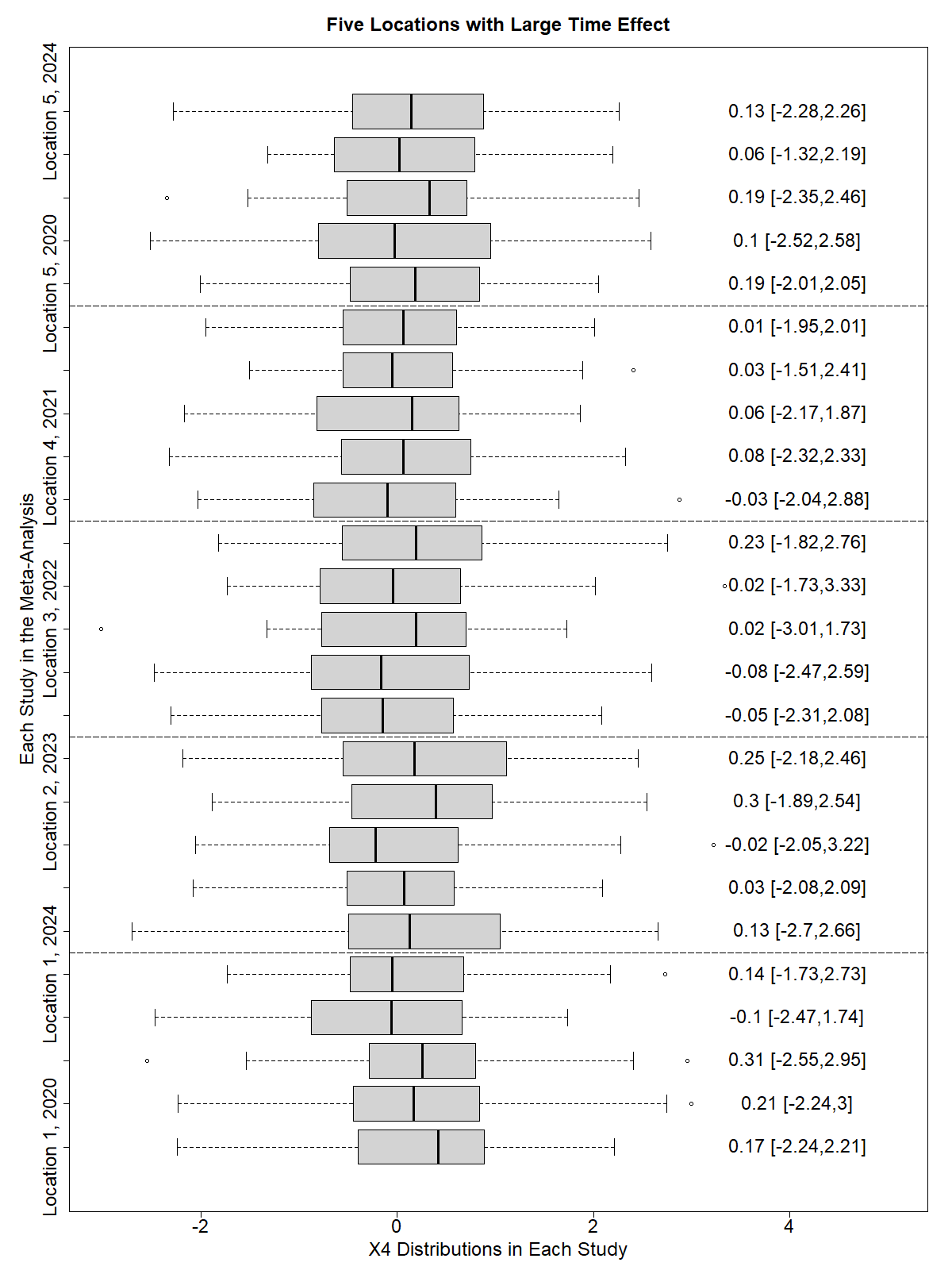}
    \includegraphics[width=5cm]{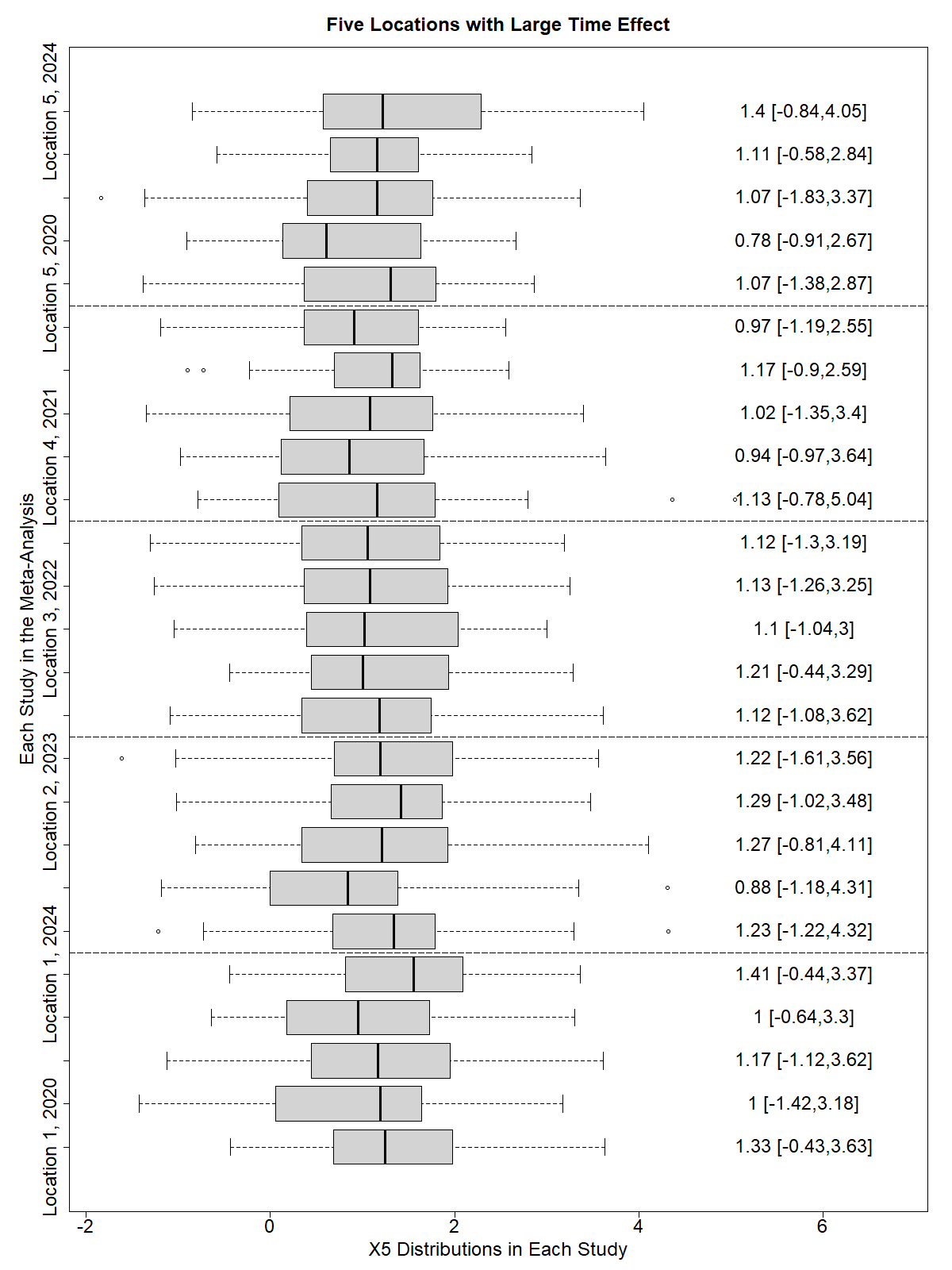}
\end{figure}

\newpage

\section*{Appendix C - Multivariate Data Generation Process}\label{appC}

Three Covariate Data Generation Process: This illustrates how each variable is generated and how the true values of the estimators are generated.\newline
\par

\begin{equation}
    \left(\begin{array}{c} Y \\ X_1 \\ X_2 \\ X_3 \end{array}\right) \sim NIID \left(\left(\begin{array}{c} \mu_Y \\ 1 \\ 0.5 \\ 1.5 \end{array}\right),\left(\begin{array}{cccc} 1 & 0.5 & 0.25 & 0.2 \\ 0.5 & 1 & 0 & 0.1 \\ 0.25 & 0 & 1 & 0.1 \\ 0.2 & 0.1 & 0.1 & 1 \end{array}\right)\right)
    \label{EQ20}
\end{equation}

$$\left(\begin{array}{c} \beta_1 \\ \beta_2 \\ \beta_3 \end{array}\right) = \left(\begin{array}{c} 0.48724 \\ 0.23724 \\ 0.12755 \end{array}\right) = \left(\begin{array}{ccc} 1 & 0 & 0.1 \\ 0 & 1 & 0.1 \\ 0.1 & 0.1 & 1 \end{array}\right)^{-1}\left(\begin{array}{c} 0.5 \\ 0.25 \\ 0.2 \end{array}\right)$$

$$\sigma^2=0.67156=1-\left(\begin{array}{ccc} 0.5 & 0.25 & 0.2 \end{array}\right)\left(\begin{array}{ccc} 1 & 0 & 0.1 \\ 0 & 1 & 0.1 \\ 0.1 & 0.1 & 1 \end{array}\right)^{-1} \left(\begin{array}{c} 0.5 \\ 0.25 \\ 0.2 \end{array}\right)$$\newline

\noindent Five Covariate Data Generation Process: This illustrates how each variable is generated and how the true values of the estimators are generated.\newline
\par

\begin{equation}
    \left(\begin{array}{c} Y \\ X_1 \\ X_2 \\ X_3 \\ X_4 \\ X_5 \end{array}\right) \sim NIID \left(\left(\begin{array}{c} \mu_Y \\ 1 \\ 0.5 \\ 1.5 \\ 0.1 \\ 1.1 \end{array}\right),\left(\begin{array}{cccccc} 1 & 0.5 & 0.25 & 0.2 & 0.25 & 0.5 \\ 0.5 & 1 & 0 & 0.1 & 0.2 & 0.3 \\ 0.25 & 0 & 1 & 0.1 & 0 & 0.2 \\ 0.2 & 0.1 & 0.1 & 1 & 0 & 0.5 \\ 0.25 & 0.2 & 0 & 0 & 1 & 0.1 \\ 0.5 & 0.3 & 0.2 & 0.5 & 0.1 & 1 \end{array}\right)\right)
    \label{EQ21}
\end{equation}

$$\left(\begin{array}{c} \beta_1 \\ \beta_2 \\ \beta_3 \\ \beta_4 \\ \beta_5 \end{array}\right)=\left(\begin{array}{c} 0.36853 \\ 0.18221 \\ -0.03271 \\ 0.14076 \\ 0.35527 \end{array}\right)\left(\begin{array}{ccccc} 1 & 0 & 0.1 & 0.2 & 0.3 \\ 0 & 1 & 0.1 & 0 & 0.2 \\ 0.1 & 0.1 & 1 & 0 & 0.5 \\ 0.2 & 0 & 0 & 1 & 0.1 \\ 0.3 & 0.2 & 0.5 & 0.1 & 0 \end{array}\right)^{-1} \left(\begin{array}{c} 0.5 \\ 0.25 \\ 0.2 \\ 0.25 \\ 0.5 \end{array}\right)$$

$$\sigma^2=0.5639=1-\left(\begin{array}{ccccc} 0.5 & 0.25 & 0.2 & 0.25 & 0.5 \end{array}\right)\left(\begin{array}{ccccc} 1 & 0 & 0.1 & 0.2 & 0.3 \\ 0 & 1 & 0.1 & 0 & 0.2 \\ 0.1 & 0.1 & 1 & 0 & 0.5 \\ 0.2 & 0 & 0 & 1 & 0.1 \\ 0.3 & 0.2 & 0.5 & 0.1 & 0 \end{array}\right)^{-1} \left(\begin{array}{c} 0.5 \\ 0.25 \\ 0.2 \\ 0.25 \\ 0.5 \end{array}\right)$$

\newpage

\section*{Appendix D - Result Figures}\label{figures}

\subsection*{Less versus More Covariates}

\begin{figure}[htp!]
    \centering
    \includegraphics[width=16cm]{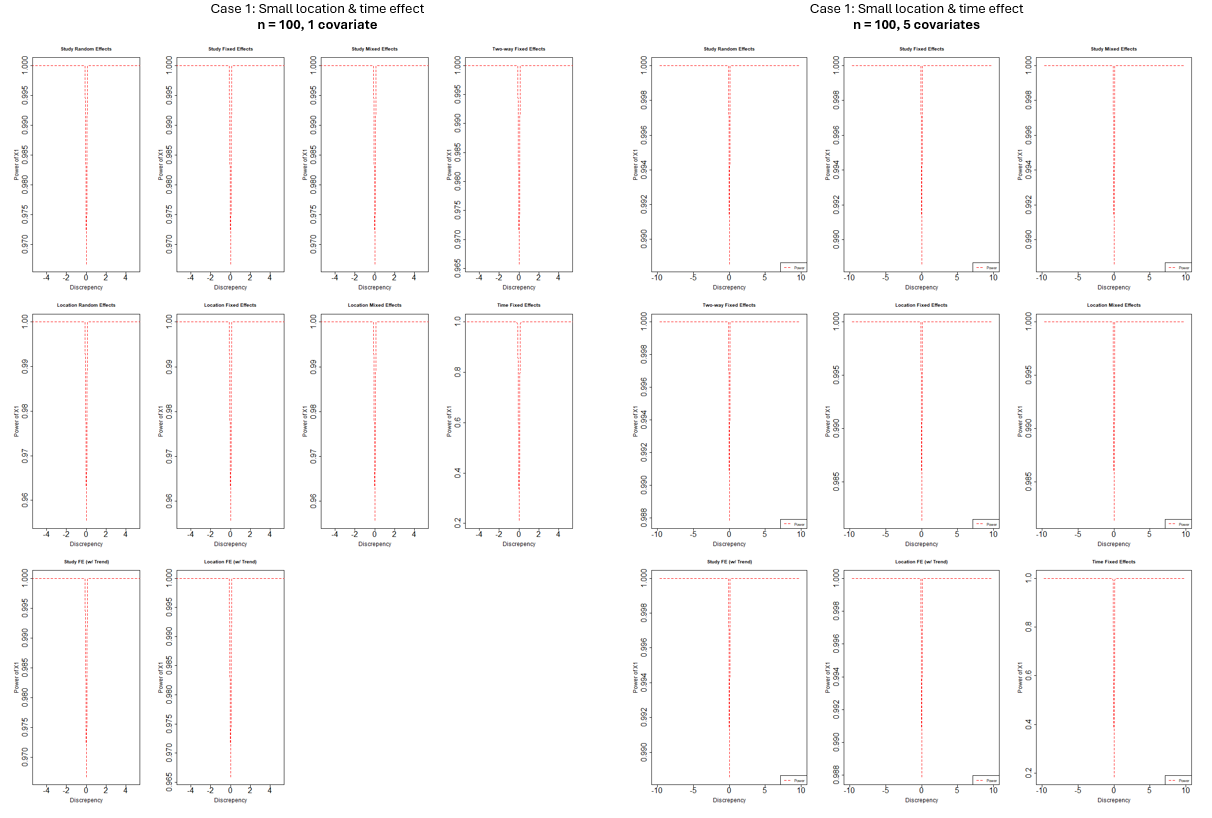}
    \caption{Sample Power Curves: These illustrate the difference of when there are one versus five covariates. Both show Case 1 \& $n = 100$. The x-axis shows the power to detect a range of discrepancies from the null hypothesis.}
    \label{resultFigure1}
\end{figure}

\begin{figure}[htp!]
    \centering
    \includegraphics[width=16cm,height=6cm]{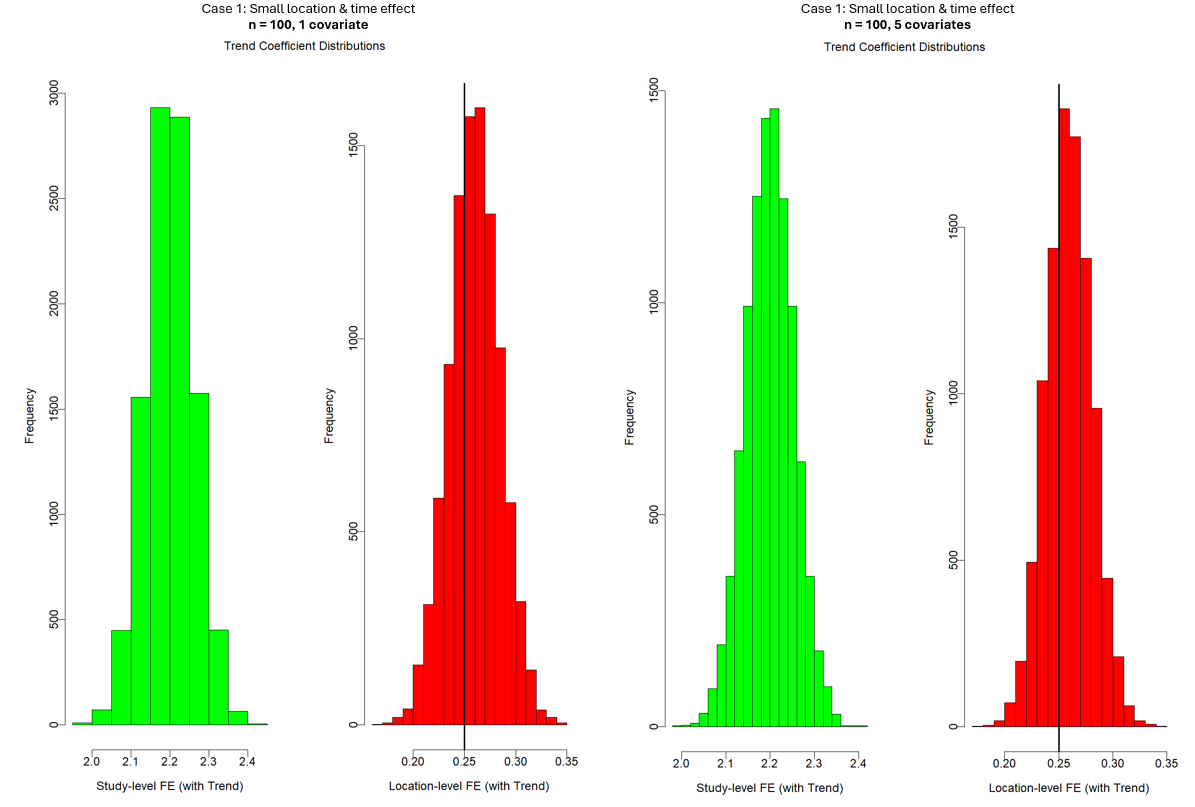}
    \caption{Sample Trend Estimates: These illustrate the difference between when there is one versus five covariates for Case 1 \& $n = 100$. Both show the true trend value as a vertical line.}
    \label{resultFigure3}
\end{figure}

\newpage

\begin{figure}[htp!]
    \centering
    \includegraphics[width=7.77cm]{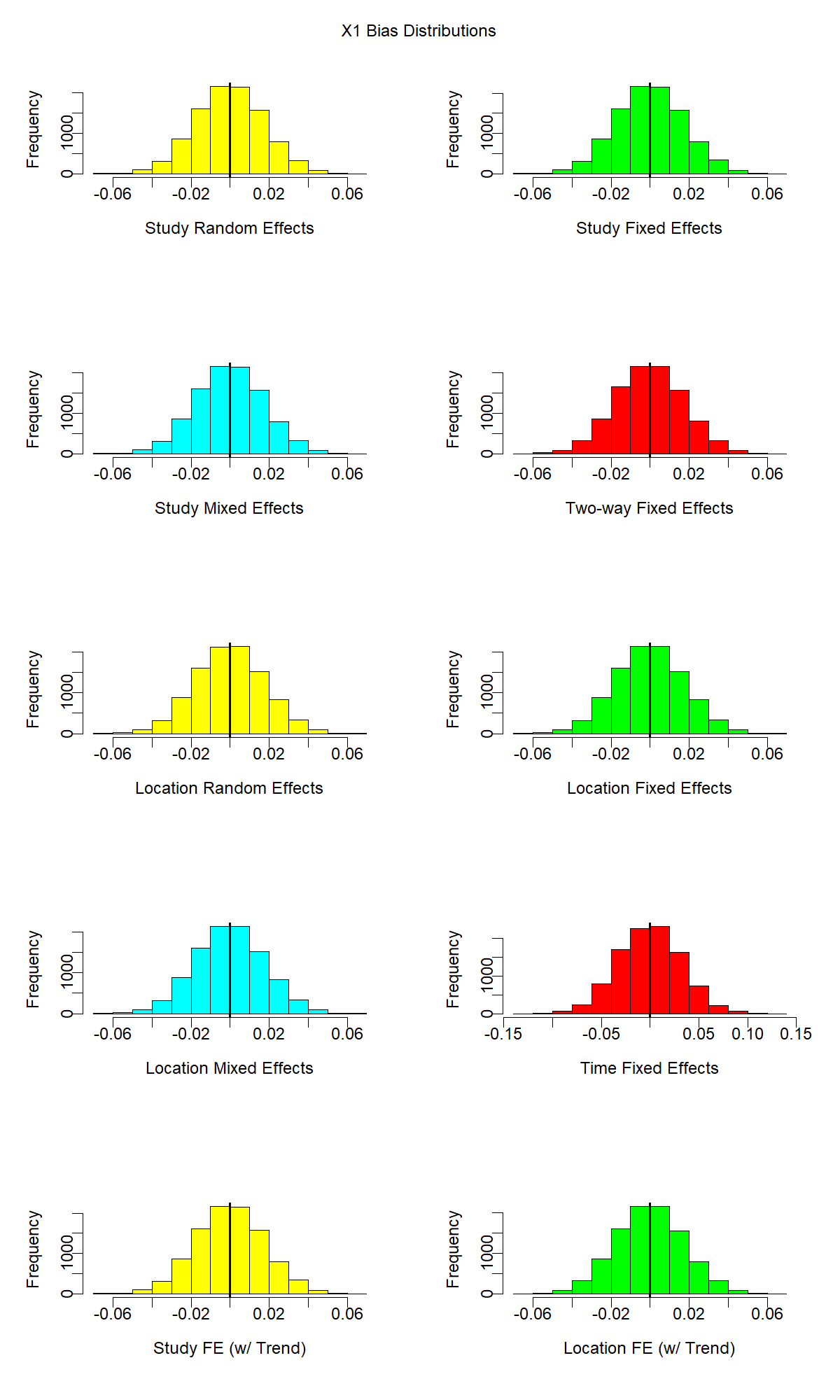}
    \includegraphics[width=7.77cm]{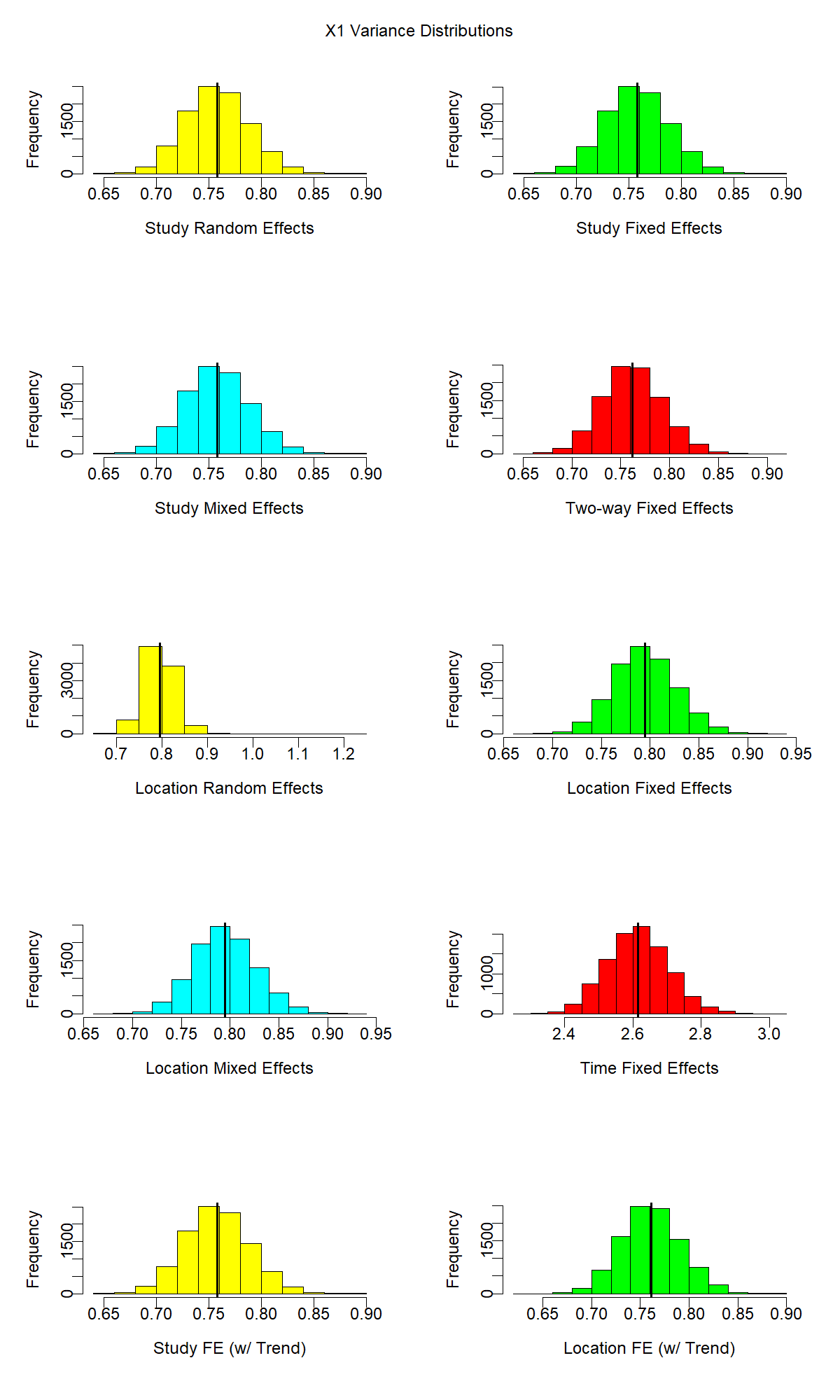}
    \caption{Sample X1 Bias \& Variance Distributions: The estimators bias and variance and MSE are respectively shown left to right. This is for when there is one covariate and can be compared with the figure below which is for when there are five covariates. Both show Case 1 \& $n = 100$.}
    \label{resultFigure2}
\end{figure}
\begin{figure}[htp!]
    \centering
    \includegraphics[width=7.77cm]{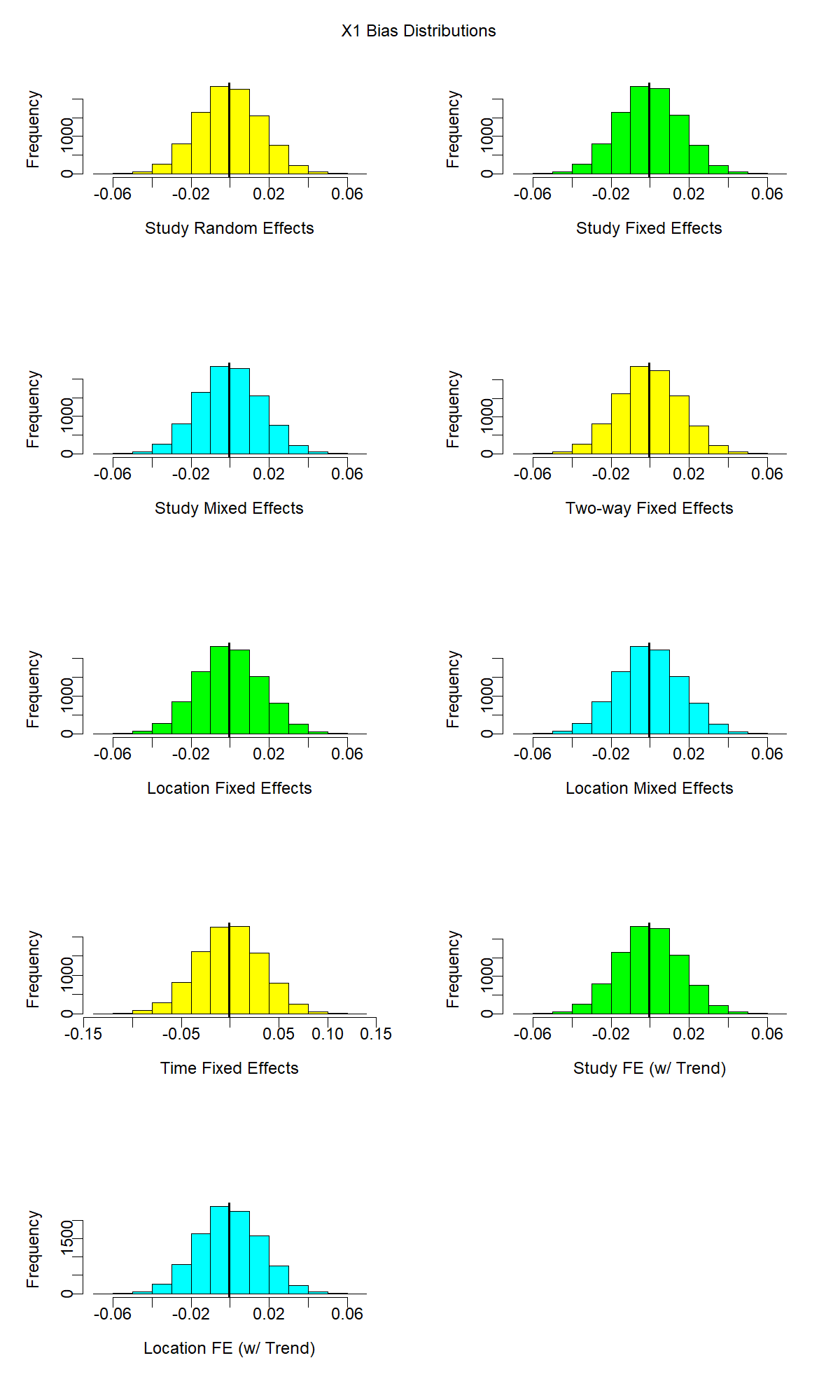}
    \includegraphics[width=7.77cm]{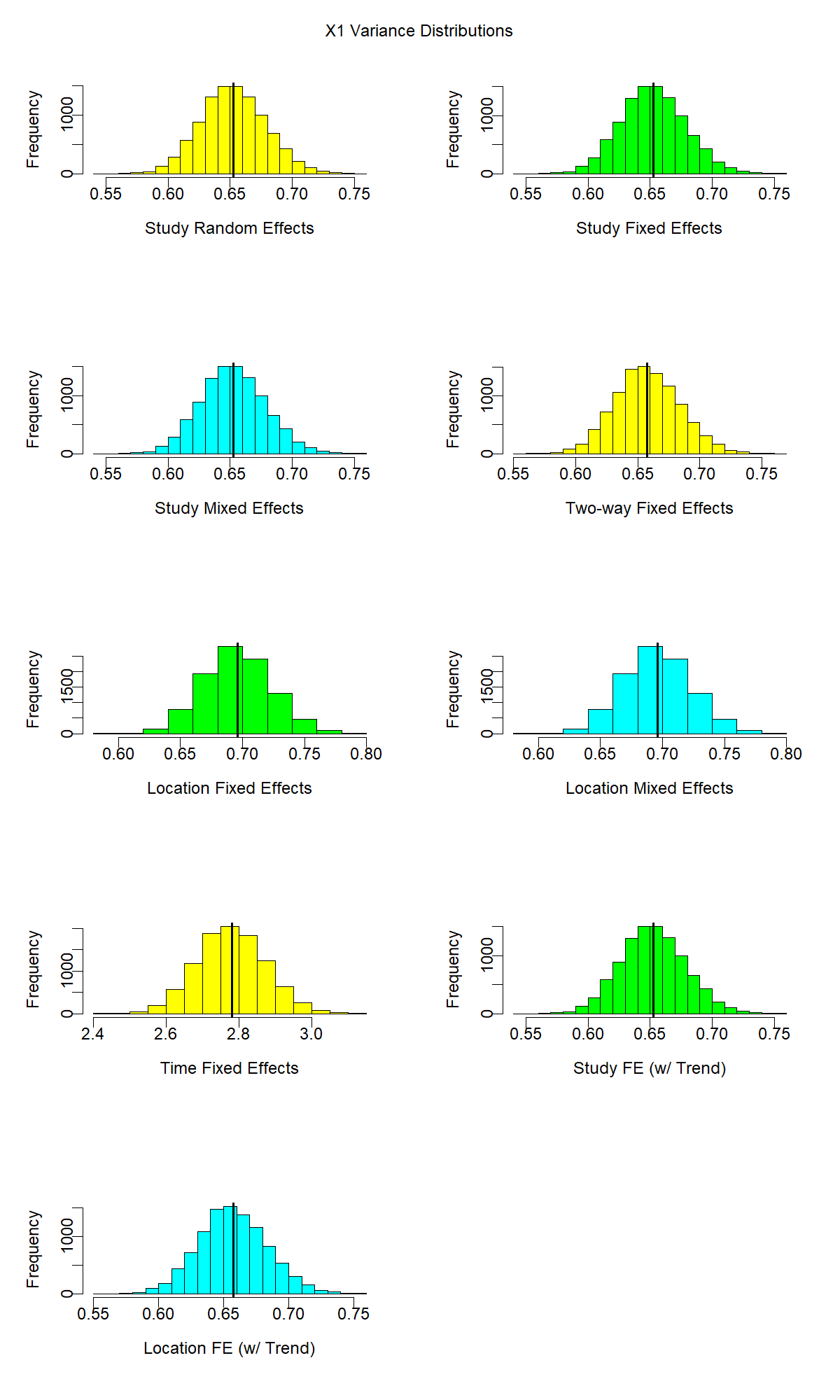}
    \caption{Sample X1 Bias \& Variance Distributions: The estimators bias and variance are respectively shown left to right. This is for when there is five covariates and can be compared with the figure above which is for when there is one covariate. Both show Case 1 \& $n = 100$.}
    \label{resultFigure2.5}
\end{figure}

\clearpage

\subsection*{Smaller versus Larger Time Effect}

\begin{figure}[htp!]
    \centering
    \includegraphics[width=16cm]{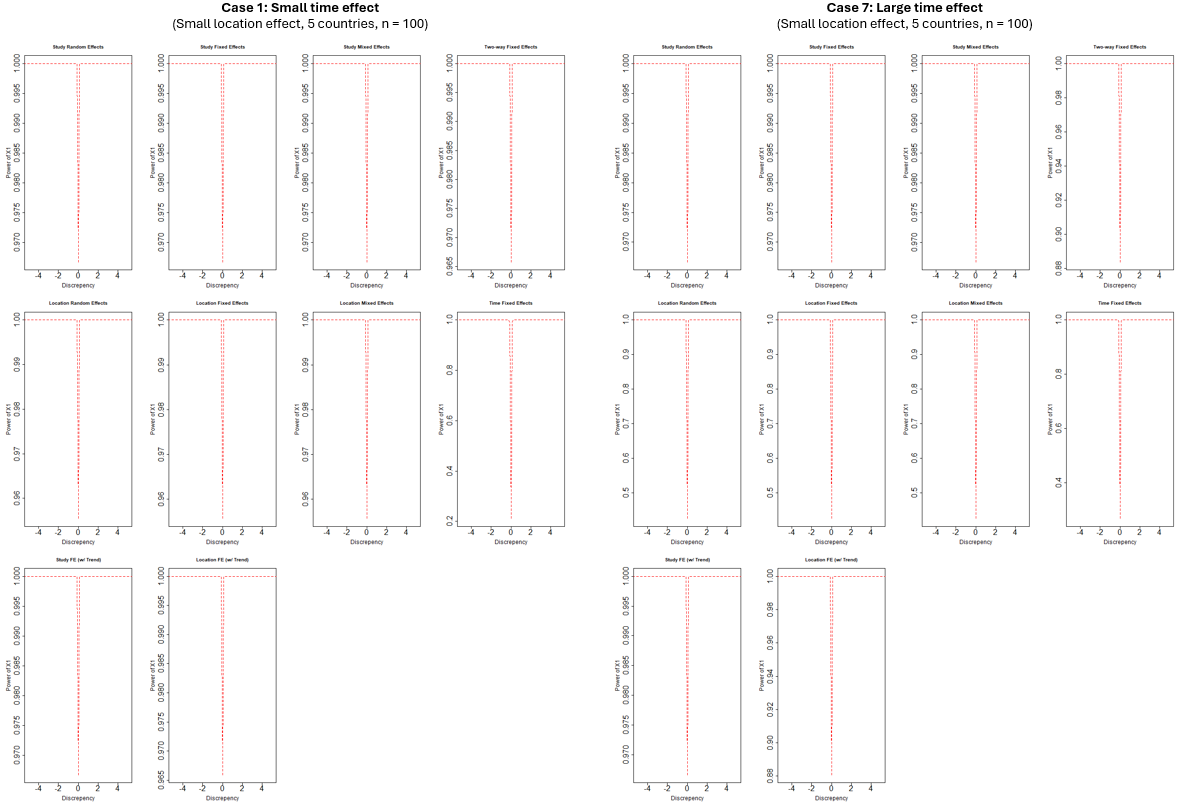}
    \caption{Sample Power Curves: These illustrate a small versus large time effect (respectively Case 1 on the left and Case 7 on the right). Both show $n = 100$ for 1 covariate, the x-axis shows the power to detect a range of discrepancies from the null hypothesis.}
    \label{resultFigure4}
\end{figure}

\begin{figure}[htp!]
    \centering
    \includegraphics[width=16cm,height=6cm]{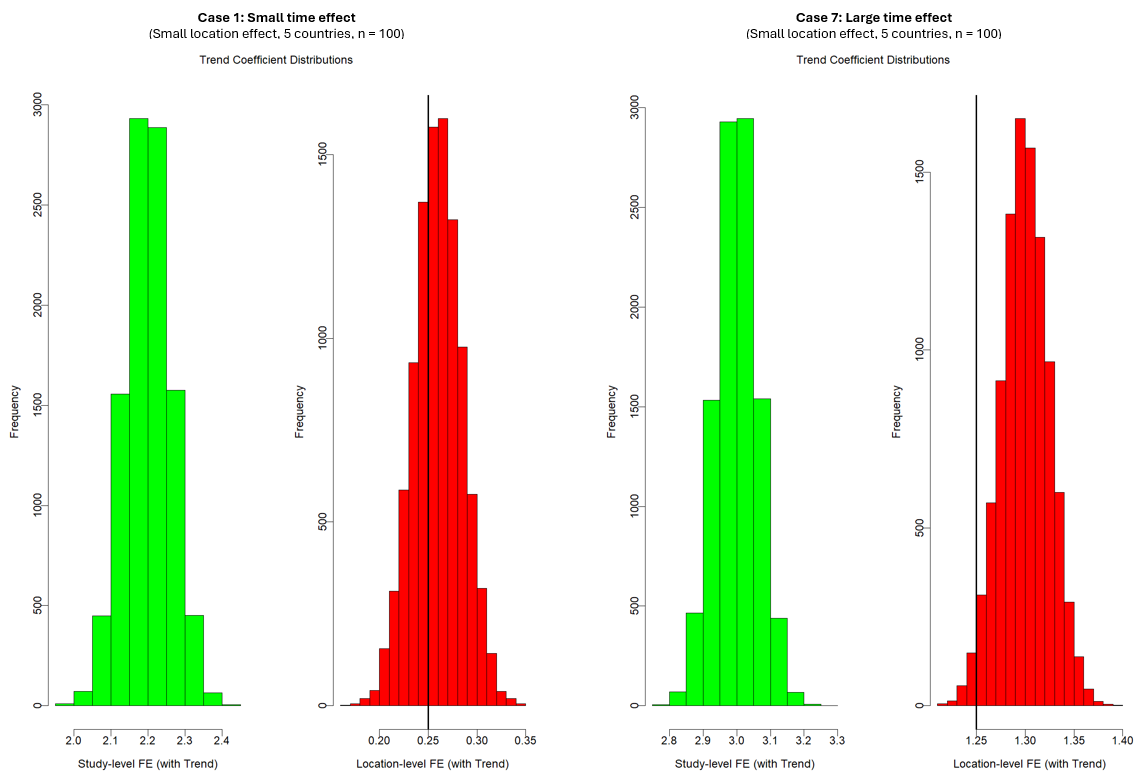}
    \caption{Sample Trend Estimates: These illustrate the difference between when there is a small versus large time effect (respectively corresponding to Case 1 on the left and Case 7 on the right). Both show the true trend value as a vertical line for 1 covariate when $n = 100$}
    \label{resultFigure6}
\end{figure}

\newpage

\begin{figure}[htp!]
    \centering
    \includegraphics[width=7.77cm]{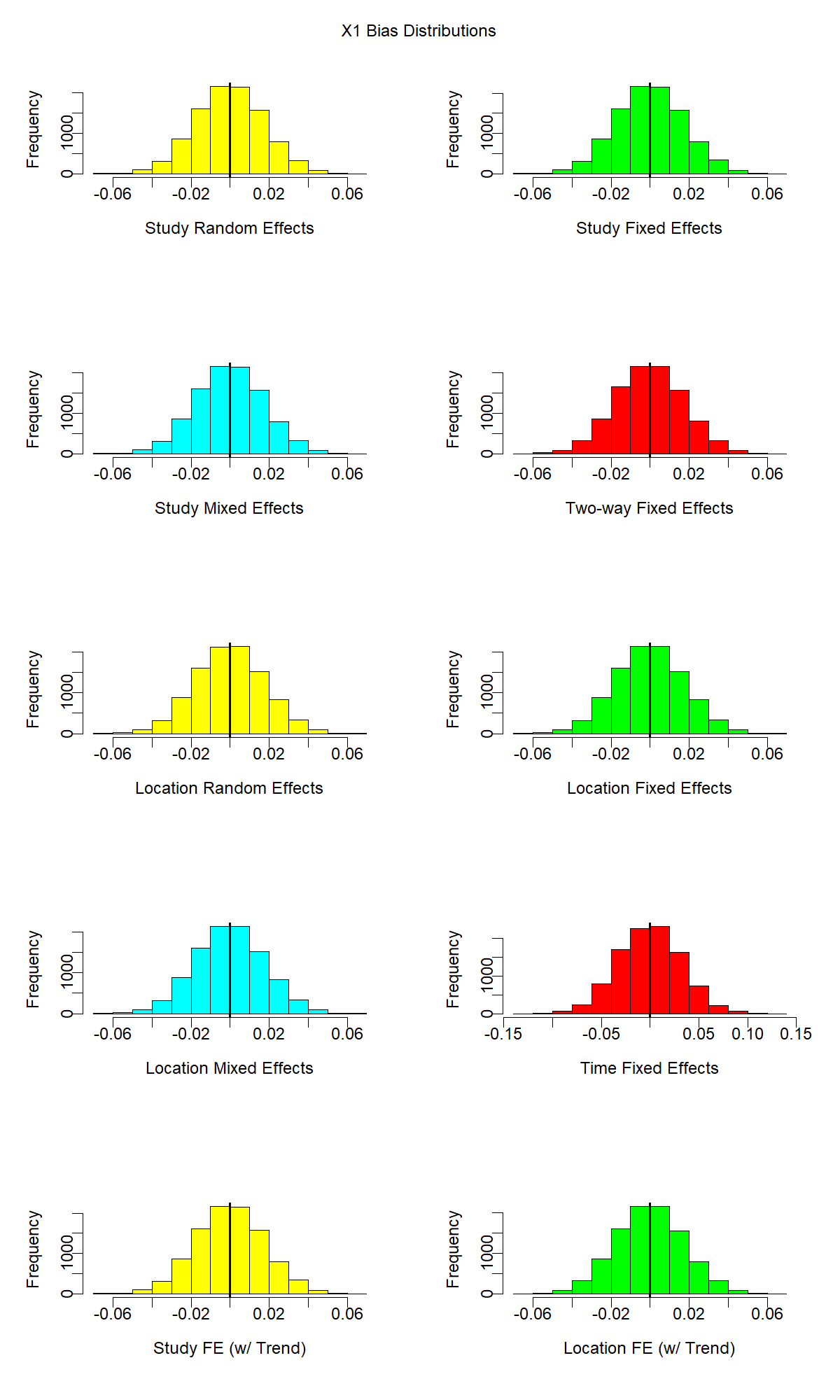}
    \includegraphics[width=7.77cm]{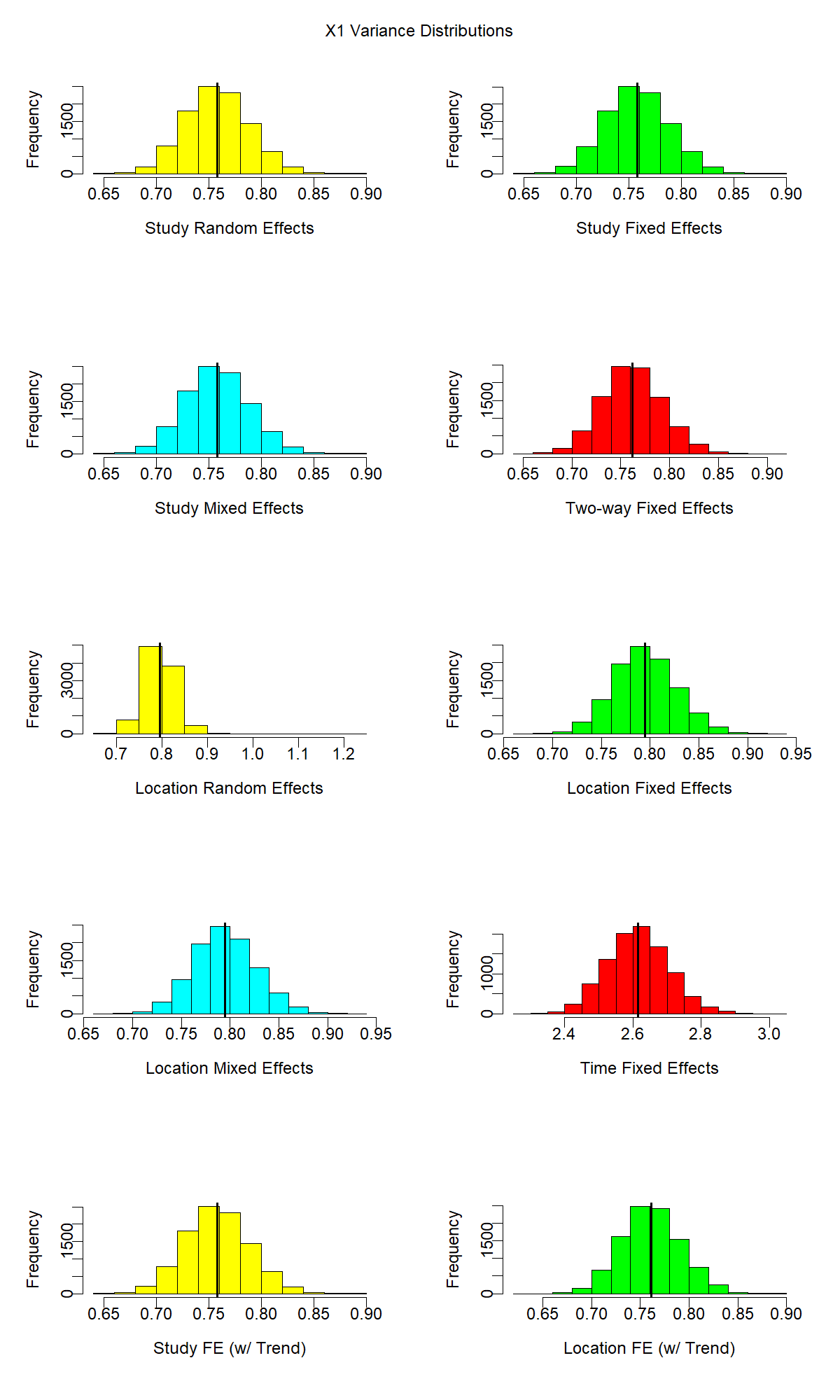}
    \caption{Sample X1 Bias \& Variance Distributions: The estimators bias and variance are respectively shown left to right. This is for when there is a small time effect (Case 1) and can be compared with the figure below which is for a large time effect (Case 7). Both show $n = 100$ for 1 covariate.}
    \label{resultFigure5}
\end{figure}
\begin{figure}[htp!]
    \centering
    \includegraphics[width=7.77cm]{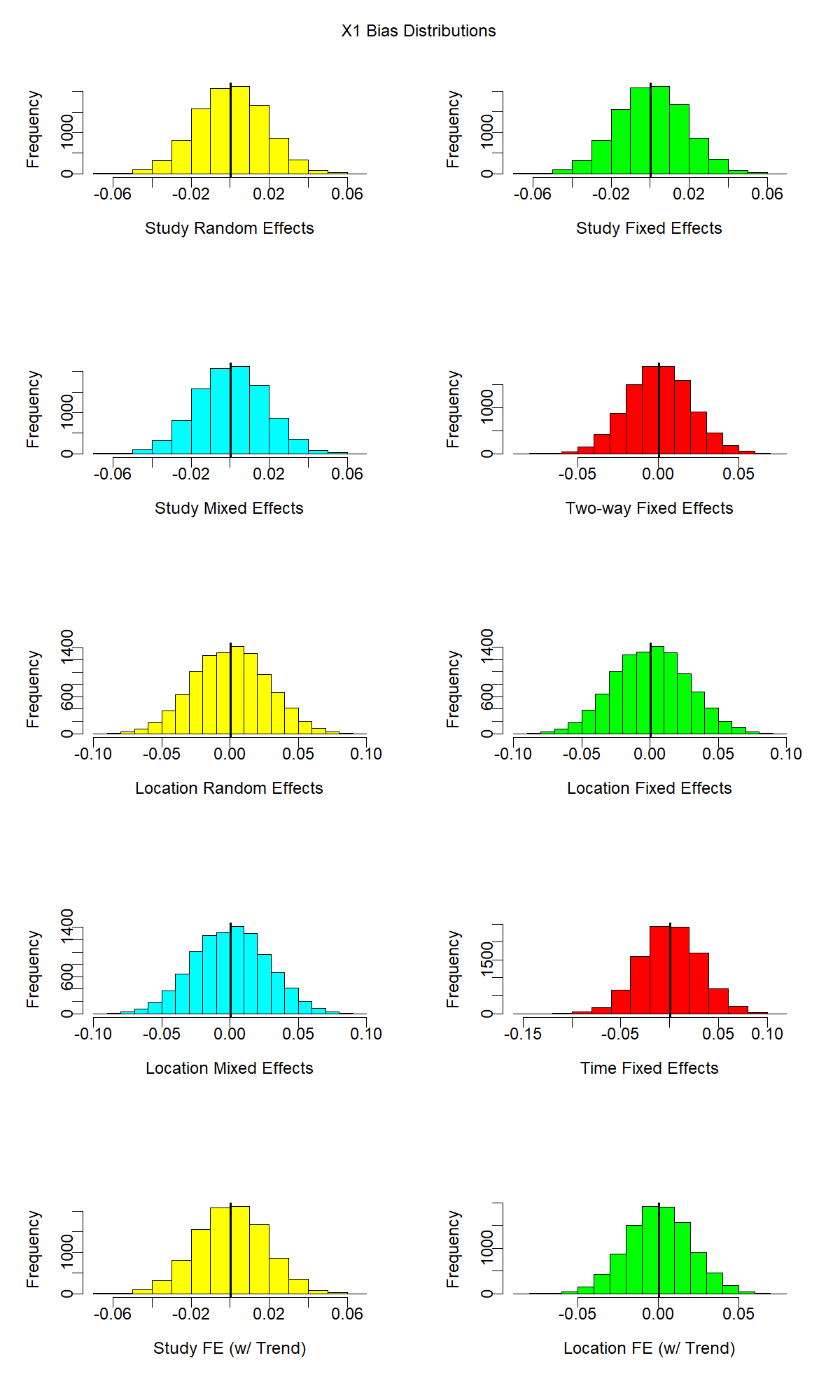}
    \includegraphics[width=7.77cm]{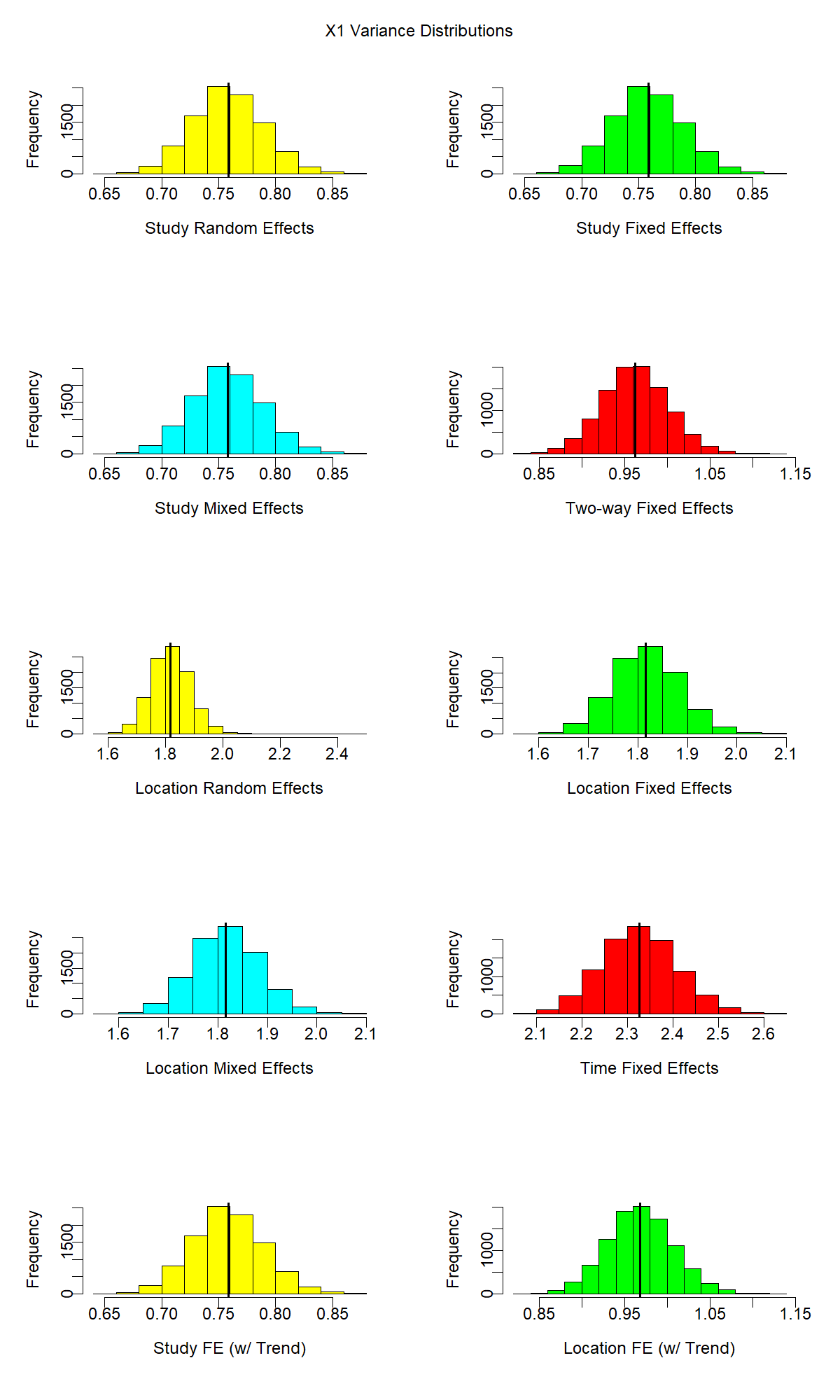}
    \caption{Sample X1 Bias \& Variance Distributions: The estimators bias and variance are respectively shown left to right. This is for when there is a large time effect (Case 7) and can be compared with the figure above which is for a small time effect (Case 1). Both show $n = 100$ for 1 covariate.}
    \label{resultFigure5.5}
\end{figure}

\clearpage

\subsection*{Smaller versus Larger Location Effect}

\begin{figure}[htp!]
    \centering
    \includegraphics[width=16cm]{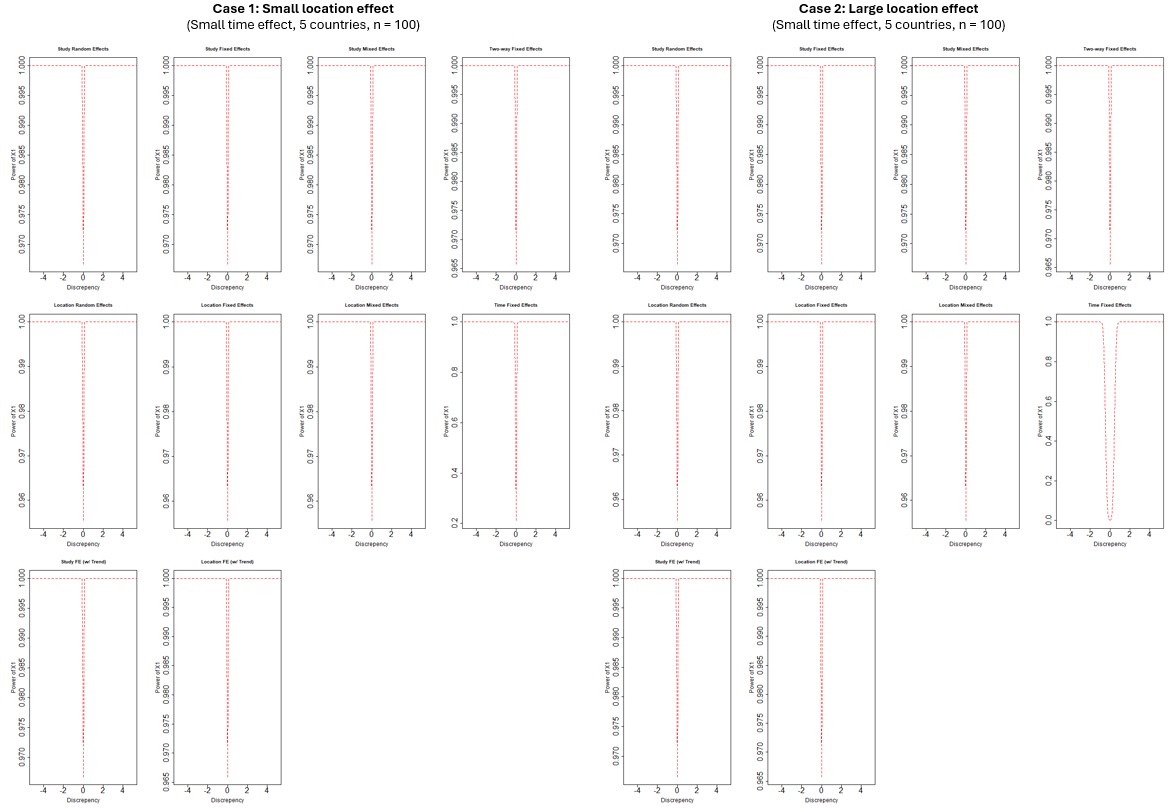}
    \caption{Sample Power Curves: These illustrate a small versus large location effect (respectively Case 1 on the left and Case 2 on the right). Both show $n = 100$ for 1 covariate, the x-axis shows the power to detect a range of discrepancies from the null hypothesis.}
    \label{resultFigure10}
\end{figure}

\begin{figure}[htp!]
    \centering
    \includegraphics[width=16cm,height=6cm]{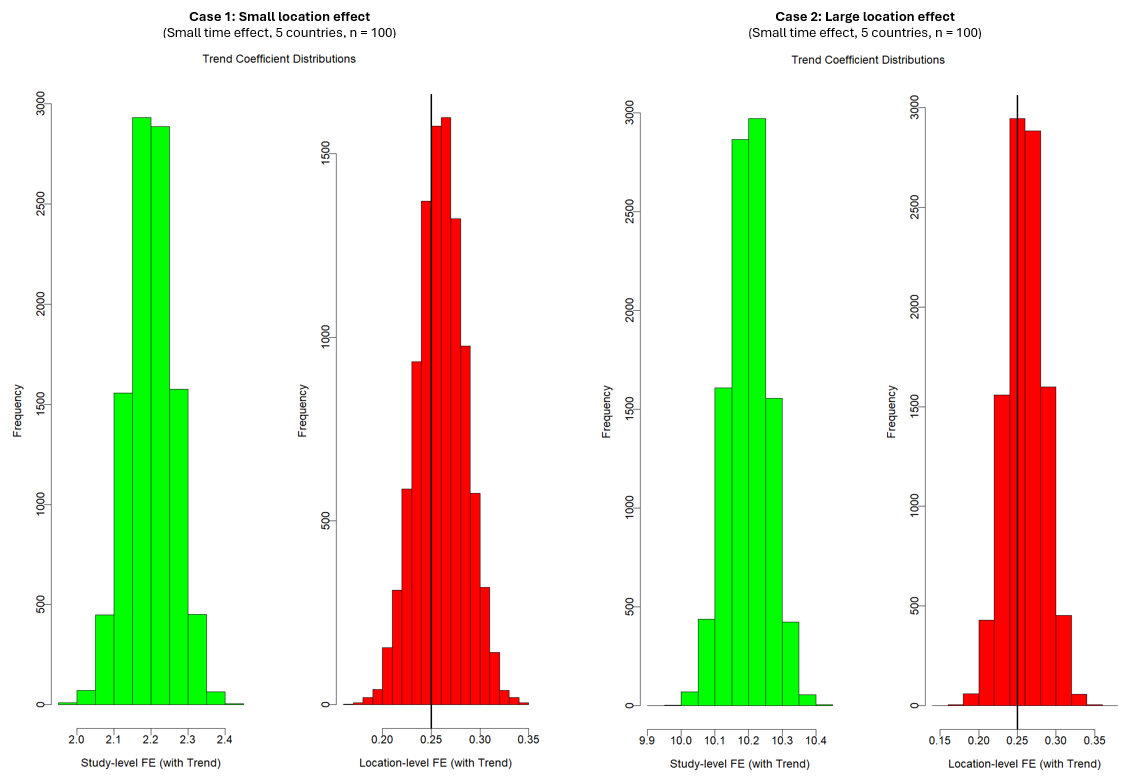}
    \caption{Sample Trend Estimates: These illustrate the difference between when there is a small versus large location effect (respectively corresponding to Case 1 on the left and Case 2 on the right). Both show the true trend value as a vertical line for 1 covariate when $n = 100$}
    \label{resultFigure12}
\end{figure}

\newpage

\begin{figure}[htp!]
    \centering
    \includegraphics[width=7.77cm]{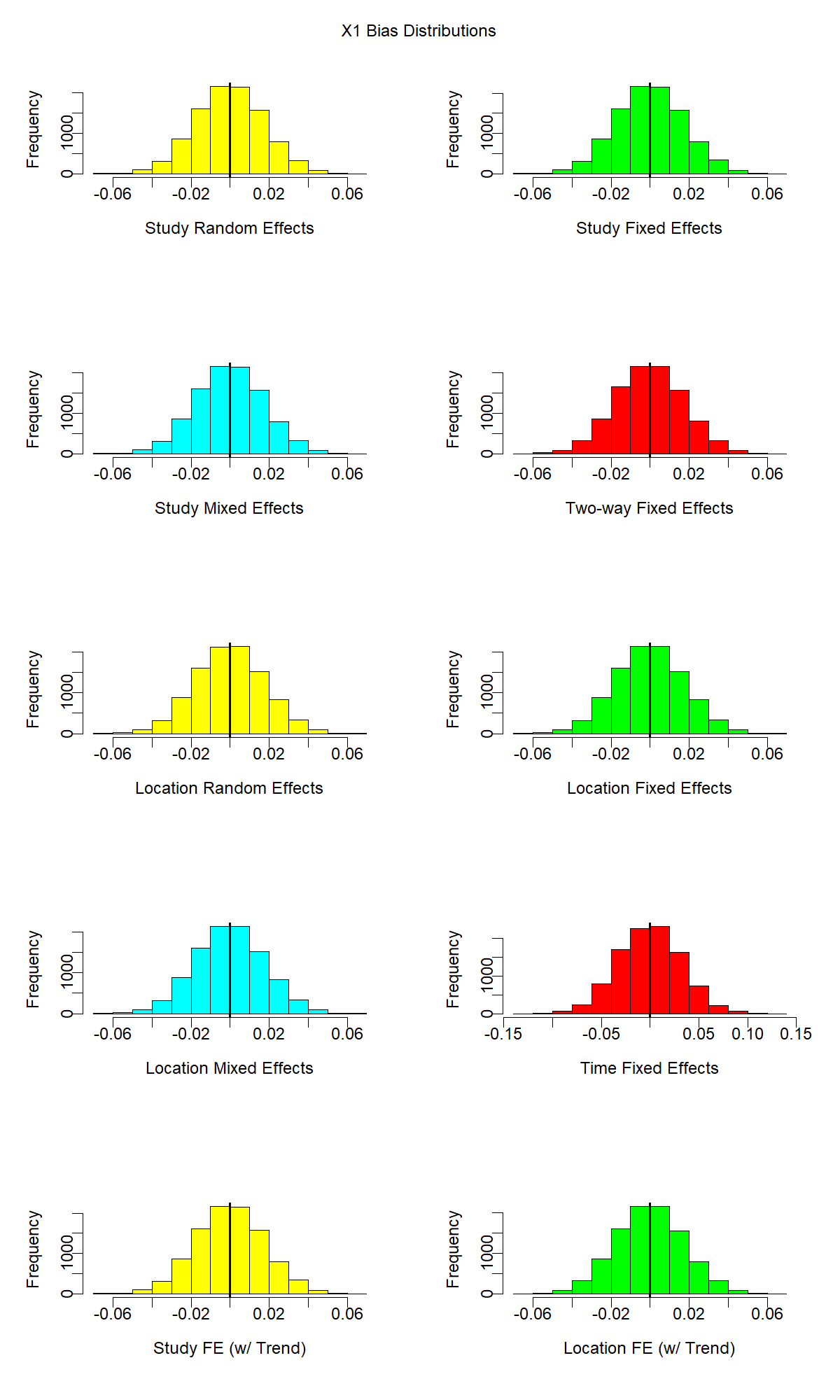}
    \includegraphics[width=7.77cm]{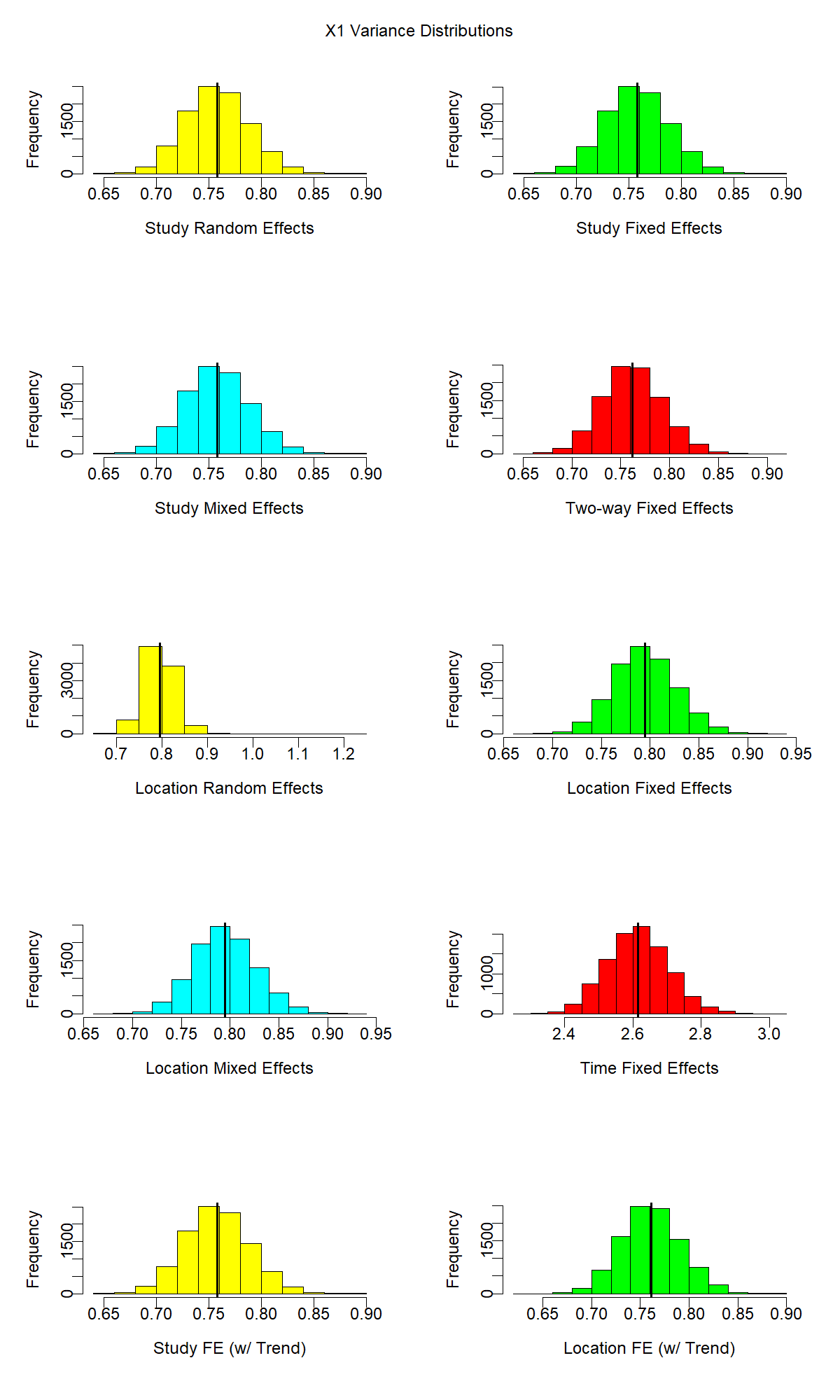}
    \caption{Sample X1 Bias \& Variance Distributions: The estimators bias and variance are respectively shown left to right. This is for when there is a small location effect (Case 1) and can be compared with the figure below which is for a large location effect (Case 2). Both show $n = 100$ for 1 covariate.}
    \label{resultFigure11}
\end{figure}
\begin{figure}[htp!]
    \centering
    \includegraphics[width=7.77cm]{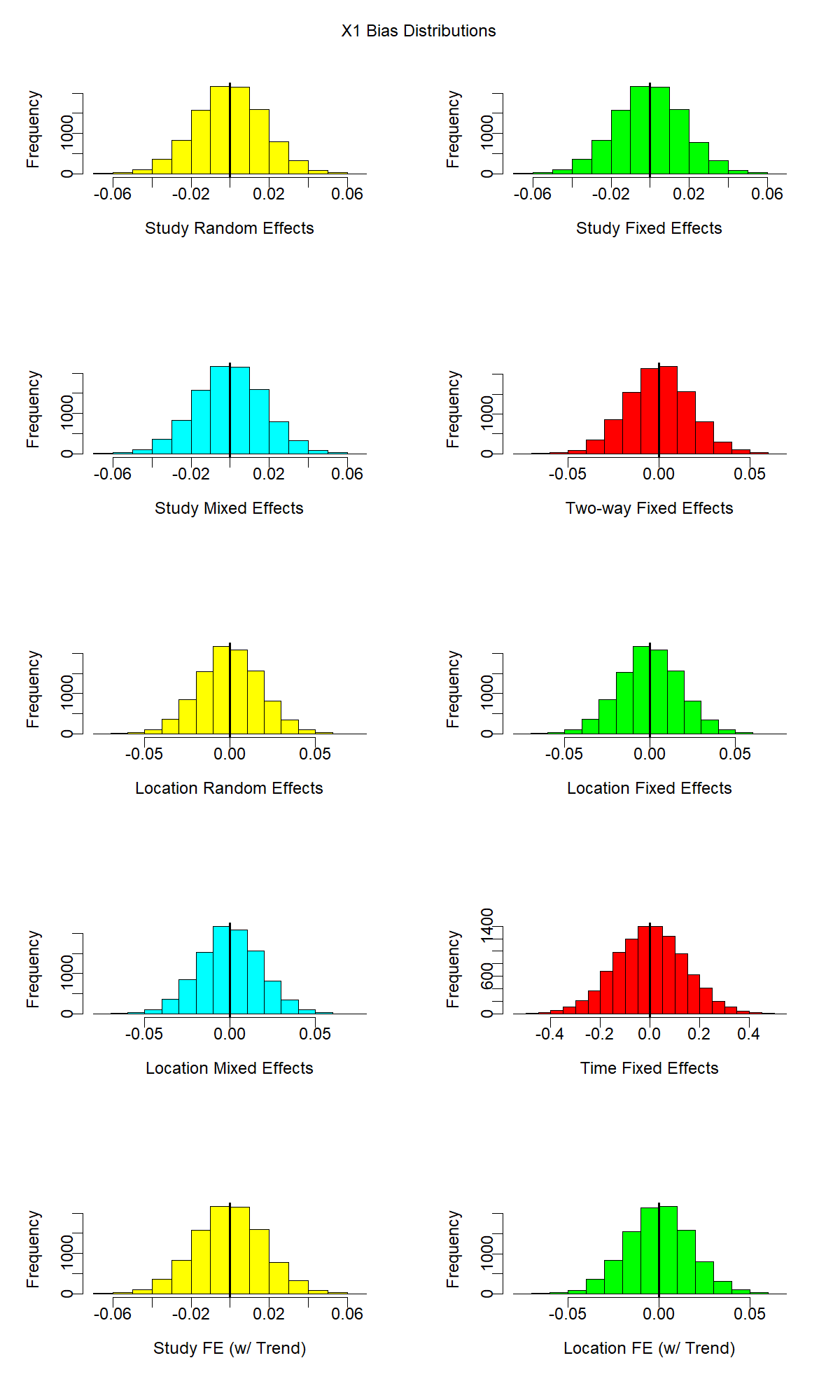}
    \includegraphics[width=7.77cm]{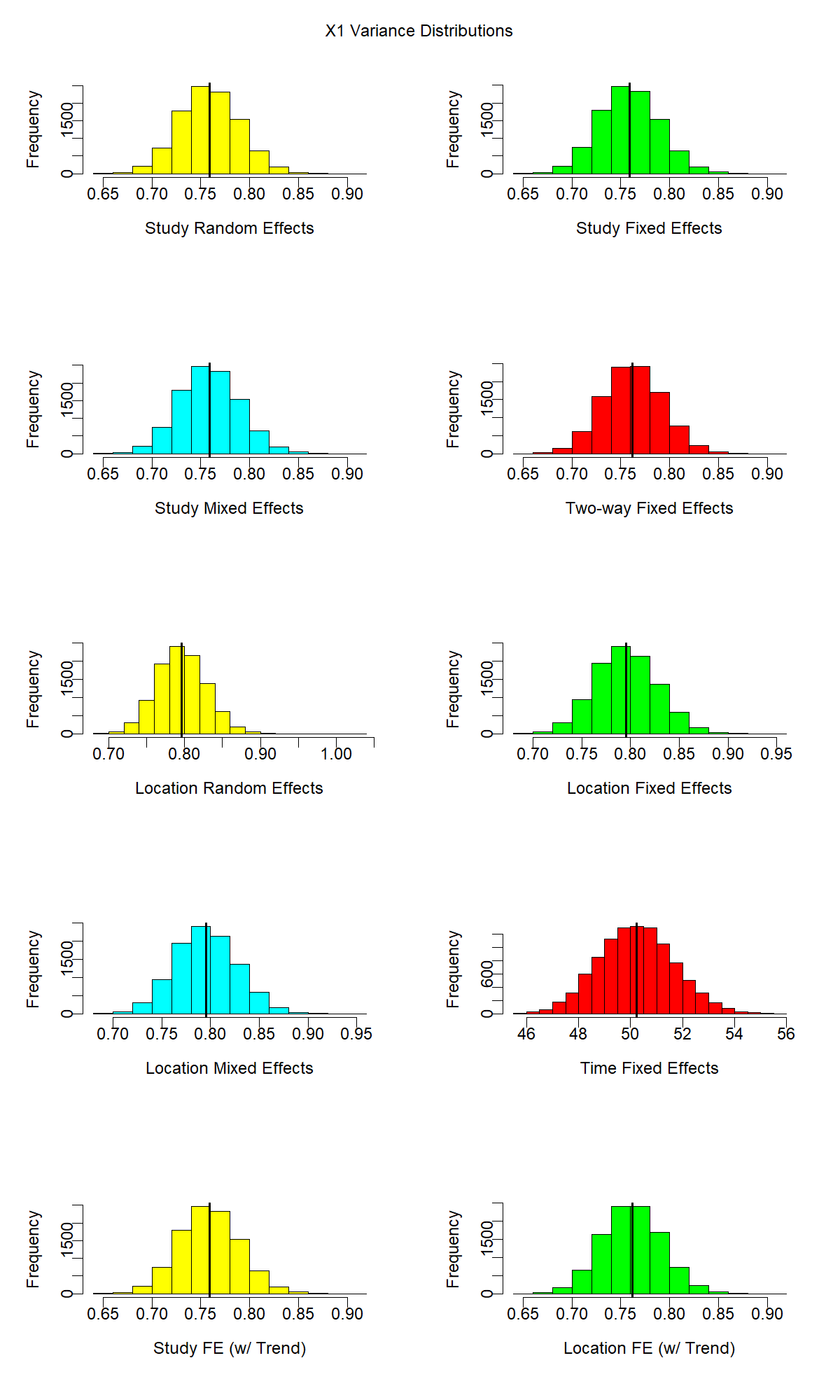}
    \caption{Sample X1 Bias \& Variance Distributions: The estimators bias and variance are respectively shown left to right. This is for when there is a large location effect (Case 2) and can be compared with the figure above which is for a small location effect (Case 1). Both show $n = 100$ for 1 covariate.}
    \label{resultFigure11.5}
\end{figure}

\clearpage

\subsection*{Less versus More Number of Locations}

\begin{figure}[htp!]
    \centering
    \includegraphics[width=16cm]{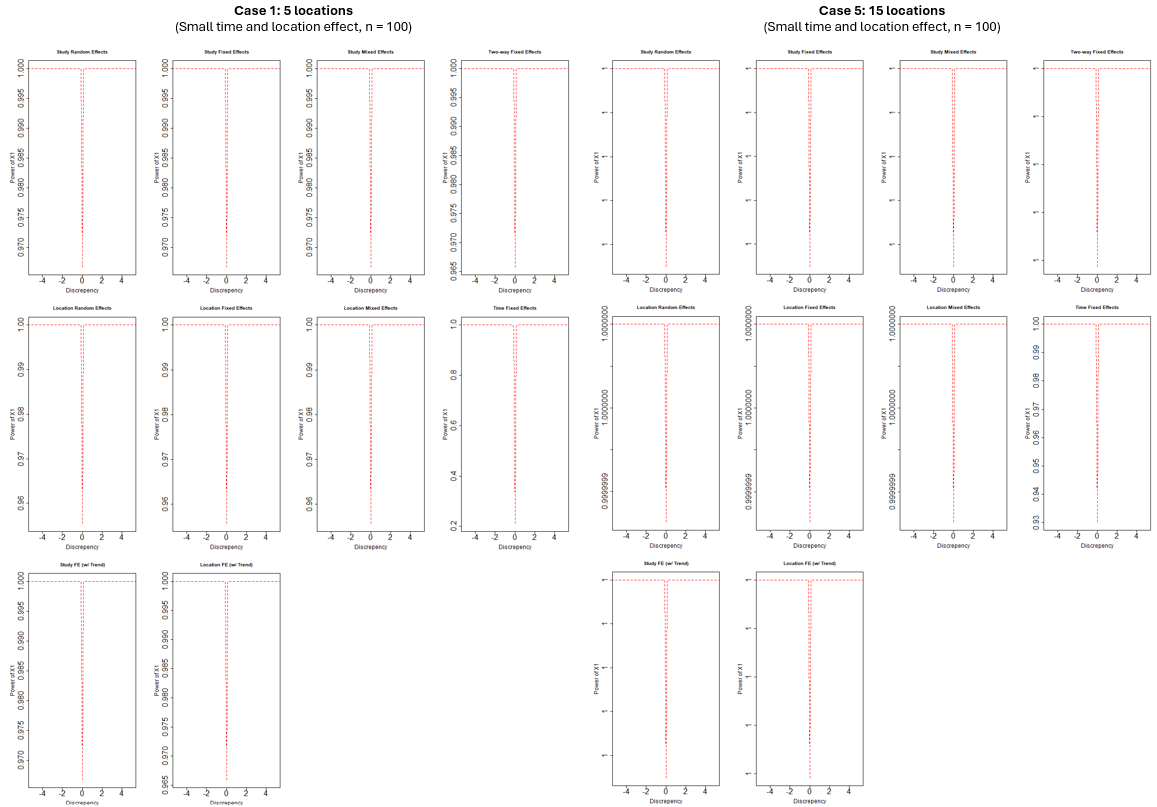}
    \caption{Sample Power Curves: These illustrate less versus more locations (respectively corresponding to Case 1 on the left and Case 5 on the right). Both show $n = 100$ for 1 covariate, the x-axis shows the power to detect a range of discrepancies from the null hypothesis.}
    \label{resultFigure7}
\end{figure}

\begin{figure}[htp!]
    \centering
    \includegraphics[width=16cm,height=6cm]{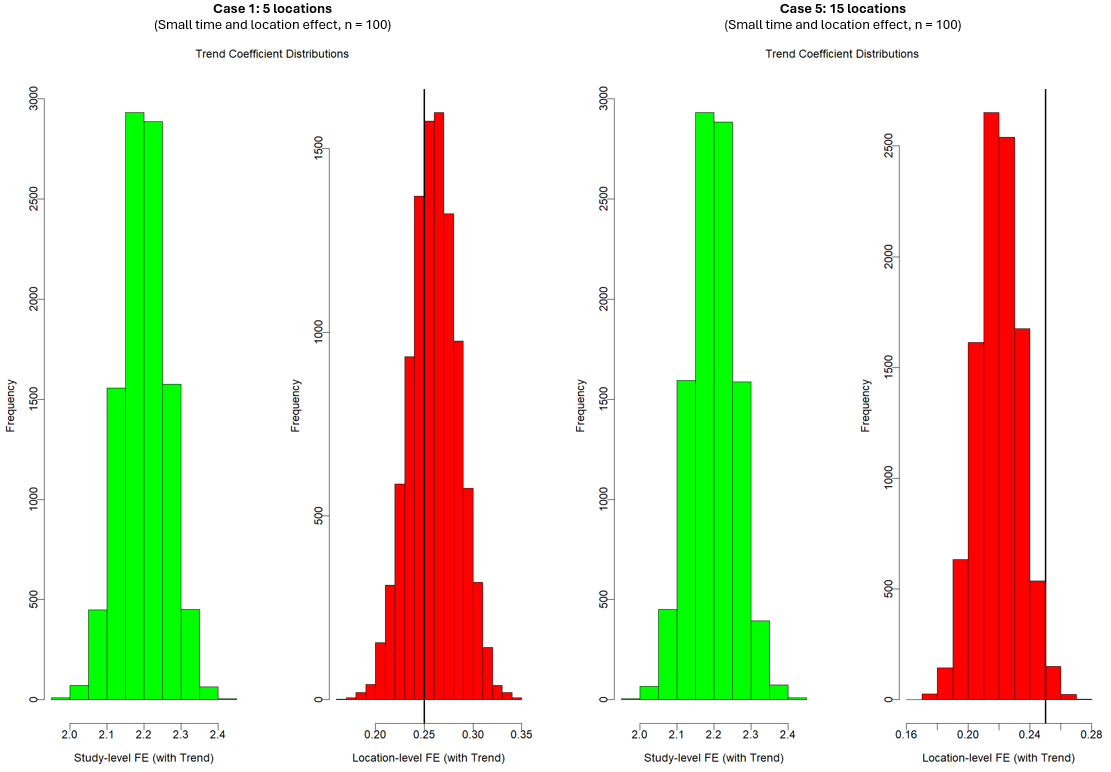}
    \caption{Sample Trend Estimates: These illustrate less versus more locations (respectively corresponding to Case 1 on the left and Case 5 on the right). Both show the true trend value as a vertical line for 1 covariate when $n = 100$.}
    \label{resultFigure9}
\end{figure}

\newpage

\begin{figure}[htp!]
    \centering
    \includegraphics[width=7.77cm]{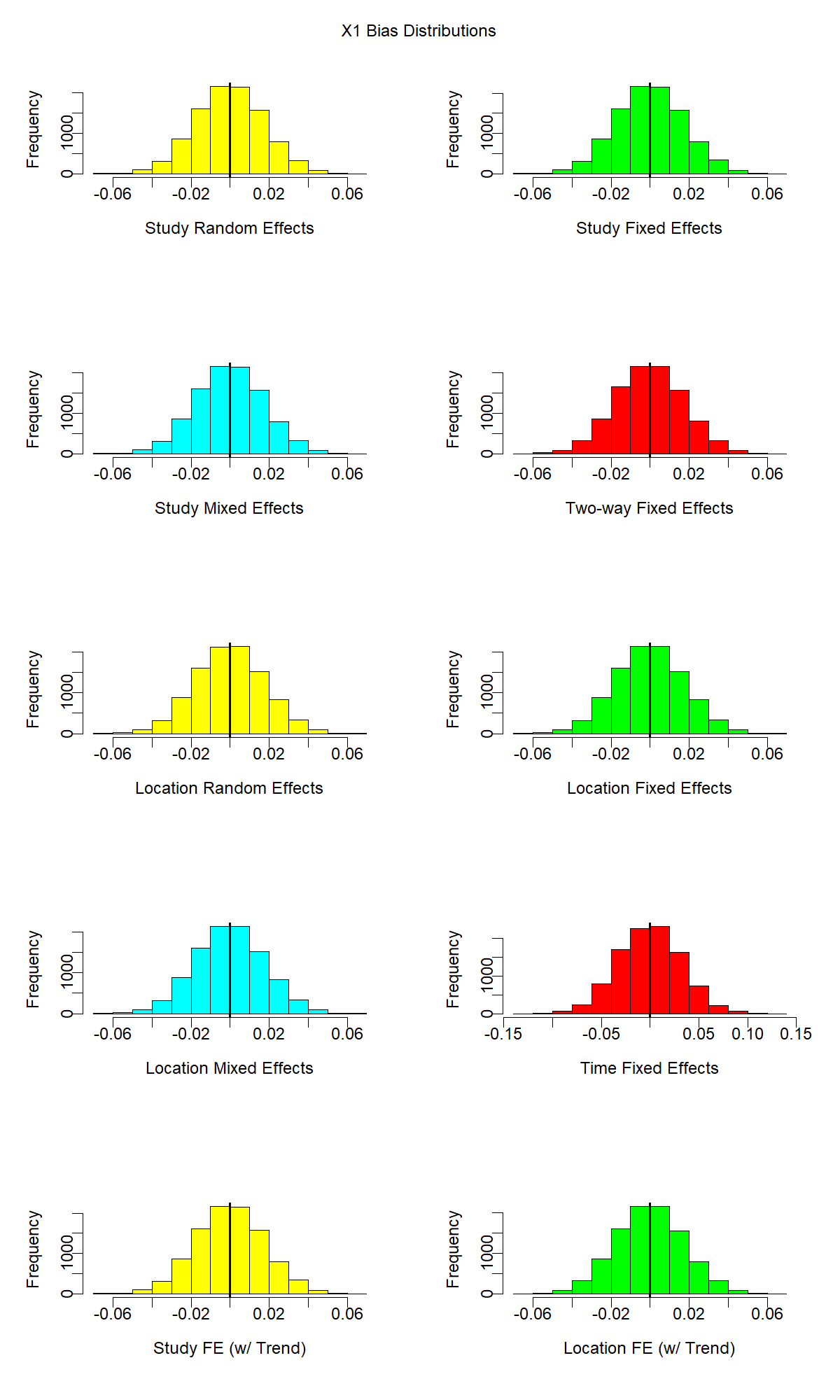}
    \includegraphics[width=7.77cm]{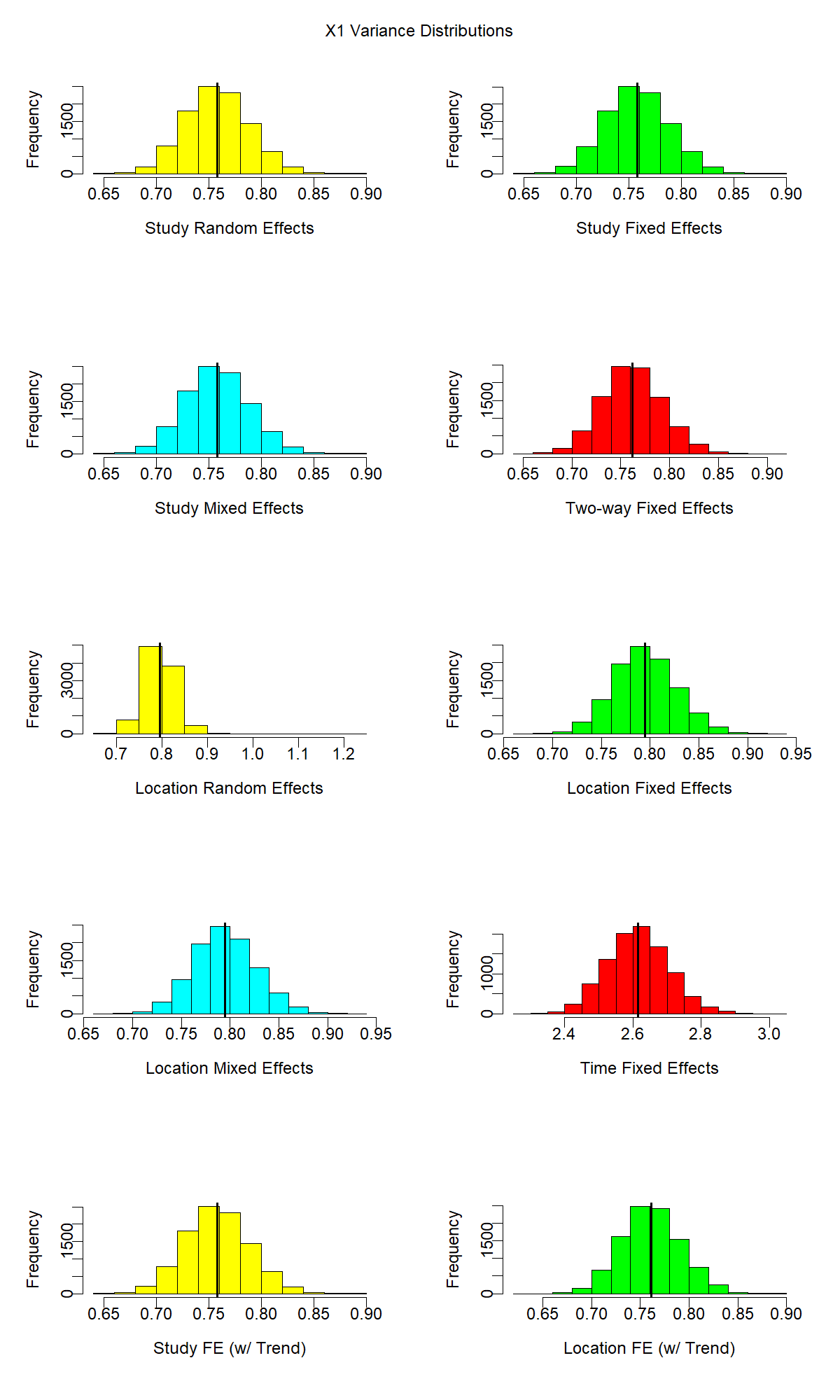}
    \caption{Sample X1 Bias \& Variance Distributions: The estimators bias and variance are respectively shown left to right. This is for when there are less locations (Case 1) and can be compared with the figure below which is for when there are more locations (Case 5). Both show $n = 100$ for 1 covariate.}
    \label{resultFigure8}
\end{figure}
\begin{figure}[htp!]
    \centering
    \includegraphics[width=7.77cm]{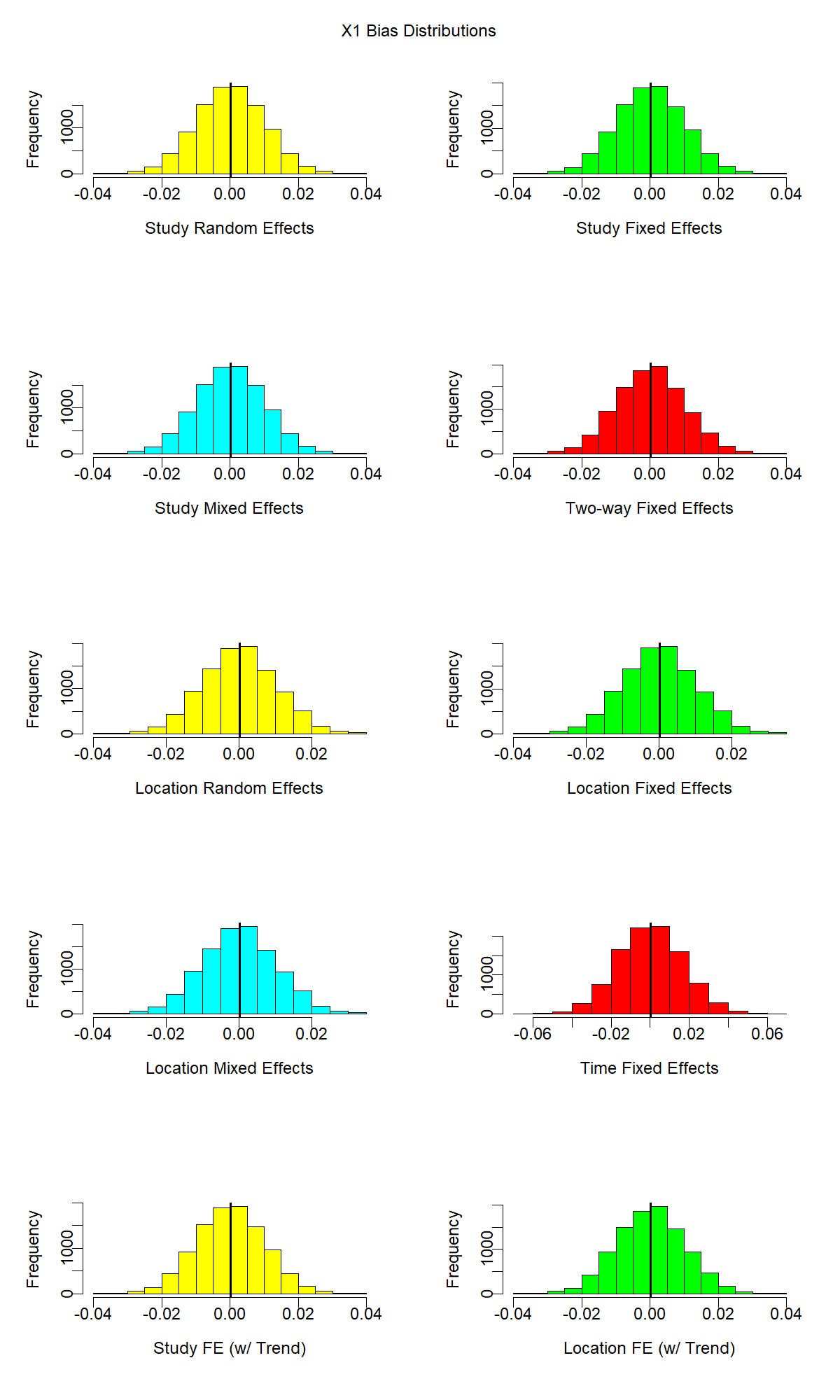}
    \includegraphics[width=7.77cm]{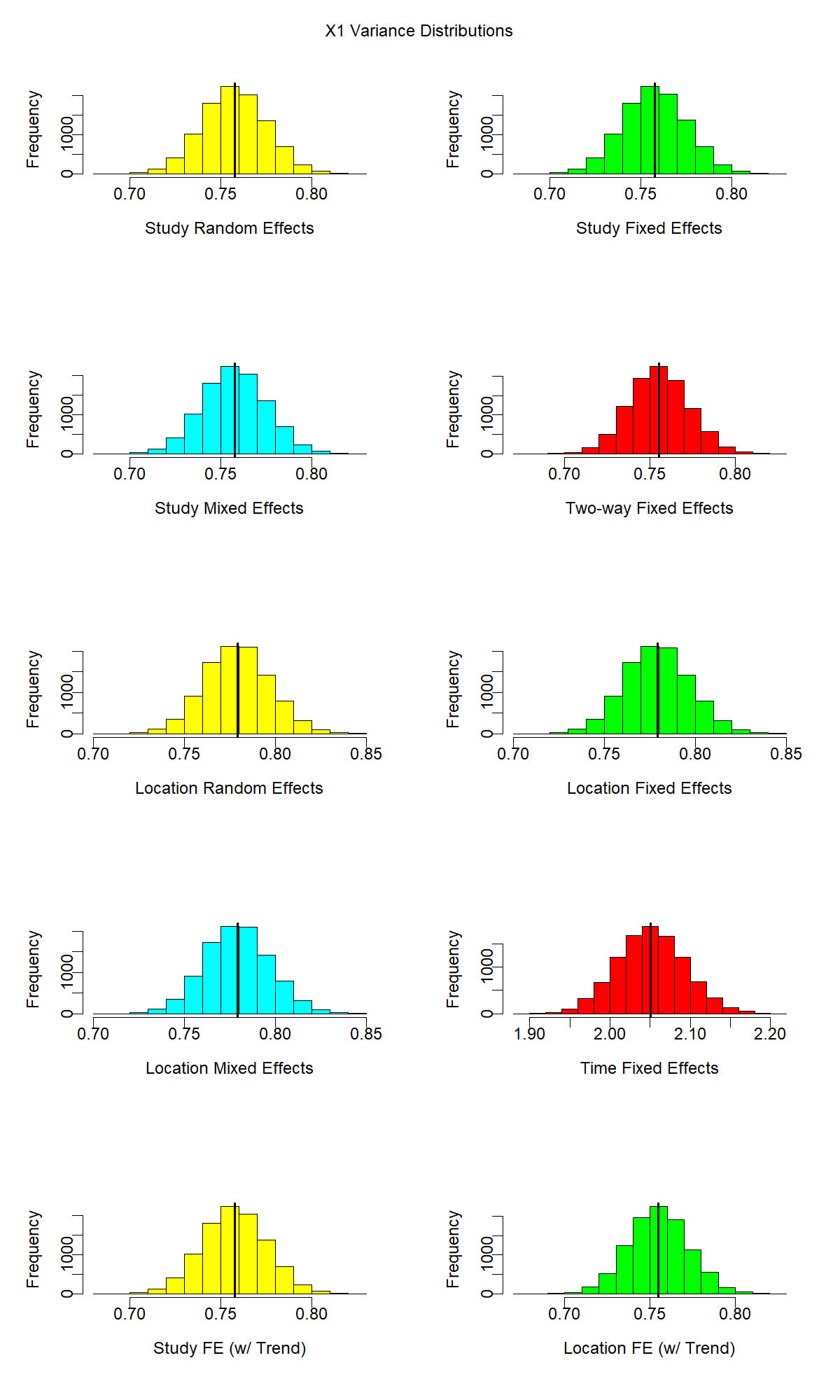}
    \caption{Sample X1 Bias \& Variance Distributions: The estimators bias and variance are respectively shown left to right. This is for when there are more locations (Case 5) and can be compared with the figure above which is for when there are less locations (Case 1). Both show $n = 100$ for 1 covariate.}
    \label{resultFigure8.5}
\end{figure}

\clearpage

\section*{Appendix E - Result Tables}\label{appE}

\begin{table}[hbt!] 
    \centering
    \caption{Heterogeneity Effect on Joint Heterogeneity Models: Shows how changes in time or location heterogeneity changes resulting performance criteria in simulation ($-$ means no change)}
    \begin{tabular}{|l|c|c|c|c|}
        \hline
        & & Power & Estimator Bias & Trend Accuracy\\
        \hline
        $\uparrow$ in Time & $ME_s$ & $-$ & $-$ & \\
        Heterogeneity & $RE_s$ & $-$ & $-$ & \\
        & $FE_s$ & $-$ & $-$ &  \\
        & $FE_{lt}$ & $\downarrow$ & $\uparrow$ & \\
        & $FE_{l,Trend}$ & $\downarrow$ & $\uparrow$ & $\downarrow$ \\
        & $FE_{s,Trend}$ & $-$ & $-$ & $-$ \\
        \hline
        $\downarrow$ in Time & $ME_s$ & $-$ & $-$ & \\
        Heterogeneity & $RE_s$ & $-$ & $-$ & \\
        & $FE_s$ & $-$ & $-$ & \\
        & $FE_{lt}$ & $\uparrow$ & $\downarrow$ & \\
        & $FE_{l,Trend}$ & $\uparrow$ & $\downarrow$ & $\uparrow$ \\
        & $FE_{s,Trend}$ & $-$ & $-$ & $-$ \\
        \hline
        $\uparrow$ in Location & $ME_s$ & $-$ & $-$ &  \\
        Heterogeneity & $RE_s$ & $-$ & $-$ &   \\ 
        & $FE_s$ & $-$ & $-$ &  \\
        & $FE_{lt}$ & $-$ & $-$ &  \\
        & $FE_{l,Trend}$ & $-$ & $-$ & $-$ \\
        & $FE_{s,Trend}$ & $-$ & $-$ & $\downarrow$ \\
        \hline
        $\downarrow$ in Location & $ME_s$ & $-$ & $-$ & \\
        Heterogeneity & $RE_s$ & $-$ & $-$ &  \\
        & $FE_s$ & $-$ & $-$ &  \\
        & $FE_{lt}$ & $-$ & $-$ &  \\
        & $FE_{l,Trend}$ & $-$ & $-$ & $-$ \\
        & $FE_{s,Trend}$ & $-$ & $-$ & $\uparrow$ \\
        \hline
        $\uparrow$ in Locations & $ME_s$ & $\uparrow$ & $-$ & \\
         & $RE_s$ & $\uparrow$ & $-$ &  \\ 
        & $FE_s$ & $\uparrow$ & $-$ &  \\
        & $FE_{lt}$ & $\uparrow$ & $-$ & \\
        & $FE_{l,Trend}$ & $\uparrow$ & $-$ &  $\downarrow$\\
        & $FE_{s,Trend}$ & $\uparrow$ & $-$ & $-$\\
        \hline
        $\downarrow$ in Locations & $ME_s$ & $\downarrow$ & $-$ &  \\
        & $RE_s$ & $\downarrow$ & $-$ & \\
        & $FE_s$ & $\downarrow$ & $-$ &  \\
        & $FE_{lt}$ & $\downarrow$ & $-$ &   \\
        & $FE_{l,Trend}$ & $\downarrow$ & $-$ & $\uparrow$ \\
        & $FE_{s,Trend}$ & $\downarrow$ & $-$ & $-$ \\
        \hline
    \end{tabular}
    \label{table4}
\end{table}
\newpage

\begin{table}[hbt!] 
    \centering
    \caption{Heterogeneity Effect on Non-Joint Heterogeneity Models: Shows how changes in time or location heterogeneity changes resulting performance criteria in simulation ($-$ means no change)}
    \begin{tabular}{|l|c|c|c|}
        \hline
        & & Power & Estimator Bias \\
        \hline
        $\uparrow$ in Time & $ME_l$ & $\downarrow$ & $\uparrow$\\
        Heterogeneity & $RE_l$ & $\downarrow$ & $\uparrow$\\
        & $FE_l$ & $\downarrow$  & $\uparrow$ \\
        & $FE_t$ & $-$ & $-$ \\
        \hline
        $\downarrow$ in Time & $ME_l$ & $\uparrow$ &  $\downarrow$\\
        Heterogeneity & $RE_l$ & $\uparrow$ & $\downarrow$\\
        & $FE_l$ & $\uparrow$ & $\downarrow$ \\
        & $FE_t$ & $-$ & $-$ \\
        \hline
        $\uparrow$ in Location & $ME_l$ & $-$ & $-$ \\
        Heterogeneity & $RE_l$ & $-$ & $-$ \\
        & $FE_l$ & $-$ & $-$ \\
        & $FE_t$ & $\downarrow$ & $\uparrow$ \\
        \hline
        $-$ in Location & $ME_l$ & $-$ & $-$ \\
        Heterogeneity & $RE_l$ & $-$ & $-$ \\
        & $FE_l$ & $-$ & $-$ \\
        & $FE_t$ & $\uparrow$ & $\downarrow$ \\
        \hline
        $\uparrow$ in Locations & $ME_l$ & $\uparrow$ & $-$ \\
        & $RE_l$ & $\uparrow$ & $-$ \\
        & $FE_l$ & $\uparrow$ & $-$ \\
        & $FE_t$ & $\uparrow$ & $-$ \\
        \hline
        $\downarrow$ in Locations & $ME_l$ & $\downarrow$ & $-$ \\
        & $RE_l$ & $\downarrow$ & $-$ \\
        & $FE_l$ & $\downarrow$ & $-$ \\
        & $FE_t$ & $\downarrow$ & $-$ \\
        \hline
    \end{tabular}
    \label{table5}
\end{table}

\newpage

\section*{Appendix F - Detailed Standard Meta-Analysis Methodology}\label{appF}

\subsection*{Study-level Fixed Effects}

\noindent A fixed effects model allows a regression to account for individual-specific effects (also known as cross-section level effects) such as each study, $s$, or location, $l$ \citep{wooldridge_econometric_2010,cameron2005,baltagi2005}. If there are individual-specific effects, but fixed effects are not used, then the model faces the key issue of an omitted variable bias. One way fixed effects can be implemented is with a standard regression that models each cross-section as a dummy variable, as long as one of the cross-sections is removed to avoid the dummy variable trap \citep{wooldridge_econometric_2010,cameron2005,baltagi2005}. Another way that fixed effects can be implemented is by averaging at the study-level. The fixed effects specification would be most appropriate when inferences are only made within each grouping (not between groupings) and the random effects assumption does not hold. These constraints and underlying assumptions will be reviewed later in this section, but first let us review the framework of a fixed effects model. A basic meta-analysis using fixed effects uses the following framework in Equation \ref{EQ0.3} below.

\begin{equation}
y_{s}=\theta+\varepsilon_s, \quad \varepsilon_{s} \sim N(0,\sigma_s^2)
    \label{EQ0.3}
\end{equation}

\noindent The above calculation assumes all studies estimate the same true effect size $\theta$ and the observed effect sizes ($y_s$) differ only due to within-study sampling error. $y_s$ is the observed effect size per study $s$, $\theta$ is the common true effect size, $\varepsilon_s$ is the sampling error, and $\sigma_s^2$ is the within-study variance. Equation \ref{EQ0.4} below shows how to calculate the estimator of $\theta$.

\begin{equation}
    \hat{\theta}_{FE}=\frac{\sum_{s=1}^k w_sy_s}{\sum_{s=1}^k w_s}
    \label{EQ0.4}
\end{equation}

\noindent The estimator for $\theta$ is calculated above with inverse-variance weights $w_s=\frac{1}{\sigma_s^2}$ as seen in Equation \ref{EQ0.4} below. This manuscript will use a meta-regression framework from statistics and econometrics seen below in Equation \ref{EQ1}.
\par

\begin{equation}
y_{s}=\mathbf{x}_s^\top\boldsymbol{\beta}+\boldsymbol{\alpha}_{s,FE}+\varepsilon_{s}, \quad \varepsilon_{s} \sim N(0,\sigma_s^2)
    \label{EQ1}
\end{equation}
\par 

\noindent In the above calculation, individual studies are indexed by $s=(1,...,S),$, $y_s$ is a $(s \times 1)$ vector for the variable of interest, $\mathbf{x}_s^\top=(1,x_{s1},x_{s2},...,,x_{sp})$ represents a $(s \times p)$ vector of covariates per study $s$, $\boldsymbol{\beta}=(\beta_0,\beta_1,...,\beta_p)$ represents a $(p \times 1)$ vector of effect sizes (i.e. regression coefficients), $\varepsilon_s$ is a $(s \times 1)$ vector for the error term, and $\boldsymbol{\alpha}_{s,FE}$ denotes the fixed effect where $\alpha_{s,FE}=\gamma_sD_s$ such that $D_s$ is a vector of dummy variables per study and $\gamma_s$ is a vector of fixed effects for study $s$ (which represents the difference in effect size between each study $s$). The parameter is estimated by using the weights of each group (in this case each study) as seen in Equation \ref{EQ2} below.

\begin{equation}
    \hat{\beta}=[\sum_{s=1}^S(x_{s}^\top w_{s}x_{s})]^{-1}\sum_{s=1}^S(x_{s}^\top w_{s}y_{s})
    \label{EQ2}
\end{equation}

\noindent To calculate the above equation, $\hat{\beta}$ is the estimate of $\beta$, and $N$ denotes all of the observations in the data \citep{cameron2005}. Note that this is equivalent to the typical WLS syntax in Equation \ref{EQ1.5}, but this equation shows the more detail with the grouping. The standard error of the estimated parameter is calculated by using the standard deviation of each group as seen in Equation \ref{EQ3} below 

\begin{equation}
    SE(\hat{\beta})=\sqrt{\left[\sum^S_{s=1}\frac{1}{\sigma^2_s}x_s^\top x_s\right]^{-1}_{jj}}
    \label{EQ3}
\end{equation}\newline

\noindent To calculate the above standard error estimate $SE(\hat{\beta})$, $\sigma^2_s$ denotes the study-level population variance \citep{riceIslas2018}. The subscript $jj$ refers to the $j-th$ diagonal elements of the variance-covariance matrix of the estimator, this syntax is used when there are multiple covariates. \newline
\par


\noindent This specification would be most appropriate when the random effects assumption does not hold, and heterogeneity is present at the study level. Note that there are assumptions and limitations to using fixed effects. First, fixed effects are designed to make inferences precisely on the sample of studies in the meta-analysis \citep{field01}. Unless the studies in the meta-analysis are representative of studies outside of the meta-analysis, they cannot speak to outside studies in the population. Second, fixed effects control any heterogeneity by modeling variation within each group which ignores variation between groups, so inferences cannot be made on the variation between groups \citep{altoe24}. Study-level fixed effects that are used in meta-analysess assume that by controlling for the variation of each study, the variation across time periods and between locations will implicitly be controlled for. But, this won’t capture if two studies share a time period or a location. This can be valid for certain experiment types such as identical experiments where the subjects and study focus do not change over time, but this assumed homogeneity across time and/or locations needs to be tested. This raises concern that study-level fixed effects may not fully capture time and location effects, which is concerning since these effects are likely present when grouping studies for a meta-analysis.\newline
\par

\subsection*{Study-level Random Effects}




\noindent In contrast, what if individual-specific effects are completely uncorrelated with each of the explanatory variables? This is unlikely in practice, but when the random effects assumption holds, a random effects model should be used instead of a fixed effects model \citep{wooldridge_econometric_2010,clark15,baltagi2005,hausman81,mundlak78}. Empirically the unlikelihood of the random effects assumption holding has been proven in a variety of ways. One source illustrates this with wage studies to show unobserved characteristics (where data is unavailable like natural ability) might influence wages but are completely uncorrelated with any explanatory variables like experience or education \citep{wooldridge_econometric_2010}. Another source shows how the random effects assumption does not hold in majority of empirical studies in a variety of economic fields including labor, health and education \citep{baltagi2005}. This has also been shown in a simulation study which generated simulation data that simulated both short and long panel data typical of social science data \citep{clark15}. These constraints and underlying assumptions will be reviewed later in this section, but first let us review the framework of a random effects model. \newline
\par

\noindent A standard meta-analysis most commonly uses the study-level random effects model which has the following framework. Suppose you have $S$ experimental studies where a similar but not necessarily identical experiment is performed in a variety of geographic locations and times. Note that individual studies are indexed by $s=(1,...,S)$. A basic meta-analysis using random effects accounts for between-study heterogeneity by allowing each study to have its own true effect size, drawn from a distribution centered at the overall mean effect $\mu$. It uses the following framework below in Equation \ref{EQ4.4}. 

\begin{equation}
    y_s = \theta_s + \varepsilon_s, \quad \theta_s = \mu + u_s, \quad u_s \sim N(0, \tau^2)
    \label{EQ4.4}
\end{equation}

$$\Rightarrow \quad y_s = \mu + u_s + \varepsilon_s, \quad u_s \sim N(0, \tau^2), \quad \varepsilon_s \sim N(0, \sigma_i^2)$$

\noindent The above equation includes $y_s$ is the observed effect size per study $s$, $\theta_s$ is the common true effect size, $\varepsilon_s$ is the sampling error, and $\sigma_s^2$ is the within-study variance. Equation \ref{EQ4.5} below shows how to calculate the estimator for $\mu$.

\begin{equation}
    \hat{\mu}_{\text{RE}} = \frac{\sum_{s=1}^k w_s^* y_s}{\sum_{s=1}^k w_s^*}
    \label{EQ4.5}
\end{equation}

\noindent The estimator for $\mu$ is calculated with random-effect weights $w_s=\frac{1}{\sigma_s^2+\tau^2}$. This manuscript will use a meta-regression framework from statistics and econometrics seen below in Equation \ref{EQ5}. 

\begin{equation}
y_{s}=\mathbf{x}_s^\top\boldsymbol{\beta}+\boldsymbol{\alpha}_{s,RE}+\varepsilon_{s}, \quad \varepsilon_{s} \sim N(0,\sigma_s^2)
    \label{EQ5}
\end{equation}

\noindent Individual studies are indexed by $s=(1,…,S)$, each subject's variable of interest is a $(s \times 1)$ vector $y_s$, $\mathbf{x}_s^\top=(1,x_{s1},x_{s2},...,,x_{sp})$ represents a $(s \times p)$ vector of covariates per study $s$, and $\boldsymbol{\beta}=(\beta_0,\beta_1,...,\beta_p)$ represents a $(p \times 1)$ vector of effect sizes (i.e. regression coefficients). The random effect, $\boldsymbol{\alpha}_{s,RE}$, is drawn from some distribution $(0,\sigma_\alpha^2)$ and the subject-level error term, $\varepsilon_s$, is drawn from $(0,\sigma_\varepsilon^2)$. This specification would be appropriate if each experimental study were viewed as a random draw (or a representative sample) of the larger population \citep{wooldridge_econometric_2010,cameron2005}. For this method, the parameter estimation and estimation strategy is similar to study-level fixed effects in Appendix G (please see fixed effects in Appendix G for further details). Like in Equation \ref{EQ1}, $\hat{\beta}$ is the estimate of $\beta$, $\bar{x}_s$ is the study-level average of the covariate, $\bar{y}_s$ is the study-level average of the dependent variable, and $N$ denotes all of the observations in the data. Equation \ref{EQ6} below shows the key difference.

\begin{equation}
    \hat{\beta}=[\sum_{i=1}^N\sum_{s=1}^S(x_{is}-\hat{\lambda}\bar{x}_s)(x_{is}-\hat{\lambda}\bar{x}_s)^\top]^{-1}\sum_{i=1}^N\sum_{s=1}^S(x_{is}-\hat{\lambda}\bar{x}_s)(y_{is}-\hat{\lambda}\bar{y}_s)
    \label{EQ6}
\end{equation}

$$\hat{\lambda}=\frac{1-\sigma_\varepsilon}{(S\sigma^2_{\alpha}+\sigma^2_{\varepsilon})^{\frac{1}{2}}}$$\newline

\noindent The above equation shows that the data is quasi-demeaned with $\hat{\lambda}$ by using the study-level population variance $\sigma^2_s$ and the population-level variance of the random effect $\sigma^2_\alpha$ \citep{wooldridge_econometric_2010,cameron2005}. The calculation of the standard error of the estimated parameter in Equation \ref{EQ7} below is also similar. 

\begin{equation}
    SE(\hat{\beta})=\sqrt{\frac{1}{\sum^S_{s=1}\frac{1}{\sigma^2_s+\sigma^2_\alpha}}}
    \label{EQ7}
\end{equation}

\noindent In the above calculation $SE(\hat{\beta})$ is the standard error estimate, and $\sigma^2_s$ is the study-level population variance. But, Equation \ref{EQ7} shows the addition of $\sigma^2_\alpha$, the population-level variance of the random effect \citep{riceIslas2018}.\newline
\par

\noindent This specification would be most appropriate when the random effects assumption does hold, heterogeneity is present at the study level, and subjects are either random draws or representative of the full population. Note that there are assumptions and limitations to using random effects compared to fixed effects. First, although the random effects model may be appropriate for experimental data that is randomly sampled, even properly designed experiments are vulnerable to at least some of the subjects behaving in ways that match a variable of interest, so caution should be used. Second, study-level variation captured from the random effects parameter is assumed to implicitly capture the variation across time periods and between locations, but this will not be able to capture precise heterogeneity (such as if two studies share a time period or a location). Third, random effects are designed to make inferences on the population instead of the sample, while fixed effects are designed to make inferences precisely on the sample of studies in the meta-analysis \citep{field01}. Fourth, random effects try to estimate the heterogeneity by assuming a hypothetical distribution distribution, while fixed effects control any heterogeneity by modeling variation within each group which ignores variation between groups \citep{altoe24}. These limitations highlight why one model is not dominantly used over the other, their differences in design can lead to different inferences and handling of heterogeneity. \newline
\par

\subsection*{Study-level Mixed Effects (GLMM)}
As introduced in Section \ref{section3}, fixed or random effect models can be seen in their parent model GLMM, which essentially combines fixed and random effects. Note that GLMM is often referred to as a mixed-effects model or a mixed model. This manuscript will use a meta-regression framework from statistics and econometrics seen in Equation \ref{EQ0.2} where $y_s$ denotes a $(s \times 1)$ vector for the dependent variable, $\mathbf{x}_s^\top=(1,x_{s1},x_{s2},...,,x_{sp})$ representing a $(s \times p)$ vector of covariates per study $s$, $\boldsymbol{\beta}=(\beta_0,\beta_1,...,\beta_p)$ representing a $(p \times 1)$ vector of regression coefficients (i.e. effect sizes), $\boldsymbol{\alpha}_{s,FE}$ denotes the fixed-effects vector, $\boldsymbol{\alpha}_{s,RE}$ denotes the random-effects vector and $\varepsilon_s$ denotes the error term vector of $(s \times 1)$ \citep{viech10}. The standard error of the estimated parameter is calculated by using the standard deviation of each group as seen in Equation \ref{EQ6} for study-level random effects in Appendix G above where $SE(\hat{\beta})$ is the standard error estimate, and $\sigma^2_s$ is the study-level population variance \citep{riceIslas2018}. There is also $\sigma^2_\alpha$, which is the population-level variance of the random effect as seen in Appendix G \citep{riceIslas2018}. These calculations are essentially identical to those for study-level random effects in Appendix G because a mixed effect model is comprised of both random effects as well as fixed effects, and fixed effects provide the foundation to calculate random effects.\newline
\par



\newpage

\section*{Appendix G - Detailed Standard Meta-Analysis Methodology}\label{appG}

\subsection*{Location-level Effects: Fixed, Random, and Mixed Effects}

\noindent \textbf{Fixed Effects}\newline
\noindent Equation \ref{EQ8} below represents location-fixed effects.

\begin{equation}
y_{l}=\mathbf{x}_l^\top\boldsymbol{\beta}+\boldsymbol{\alpha}_{l,FE}+\varepsilon_{l}, \quad \varepsilon_{l} \sim N(0,\sigma_l^2)
    \label{EQ8}
\end{equation}\newline

\noindent For location level fixed-effects, each location is indexed by $l=1,...,L$, $y_l$ is a $(l \times 1)$ vector for the variable of interest, $\mathbf{x}_l^\top=(1,x_{l1},x_{l2},...,,x_{lp})$ represents a $(l \times p)$ vector of covariates per location $l$, $\boldsymbol{\beta}=(\beta_0,\beta_1,...,\beta_p)$ represents a $(p \times 1)$ vector of effect sizes (i.e. regression coefficients), and $\varepsilon_l$ is a $(l \times 1)$ vector for the error term. Note that $\boldsymbol{\alpha}_{l}$ denotes the fixed effect where $\boldsymbol{\alpha}_{l,FE}=\gamma_l D_l$ such that $D_l$ is a vector of dummy variables per location and $\gamma_l$ is a vector of fixed effects for location $l$ (which represents the difference in effect size between each location $l$). The parameter is estimated using the weights of each group (in this case, each location) as seen in Equation \ref{EQ9} below.

\begin{equation}
    \hat{\beta}=[\sum_{i=1}^S\sum_{l=1}^L(x_{il}^\top w_{il}x_{il})]^{-1}\sum_{i=1}^S\sum_{l=1}^L(x_{il}^\top w_{il}y_{il})
    \label{EQ9}
\end{equation}

\noindent In estimating the parameter above, $\hat{\beta}$ is the estimate of $\beta$ and $S$ denotes all studies in the meta-regression \citep{cameron2005}. Note that this is equivalent to the typical WLS syntax in Equation \ref{EQ1.5}, but this equation shows more detail with the grouping. The standard error of the estimated parameter is calculated using the standard deviation of each group as seen in Equation \ref{EQ10} below.

\begin{equation}
    SE(\hat{\beta})=\sqrt{\left[\sum^L_{l=1}\frac{1}{\sigma^2_l}x_l^\top x_l\right]^{-1}_{jj}}
    \label{EQ10}
\end{equation}

\noindent To calculate the above standard error estimate $SE(\hat{\beta})$, $\sigma^2_l$ denotes the location-level population variance \citep{riceIslas2018}. The subscript $jj$ refers to the $j-th$ diagonal elements of the variance-covariance matrix of the estimator, this syntax is used when there are multiple covariates. \newline
\par

\noindent To estimate the standard error above $SE(\hat{\beta})$, the location-level population variance $\sigma^2_l$ is needed \citep{riceIslas2018}. Grouping by location, $l$, instead of by study, $s$, is a key difference since variation in locations is now explicitly controlled for, instead of being implicitly controlled for by the grouping of studies. The subscript $jj$ refers to the $j-th$ diagonal elements of the variance-covariance matrix of the estimator, this syntax is used when there are multiple covariates.\newline
\par

\noindent \textbf{Random Effects}\newline
\noindent The simulation will also test location-level random effects, which is similar, but has some slight differences. This also uses Equation \ref{EQ8}, but the fixed effect is replaced with the random effect $\boldsymbol{\alpha}_{l,RE}$. This random effect is drawn from $N(0,\sigma_\alpha^2)$ and the error term $\varepsilon_l$ is $N(0,\sigma_\varepsilon^2)$ distributed. The estimation of parameters and their standard errors is the same as it was for study-level random effects in Appendix F, just the subscripts change from study-level, $s$ to location-level, $l$ as seen below in Equation \ref{EQ10.1} \& \ref{EQ10.2}. 

\begin{equation}
    \hat{\beta}=[\sum_{i=1}^S\sum_{l=1}^L(x_{il}-\hat{\lambda}\bar{x}_l)(x_{il}-\hat{\lambda}\bar{x}_l)^\top]^{-1}\sum_{i=1}^S\sum_{l=1}^L(x_{il}-\hat{\lambda}\bar{x}_l)(y_{il}-\hat{\lambda}\bar{y}_l)
    \label{EQ10.1}
\end{equation}

$$\hat{\lambda}=\frac{1-\sigma_\varepsilon}{(L\sigma^2_{\alpha}+\sigma^2_{\varepsilon})^{\frac{1}{2}}}$$\newline

\begin{equation}
    SE(\hat{\beta})=\sqrt{\frac{1}{\sum^L_{l=1}\frac{1}{\sigma^2_l+\sigma^2_\alpha}}}
    \label{EQ10.2}
\end{equation}\newline

\noindent \textbf{Mixed Effects}\newline
\noindent As introduced in Section \ref{section3}, a mixed effects model (sometimes referred to as a GLMM) simply combines fixed and random effects together. This manuscript will use a meta-regression framework from statistics and econometrics seen in Equation \ref{EQ10.3} below

\begin{equation}
y_{l}=\mathbf{x}_l^\top\boldsymbol{\beta}+\boldsymbol{\alpha}_{l,FE}+\boldsymbol{\alpha}_{l,RE}+\varepsilon_{l}, \quad \varepsilon_{l} \sim N(0,\sigma_l^2)
    \label{EQ10.3}
\end{equation}

\noindent In the equation above, $y_l$ denotes a $(l \times 1)$ vector for the dependent variable, $\mathbf{x}_l^\top=(1,x_{l1},x_{l2},...,,x_{lp})$ representing a $(l \times p)$ vector of covariates per location $l$, $\boldsymbol{\beta}=(\beta_0,\beta_1,...,\beta_p)$ representing a $(p \times 1)$ vector of regression coefficients (i.e. effect sizes), $\boldsymbol{\alpha}_{l,FE}$ denotes the fixed effects vector, $\boldsymbol{\alpha}_{l,RE}$ denotes the random effects vector, and $\varepsilon_l$ denotes the $(l \times 1)$ error term vector \citep{viech10}. Estimating the parameters and their standard errors will be the same as it is for location-level random effects for the same reasons articulated for study-level mixed effects in Appendix F (please see Equation \ref{EQ10.1} \& \ref{EQ10.2}). This is because a mixed effect model is comprised of both random effects as well as fixed effects, and fixed effects provide the foundation to calculate random effects.\newline
\par

\subsection*{Time-level Fixed Effects}

Estimation of parameters and their standard errors are the same as it was in Section \ref{section4.1} and for study-level fixed effects in Appendix F. The parameter is estimated using the weight of each group (in this case each time) as seen in Equation \ref{EQ12} below.

\begin{equation}
    \hat{\beta}=[\sum_{i=1}^S\sum_{t=1}^T(x_{it}^\top w_{it}x_{it})]^{-1}\sum_{i=1}^S\sum_{t=1}^T(x_{it}^\top w_{it}y_{it})
    \label{EQ12}
\end{equation}

\noindent To estimate the parameter, $\hat{\beta}$ is the estimate of $\beta$ and $S$ denotes all studies in the meta-regression \citep{cameron2005}. Note that this is equivalent to the typical WLS syntax in Equation \ref{EQ1.5}, but this equation shows the more detail with the grouping. The standard error of the estimated parameter is calculated using the standard deviation of each group as seen in Equation \ref{EQ13} below. 

\begin{equation}
    SE(\hat{\beta})=\sqrt{\left[\sum^T_{t=1}\frac{1}{\sigma^2_t}x_t^\top x_t\right]^{-1}_{jj}}
    \label{EQ13}
\end{equation}

\noindent To estimate the standard error $SE(\hat{\beta})$, the time-level population variance is needed $\sigma^2_t$ \citep{riceIslas2018}. Grouping by time, $t$ instead of by study, $s$, is a key difference since variation in time is now explicitly controlled for, instead of being implicitly controlled for by the grouping of studies. The subscript $jj$ refers to the $j-th$ diagonal elements of the variance-covariance matrix of the estimator, this syntax is used when there are multiple covariates.\newline
\par

\subsection*{Combined Location and Time-level Fixed Effects}

The estimation of parameters and their standard errors will be calculated similarly as in Section \ref{section4.1} \& \ref{section4.2} with the same estimation strategy. Both indices need to be paired together since each observation is observed at a specific location and time period. The parameter is estimated using the weight of both groupings (location and time) as seen in Equation \ref{EQ15} below.

\begin{equation}
    \hat{\beta}=[\sum_{i=1}^S\sum_{l=1}^L\sum_{t=1}^T(x_{ilt}^\top w_{ilt}x_{ilt})]^{-1}\sum_{i=1}^S\sum_{l=1}^L\sum_{t=1}^T(x_{ilt}^\top w_{ilt}y_{ilt})
    \label{EQ15}
\end{equation}

\noindent In the above equation, $\hat{\beta}$ is the estimate of $\beta$ and $S$ denotes all studies in the meta-regression \citep{cameron2005}. Note that this is equivalent to the typical WLS syntax in Equation \ref{EQ1.5}, but this equation shows the more detail with the grouping. The standard error of the estimated parameter is calculated using the standard deviation of both groups as seen in Equation \ref{EQ16} below. 

\begin{equation}
    SE(\hat{\beta})=\sqrt{\frac{1}{\sum^L_{l=1}\sum^T_{t=1}\frac{1}{\sigma^2_l+\sigma^2_t}}}
    \label{EQ16}
\end{equation}

\noindent In the above equation, $SE(\hat{\beta})$ is the standard error estimate, $\sigma^2_l$ is the location-level population variance, and $\sigma^2_l$ is the time-level population variance \citep{baltagi2005,cameron2005}.\newline
\par

\newpage

\section*{Appendix H - Details on MSE \& Similar Measures}\label{appH}

\noindent As mentioned in Section \ref{section5.2.2}, a common metric the meta-analysis simulation literature measures bias is with mean squared error (MSE) $\frac{1}{n}\sum^N_{i=1}(\hat{\beta}_i-\beta_i)^2$. But, there are other variations which are rare in this literature, but well circulated in statistics and econometrics literature. These include mean absolute error, mean percent error, or mean absolute percent error (respectively MAE, MPE or MAPE) \citep{gneiting11,west06,timmermann06,maronna19,hyndman21}. Note that estimator bias is used instead of these variations because our simulation uses all the in-sample data to estimate, so these would just calculate model performance instead of the intended bias. While MSE is designed to penalize larger deviations from the true value, MAE is designed without penalty via $\frac{1}{n}\sum^N_{i=1}|\hat{\beta}_i-\beta_i|$. Section \ref{section5.1} shows that our simulation is not designed to have outliers, but if it did, MAE would be more appropriate to use than MSE due to the sensitivity of MSE to outliers \citep{maronna19,muller17,hyndman21,french87}. MSE is scale dependent, so if this is a concern it may be more appropriate to use MPE which is scale independent and measures the percentage difference via $\frac{1}{n}\sum^N_{i=1}\frac{\hat{\beta}_i-\beta_i}{\beta_i}$ \citep{gneiting11,west06,diebold96}. MAPE is simply the absolute value of the percentage difference from MPE which in addition to being scale independent it focuses on the magnitude of prediction error instead of over/under prediction like MPE \citep{gneiting11,west06,timmermann06}.\newline
\par

\end{document}